\documentclass[aps,reprint,superscriptaddress,nofootinbib]{revtex4-2}
\usepackage[utf8]{inputenc}
\usepackage{amsmath}
\usepackage{amssymb}
\usepackage{physics}
\usepackage{bbold}
\usepackage{graphicx}
\usepackage{multirow}
\usepackage{hyperref}
\hypersetup{colorlinks=true, citecolor=blue}
\usepackage{cleveref}

\begin{document}

\title{Classical Splitting
    of Parametrized Quantum Circuits
}

\author{Cenk T\"uys\"uz}
\email{cenk.tueysuez@desy.de}
\affiliation{Deutsches Elektronen-Synchrotron (DESY), Platanenallee 6, 15738
Zeuthen, Germany}
\affiliation{Instit\"ut für Physik, Humboldt-Universit\"at zu Berlin,
Newtonstr. 15, 12489 Berlin, Germany}
\author{Giuseppe Clemente}
\affiliation{Deutsches Elektronen-Synchrotron (DESY), Platanenallee 6, 15738
Zeuthen, Germany}
\author{Arianna Crippa}
\affiliation{Deutsches Elektronen-Synchrotron (DESY), Platanenallee 6, 15738
Zeuthen, Germany}
\affiliation{Instit\"ut für Physik, Humboldt-Universit\"at zu Berlin,
Newtonstr. 15, 12489 Berlin, Germany}
\author{Tobias Hartung}
\affiliation{Department of Mathematical Sciences, 4 West, University of Bath,
Claverton Down, Bath BA2 7AY, UK}
\affiliation{Computation-Based Science and Technology Research Center, The
Cyprus Institute, 20 Kavafi Street, 2121 Nicosia, Cyprus}
\author{Stefan K\"uhn}
\affiliation{Computation-Based Science and Technology Research Center, The
Cyprus Institute, 20 Kavafi Street, 2121 Nicosia, Cyprus}
\author{Karl Jansen}
\affiliation{Deutsches Elektronen-Synchrotron (DESY), Platanenallee 6, 15738
Zeuthen, Germany}

\begin{abstract}
    Barren plateaus appear to be a major obstacle to using variational
    quantum algorithms to simulate large-scale quantum systems or replace 
    traditional machine learning algorithms. They can be caused by multiple
    factors such as expressivity, entanglement, locality of observables, or
    even hardware noise. We propose classical splitting of ans\"atze or
    parametrized quantum circuits to avoid barren plateaus. Classical splitting 
    is realized by  splitting an $N$ qubit ansatz to multiple ans\"atze that consists of
    $\mathcal{O}(\log N)$ qubits. We show that such an ansatz can be used to avoid barren 
    plateaus. We support our results with numerical experiments and perform
    binary classification on classical and quantum datasets. Then, we propose an
    extension of the ansatz that is compatible with variational quantum simulations.
    Finally, we discuss a speed-up for gradient-based optimization and hardware
    implementation, robustness against noise and parallelization, making classical
    splitting an ideal tool for noisy intermediate scale quantum (NISQ) applications.
\end{abstract}

\maketitle

\section{Introduction}

Variational quantum algorithms (VQAs)\cite{cerezo_variational_2021} are
promising tools to solve a wide range of problems, such as finding the ground
state of a given hamiltonian via the variational quantum eigensolver
(VQE)\cite{Peruzzo2014}, solving combinatorial
optimization problems with the quantum approximate optimization
algorithm (QAOA)\cite{farhi_quantum_2014} or solving classification problems
using quantum neural networks \cite{farhi_classification_2018}.

VQAs are suitable for noisy intermediate scale quantum
(NISQ)~\cite{Preskill2018} hardware as they can be implemented with a small number of layers and
gates for simple tasks. However, a scalability problem arises with the increasing
number of qubits, hindering a possible advantage. VQAs rely on a
classical optimization loop that updates the parameters of the ansatz
iteratively until a condition on the cost function is satisfied. Classical
optimizers use the information on the parametrized cost landscape to find the
minimum. The updates on the parameters move the ansatz to a lower
point on the cost surface. In 2018, McClean et al.\ showed that the cost
landscape flattens with the increasing number of qubits, making it exponentially
harder to find the solution for the optimizer~\cite{mcclean_barren_2018}. The
flattening was first observed by looking at the distribution of gradients
across the parameter space, and the problem was named barren plateaus (BPs).
A VQA is said to have a BP if its gradients decay exponentially with
respect to one of its hyper-parameters, such as the number of qubits or layers.

Since the discovery of the BP problem, there has been significant progress that
improved our understanding of what causes BPs and several methods to avoid
them have been proposed. It has been shown that
noise~\cite{wang_noise-induced_2021}, 
entanglement~\cite{ortiz_marrero_entanglement-induced_2021}, and the locality of
the observable~\cite{cerezo_cost_2021} play an essential role for determining
whether an ansatz will exhibit BPs. It has also been shown that the choice of
ansatz (e.g.\ expressivity) of the circuit is one of the decisive factors that impact
BPs~\cite{holmes_connecting_2022}. For instance, the absence of BPs has been shown
for quantum convolutional neural networks
(QCNN)~\cite{cong_quantum_2019,Pesah2020} and tree tensor networks
(TTN)~\cite{Grant2018, zhao_analyzing_2021}. On the other hand, the hardware
efficient ansatz (HEA)~\cite{mcclean_barren_2018,zhao_analyzing_2021,
kandala_hardware-efficient_2017} and matrix product states
(MPS)~\cite{zhao_analyzing_2021} have been shown to have BPs.

One of the essential discoveries showed that BPs are equivalent to cost concentration
and narrow gorges~\cite{arrasmith_equivalence_2021}. This implies that BPs are
not only a result of the exponentially decaying gradient but also of the cost
function itself, and they can be identified by analyzing random points
on the cost surface. As a result, gradient-free
optimizers are also prone to BPs and do not offer a way to circumvent this
problem~\cite{arrasmith_effect_2021}.

Many methods have been suggested to mitigate BPs in the literature. Some of
these methods suggest to use different ans\"atze or cost 
functions~\cite{wu_mitigating_2021, zhang_toward_2021}, determining a better
initial point to start the optimization~\cite{Grant2019,liu_parameter_2021,
rad_surviving_2022,zhang_gaussian_2022}, determining the step size during the
optimization based on the ansatz~\cite{sack_avoiding_2022}, correlating
parameters of the ansatz (e.g., restricting the directions of rotation)~\cite{volkoff_large_2021,holmes_connecting_2022},
or combining multiple methods~\cite{patti_entanglement_2021,
broers_optimization_2021}.

In this work, we propose a novel idea in which we claim that
if any ansatz of $N$ qubits is classically separated to a set of ans\"atze with
$\mathcal{O}(\log N)$ qubits, the new ansatz will not exhibit Barren Plateaus. This
work is not the first proposal in the literature that considers partitioning an 
ansatz. However, our proposal is significantly different. Most work in 
the literature first considers an ansatz and then emulates the result of that
ansatz through many ans\"atze (exponentially many in general) with less number of qubits (which increases the
effective size of quantum simulations) using gate
decompositions, entanglement forging, divide and conquer or other 
methods~\cite{bravyi_trading_2016, peng_simulating_2020, tang_cutqc_2021,perlin_quantum_2021, eddins_doubling_2021, saleem_quantum_2021, fujii_deep_2022, marshall_high_2022, tang_cutqc_2021}. On the other hand, this work proposes using
ans\"atze that are classically split, meaning that there are no two-qubit gate
operations between the subcircuits before splitting. This way, there is no need for gate
decompositions or other computational steps. Our results show that this approach
provides many benefits such as better trainability, robustness  against noise and
faster implementation on NISQ devices.

In the remainder of the paper, we start by giving an analytical illustration of the method in Section~\ref{sec:bp}. Then, we provide numerical evidence for our
claim in Section~\ref{sec:numerical} and extend our results to practical use
cases by comparing binary classification performance of classical splitting for
classical and quantum data. Next, we propose an extension of the classical
splitting ansatz and perform experiments to simulate the ground state of the 
transversal-field ising hamiltonian. Finally, we discuss the advantages of employing 
classical splitting, make comments on future directions in 
Section~\ref{sec:discussion} and give an outlook in Section~\ref{sec:conclusion}.

\section{Avoiding Barren Plateaus}
\label{sec:bp}

Barren plateaus (BPs) can be identified by investigating how the gradients of an
ansatz scale with respect to a parameter. Here, we will start with the notation of
McClean et al.\ and extend it to classical splitting~\cite{mcclean_barren_2018}.
The ansatz is composed of consecutive parametrized ($V$) and non-parametrized
entangling ($W$) layers. We define 
$U_l(\theta_l) = \mbox{exp}(-i\theta_l V_l)$, where $V_l$ is a Hermitian
operator and $W_l$ is a generic unitary operator. Then the ansatz can be
expressed with a multiplication of layers,

\begin{equation}
    U(\boldsymbol{\theta}) = \prod_{l=1}^{L} U_l(\theta_l)W_l.
\end{equation}

Then, for an observable $O$ and input state of $\rho$, the cost is given as

\begin{equation}
    C(\boldsymbol{\theta})
    = \mbox{Tr}[OU(\boldsymbol{\theta})\rho U^\dagger(\boldsymbol{\theta})].
\end{equation}

The ansatz can be separated into two parts to investigate a certain layer, such
that $U_- \equiv \prod_{l=1}^{j-1} U_l(\theta_l)W_l$ and
$U_+ \equiv \prod_{l=j}^{L} U_l(\theta_l)W_l$. Then, the gradient of the
$j^{\rm th}$ parameter can be expressed as

\begin{equation}
    \partial_{j} C(\boldsymbol{\theta})
    = \frac{\partial C(\boldsymbol{\theta})}{\partial{\theta_j}}
    = i\,\mbox{Tr}[[V_j,U_+^\dagger OU_+]U_- \rho U_-^\dagger].
\end{equation}

The expected value of the gradients can be computed using the Haar measure.
Please
see Appendix~\ref{app:bp} for more details on the Haar measure, unitary
t-designs and details of the proofs in this section. If we assume the ansatz
$U(\theta)$ forms a unitary 2-design, then this implies that
$\langle \partial_{k}C(\boldsymbol{\theta}) \rangle
=0$~\cite{mcclean_barren_2018}. Since the average value of the gradients are
centered around zero, the variance of the distribution, which is defined as,

\begin{equation}
    \mbox{Var}[\partial_{k}C(\boldsymbol{\theta})] 
    = \langle (\partial_{k}C(\boldsymbol{\theta}))^2 \rangle
    - \langle \partial_{k}C(\boldsymbol{\theta}) \rangle^2,
\end{equation}

\noindent can inform us about the size of the gradients. The variance of the
gradients of the $j^{\rm th}$ parameter of the ansatz, where $U_-$ and $U_+$ are both
assumed to be unitary 2-designs, and the number of qubits is $N$, is given
as~\cite{mcclean_barren_2018,holmes_connecting_2022},

\begin{equation}
    {\rm Var}[\partial_{j} C(\boldsymbol{\theta})]
    \approx \mathcal{O}\left( \frac{1}{2^{6N}} \right).
    \label{eq:bp}
\end{equation}

This means that for a unitary 2-design the
gradients of the ansatz vanish exponentially with respect to the number of
qubits $N$. Details of this proof is provided in Appendix~\ref{app:bp}. Now, let us
consider the classical splitting (CS) case. We split the ansatz
$U(\boldsymbol{\theta})$ to $k$ many $m$-qubit ans\"atze, where we assume without loss of
generality that $N=k \times m$. Then, we introduce a new notation for each classically split layer, 

\begin{equation}
U_l^i(\theta_l^i) = e^{-i\theta_l^i V_l^i} W_l^i,
\end{equation}

\noindent where index $l$ determines the layer and index $i$ determines which
sub-circuit it belongs to. This notation combines the parametrized and entangling gates under $U_l^i$. Then, the overall CS ansatz can be be expressed as,

\begin{figure*}[!t]
    \centering
    \includegraphics[width=\linewidth]{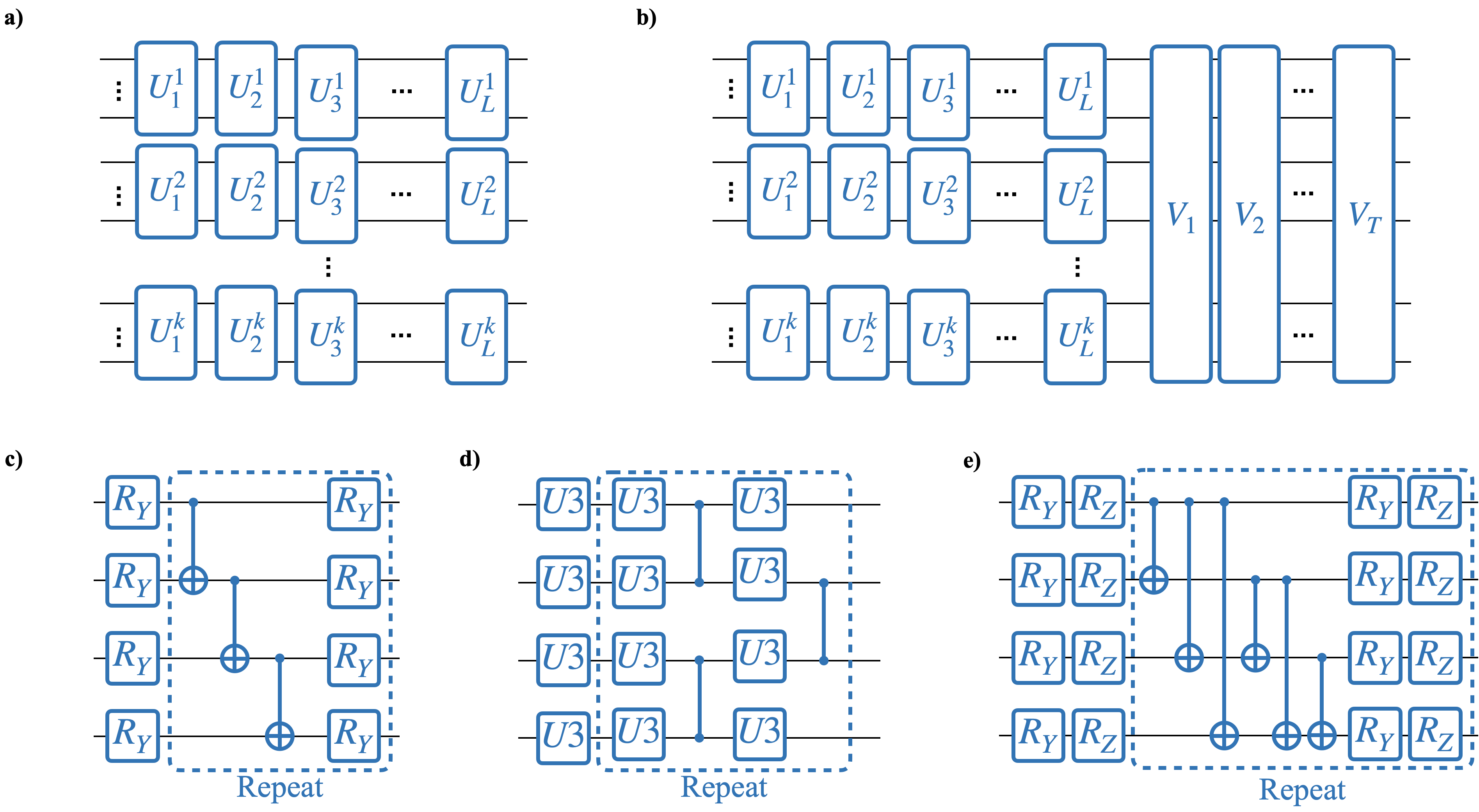}
    \caption{All types of ans\"atze used in this work. (a) An $N$-qubit generic
    ansatz consisting of $L$ layers of the parametrized unitary $U$ are
    separated in to $k=N/m$ many $m$-qubit ans\"atze. This ansatz will be referred
    to as the classically split (CS) ansatz. The standard ansatz 
    can be recovered by setting $m=N$. (b) Extended classically split (ECS) ansatz.
    This is an extension to the CS ansatz. First $L$ layers of
    the ansatz consists of $k=N/m$ many m qubit $U$ blocks. Then, $T$ layers
    of $N$ qubit $V$ layers are applied. (c) A simple ansatz that consists of $R_Y$ rotation gates
    and $CX$ gates connected in a ``ladder" layout. (d) Hardware
    Efficient Ansatz (HEA) that is used to produce the quantum dataset. Parameters of
    the first column of U3 gates are sampled from a uniform distribution
    $\in [-1,1]$, while the rest of the parameters are provided by the
    dataset~\cite{schatzki_entangled_2021}. (e) EfficientSU2 ansatz with
    ``full" entangler layers~\cite{treinish_qiskit_2022}.
        }
    \label{fig:all-ansatze}
\end{figure*}

\begin{equation}
    U(\boldsymbol{\theta}) 
    = \prod_{l=1}^{L} \bigotimes_{i = 1}^{k} U_l^i(\theta_l^i)
    = \bigotimes_{i = 1}^{k} \prod_{l=1}^{L} U_l^i(\theta_l^i)
    = \bigotimes_{i = 1}^{k} U^i(\boldsymbol{\theta^i}).
\end{equation}

The CS ansatz can be seen in Fig.~\ref{fig:all-ansatze}\textcolor{red}{a}.
Next, we will assume the observable and the input state to be classically split, 
such that they both can be expressed as a tensor product of $m$-qubit
observables or states. This assumption restricts our proof to be valid only for
$m$-local quantum states and $m$-local observables. It is important to note here
that we use a definition that is different from the literature throughout the paper.
For this proof, an $m$-local observable is an observable such that there are no
operators that act on overlapping groups of m qubits. A generic $m$-local observable
can be expressed as,

\begin{equation}
    O_{\mbox{m-local}}
    = \sum_{i=1}^{k} O_i \otimes \mathbb{1}_{\bar{i}}
    = \sum_{i=1}^{k} \bigotimes_{j=1}^{k} \left( O_i - \mathbb{1} \right)
    \delta_{i,j}+\mathbb{1},
    \label{eq:obs-m-local}
 \end{equation}

\noindent where $O_i$ is an observable over the qubits
$\{(i-1)m+1,(i-1)m+2,...,im\}$, and $\bar{i}$ represents the remaining $N-m$
qubits. Then, the cost function becomes;

\begin{equation}
    \begin{aligned}
    C(\boldsymbol{\theta})
    &= \sum_{i=1}^{k} \mbox{Tr}[\bigotimes_{j = 1}^{k} 
    \left( \left( O_i - \mathbb{1} \right) \delta_{i,j}+\mathbb{1} \right)
    U^j(\boldsymbol{\theta}^j) \rho_j U^{j\dagger}(\boldsymbol{\theta}^j)]\\
    &= \sum_{i=1}^{k} \prod_{j=1}^{k} \mbox{Tr}[\left(
    \left( O_i - \mathbb{1} \right) \delta_{i,j}+\mathbb{1}\right) 
    U^j(\boldsymbol{\theta}^j) \rho_j U^{j\dagger}(\boldsymbol{\theta}^j)]\\
    &= \sum_{i=1}^{k} \mbox{Tr}[O_i U^i(\boldsymbol{\theta}^i)
    \rho_i U^{i\dagger}(\boldsymbol{\theta}^i)].
    \end{aligned}
\end{equation}

This can be written as a simple sum,

\begin{equation}
C(\boldsymbol{\theta}) = \sum_{i=1}^{k} C^i(\boldsymbol{\theta^i}),
\end{equation}

\noindent where,

\begin{equation}
    C^i(\boldsymbol{\theta^i}) = Tr[O_i U^i(\boldsymbol{\theta^i})
    \rho_i U^{i\dagger}(\boldsymbol{\theta^i})].
    \label{eq:independent-cost}
\end{equation}

Then, the costs of each classically separated circuit are
independent of each other. The gradient of $j^{th}$ parameter of the $i^{th}$
ansatz can be written as, 

\begin{equation}
\begin{aligned}
    \partial_{i,j} C(\boldsymbol{\theta}) &= \partial_{i,j}
    C^i(\boldsymbol{\theta^i})\\ 
    &= \partial_{i,j}( Tr[O_i U^i(\boldsymbol{\theta}^i) \rho_i U^{i\dagger}
    (\boldsymbol{\theta^i})]).
\end{aligned}
\label{eq:cs_grad}
\end{equation}

Now, let us consider each ansatz $U^i(\boldsymbol{\theta}^i)$ to be a unitary
2-design. We want to choose the integer $m$ such that it scales
logarithmically in $N$. Hence, we choose $\beta$ and $\gamma$
appropriately, such that $m = \beta \log_\gamma N$ holds. Then,
if we combine Eq.~\eqref{eq:bp} with Eq.~\eqref{eq:cs_grad}, the
variance of the gradient of $j^{\rm th}$ parameter can be expressed
as 

\begin{equation}
    {\rm Var}[\partial_{j} C(\boldsymbol{\theta})]
    \approx \mathcal{O}\left( \frac{1}{2^{(6m)}} \right)
    =  \mathcal{O}\left(\frac{1}{N^{6\beta \log_\gamma 2}}\right).
    \label{eq:cs-bp}
\end{equation}

Here, the dependence on $i$ or $j$ becomes
irrelevant (a simpler choice for ansatz design would be to choose every new
ansatz to be the same), so it can be dropped for a simpler notation. Similar to
Eq.~\eqref{eq:bp} the variance scales with the dimension of the hilbert space 
(e.g. $\mathcal{O}(2^m)$). Then, the overall expression scales with,
$\mathcal{O}(N^{-6\beta \log_\gamma 2})$, where $\beta$ and $\gamma$ are 
constant (e.g. $\beta=1$ and $\gamma=2$ results in $m=\log_2N$). As a result,
the variance of the classical splitting ansatz scales with
$\mathcal{O}(\mbox{poly(N)}^{-1})$ instead of $\mathcal{O}(\mbox{exp(N)}^{-1})$.
Therefore, a CS ansatz, irrespective of its choice of gates or layout, can be used
without leading to BPs.

\section{Numerical Experiments}
\label{sec:numerical}
In this section, we report results of four numerical experiments. We investigate 
the scaling of gradients under classical splitting by computing variances over many 
samples in Section~\ref{sec:numerical-bp}. Then, we perform three experiments to
observe how classical splitting affects performance of an ansatz. This task by itself
leads to many questions as there are multitudes of metrics that one needs to compare
and as many different problems one can consider. For this purpose, we consider problems
well known in the literature, where trainability of ans\"atze plays a significant role.

First, we perform binary classification on a synthetic 
classical dataset in Section~\ref{sec:classical-training}. The dataset contains two
distributions that are called as classes. The goal is to predict the class of each
sample. We perform the same task for distribution of quantum states in
Section~\ref{sec:ntangled-training}. Then, we give practical remarks in 
Section~\ref{sec:practical}. Finally, we propose an extension to the CS
ansatz and employ it for quantum simulating the ground state of the transverse field
ising hamiltonian in Section~\ref{sec:vqe}.

For the first three experiments (Sections \ref{sec:numerical-bp} to
\ref{sec:ntangled-training}),  we consider the CS ansatz with layers that
consists of $R_Y$ rotation gates and CX entangling gates applied in a ladder
formation for each layer. This layer can be seen in 
Fig.~\ref{fig:all-ansatze}\textcolor{red}{c}. As the observable, we construct
the 1-local observable defined in Eq.~\eqref{eq:1-local-h}, where $Z_i$
represents the Pauli-Z operator applied on the $i^{th}$ qubit and
$\mathbb{1}_{\bar{i}}$ represents the identity operator applied on the rest of
the qubits.

\begin{equation}
   O = \frac{1}{N} \sum_{i=1}^{N} Z_i \otimes \mathbb{1}_{\bar{i}}
   \label{eq:1-local-h}
\end{equation}

\subsection{Barren Plateaus}
\label{sec:numerical-bp}
Barren Plateaus are typically identified by looking at the
variance of the first parameter over a set of random
samples~\cite{mcclean_barren_2018}. Recently, it has been shown that this is
equivalent to looking at the variance of samples from the difference of two
cost values evaluated at different random
points of the parameter space~\cite{arrasmith_equivalence_2021}. Since the
gradient-free optimization methods are also affected from BPs, the values of
the cost become a more inclusive indicator~\cite{arrasmith_effect_2021}. For
this reason, we will report our findings with respect to the cost, rather than
the gradients to draw a broader picture. Results with respect to the gradient
of the first parameter is presented in Appendix~\ref{app:bp-numerics} for the sake
 of completeness.

The experiments were performed using analytical gradients and expectation values, 
assuming a perfect quantum computer and infinite number of measurements, using Pennylane~\cite{bergholm_pennylane_2020} and 
Pytorch~\cite{paszke_pytorch_2019}. Variances are computed over 2000 samples, 
where the values of the parameters are randomly drawn from a uniform
distribution over $[0,2\pi]$.

\begin{figure}[!t]
    \centering
    \includegraphics[width=\linewidth]{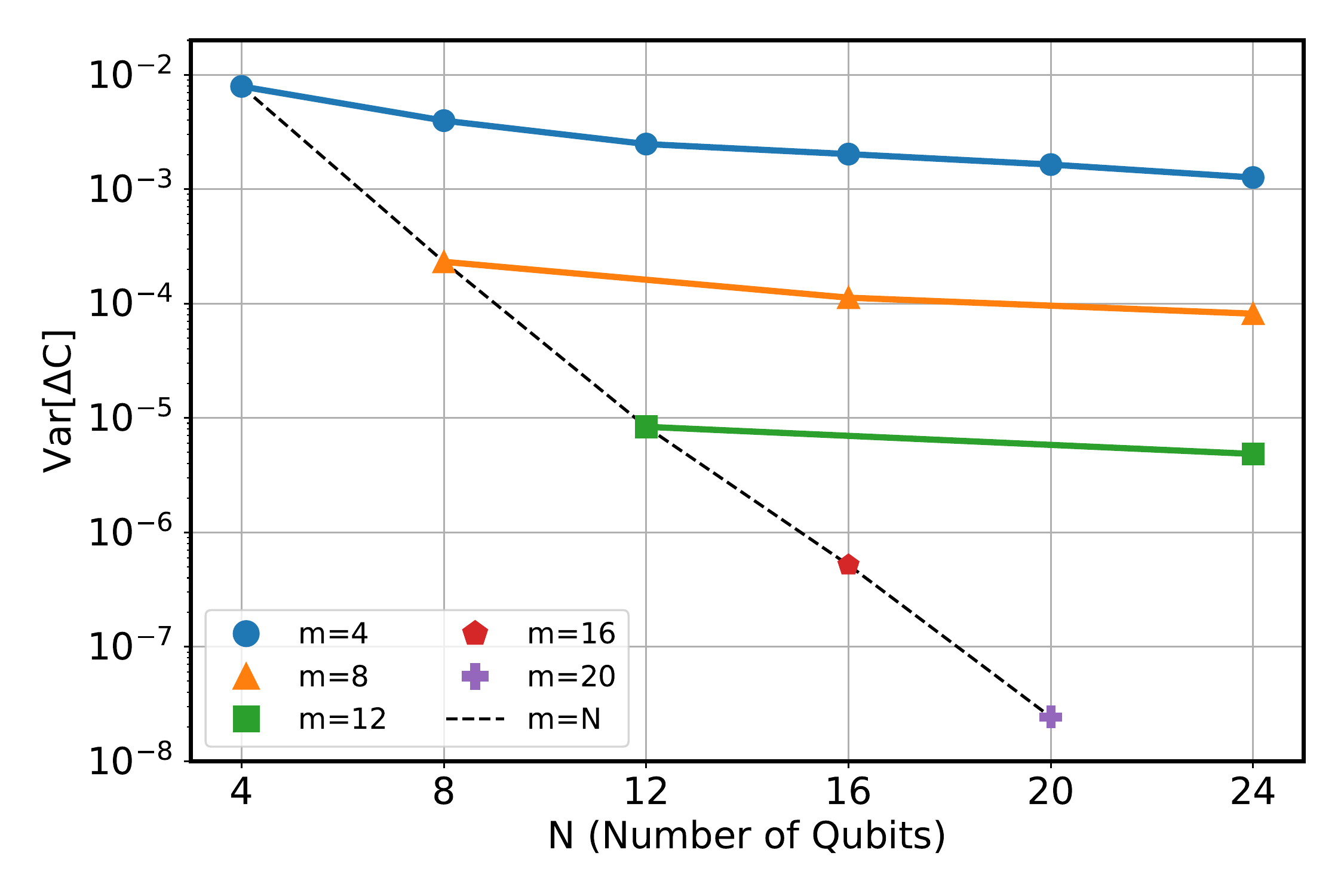}
    \caption{The variance of the change in cost vs. the number of
        qubits for varying values of m. Each color/marker represents a certain
        value of m and data points of the standard ansatz ($m=N$) is plotted with
        a dashed black line.
    }
    \label{fig:bp1}
\end{figure}

We start by presenting the variances over different values of $m$ and $N$ in
Fig.~\ref{fig:bp1}. We fix the number of layers ($L$) to $N$, so that the ansatz
exhibits BPs in the no classical splitting setting ($m=N$). The results indicate that
a constant value of $m$ resolves the exponential behaviour, as expected from
Eq.~\eqref{eq:cs-bp}. Furthermore, it is evident that larger values of $m$ can
allow the ansatz to escape BPs, given that $m$ grows slow enough (e.g.
$\mathcal{O}(\log N)$).

\begin{figure}[!t]
    \centering
    \includegraphics[width=\linewidth]{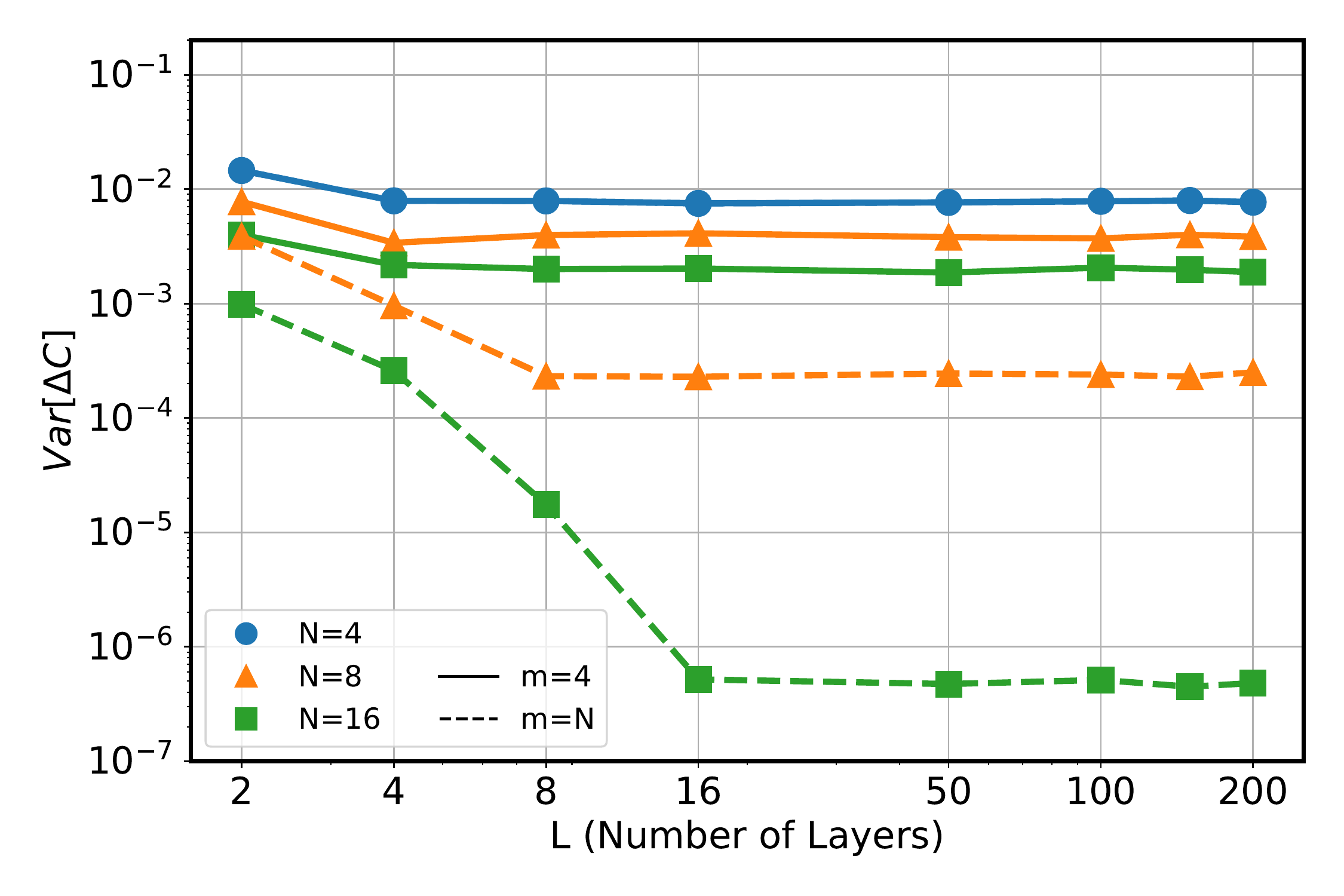}
    \caption{The variance of the change in cost vs. the number of
        layers for $m=4$ (solid lines) and $m=N$ (dashed lines) with varying number
        of qubits.}
    \label{fig:bp2}
\end{figure}

Our theoretical findings illustrate that the classical splitting can be
used to avoid BPs irrespective of the number of layers. In our first
experiment, we numerically showed that this holds when we set $L=N$. Recent
findings showed that, a transition to BPs happens at a depth of $\mathcal{O}(\log N)$
for an ansatz with a local cost
function~\cite{cerezo_cost_2021}. Therefore, there is great importance in
investigating the behaviour for larger values of $L$. For considerably low
values of $N$ (e.g. $N<32$), we can assume a constant value for $m$ (e.g.
$m=4$), such that $m$ is approximately $\mathcal{O}(\log N)$.
We present variances of two ans\"atze ($m=4$, $m=N$) for up to 200 layers and 16
qubits in Fig.~\ref{fig:bp2}. For the standard ansatz, we see a clear
transition to BPs with increasing number of layers, as
expected~\cite{cerezo_cost_2021}. On the other hand, the CS ansatz ($m=4$)
shows a robust behavior from small to large number of layers. 

These two experiments show the potential of the classical splitting in
avoiding BPs. However, the question of whether this potential can be
transferred in-to practice (e.g. binary classification performance or
quantum simulation) still lacks an answer. Next, we will be addressing
this question.

\subsection{Binary classification using a classical dataset}
\label{sec:classical-training}
In this experiment, we will continue using the same ansatz with same
assumptions to perform binary classification using a classical dataset. Our
goal here is to compare performance of the CS ansatz to the
standard case for increasing number of qubits. We need a dataset that
can be scaled for this purpose. However, datasets are typically constant in
dimension and do not offer an easy way to test the scalability in this sense.
Therefore, we employ an ad-hoc dataset that can be produced with
different number of features.

Three datasets ($N=4$, $8$ and $16$) were produced using the make\_classification 
function of scikit-learn\footnote{The classical dataset is produced for 600 data samples
with a 420/180 train/test split, a class separation value of 1.0, 2.0\% class assignment
error and no redundant or repeated features.}~\cite{pedregosa_scikit-learn_2011}. This
tool allows us to draw samples from an $N$-dimensional hypercube, where samples of each
class are clustered around the vertices. Each dataset contains 420 training and 180
testing samples. Each of the data samples were encoded using one $R_Y$ gate
per qubit, such that each ansatz uses the same number of features of the
given dataset. Please see Appendix~\ref{app:classical-dataset} for more
details on the production of the dataset and distributions of samples. 

The binary classification was performed using the expectation value over the
observable defined in Eq.~\eqref{eq:1-local-h} and the binary cross entropy
function was used as the loss function during training, such that,

\begin{equation}
    L(y, \hat{y}) = - y \log \hat{y} - (1-y) \log (1-\hat{y}),
\end{equation}

where $y$ (i.e. $y \in \{0,1\}$) is the class label of the given data sample
and $\hat{y}$ is the prediction (i.e. $\hat{y} = \mbox{Tr}[OU(\boldsymbol{\theta})\rho(x)
U^\dagger(\boldsymbol{\theta})]$, where x is the data sample)\footnote{Here, the expectation
value can have values between [-1,1], we scale it to be [0,1] to compensate for the discrepancy
between the class labels.}. The ADAM optimizer~\cite{kingma_adam_2017} with a
learning rate of 0.1 was used and all models are trained for 100 epochs using
full batch size (bs=420)\footnote{In the case of $N=m=L=16$ full batch size was
not possible due to vast memory requirement. Therefore, bs=60 was
used only for this case. In Appendix~\ref{app:classical-training}, we show that
using a smaller batch size does not affect the performance of the model
significantly.}. We report our results based on 50 runs for each setting.

\begin{figure}[!t]
    \centering
    \includegraphics[width=\linewidth]{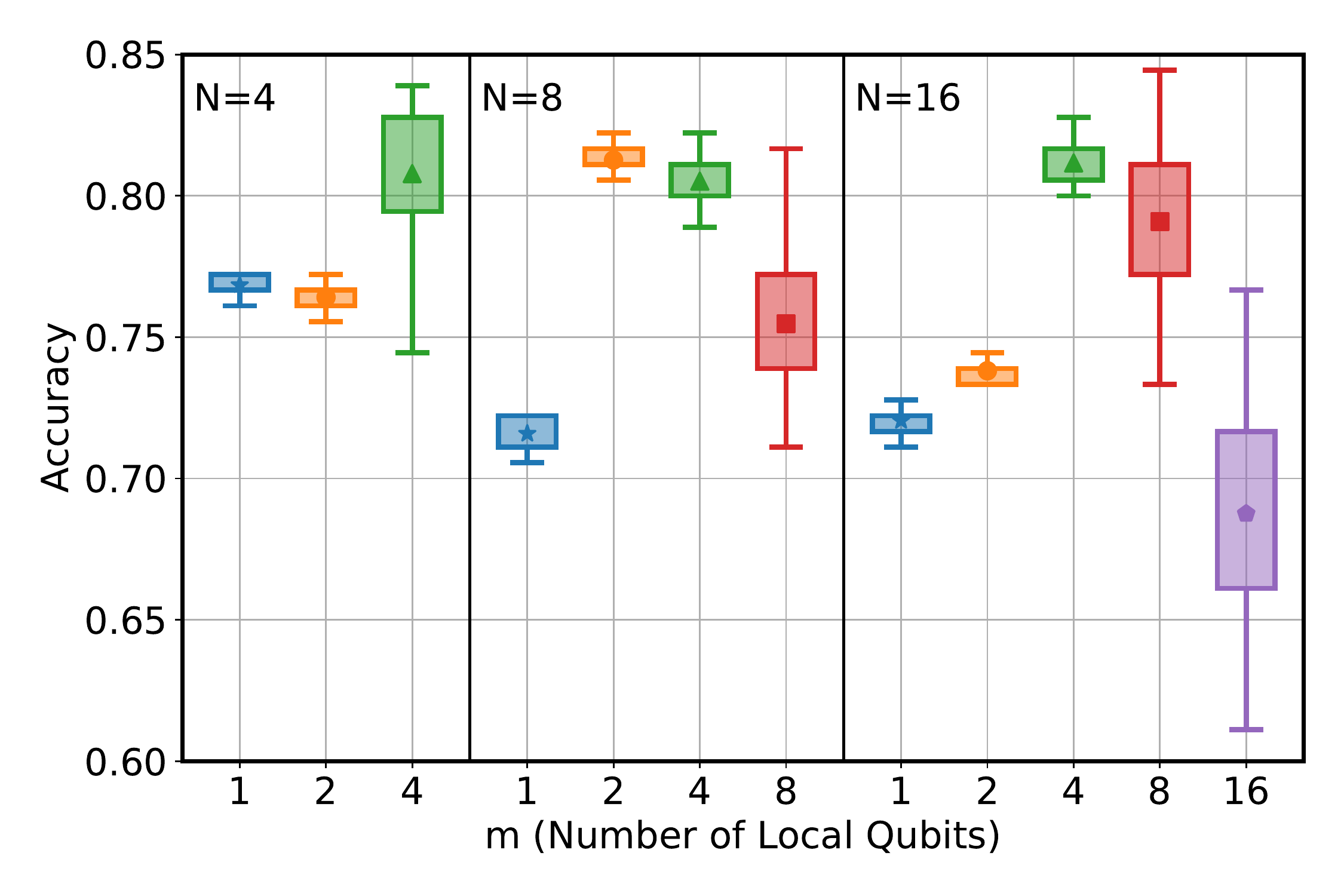}
    \caption{Box plot of the best test accuracy obtained over 50 runs plotted
        with respect to the relevant local number of qubits ($m$). Each column
        represents a problem with a different sample size (4, 8, 16).
        Each marker is placed on the median, boxes cover the range from the
        first to third quartiles and the error bars extend the quartiles by 3
        times range. Each $m$ value is plotted with a different marker and color.}
    \label{fig:classical_accuracy}
\end{figure}

Classification performance of ans\"atze for changing values of $m$ using the
three datasets are presented in Fig.~\ref{fig:classical_accuracy}. Here, the
results show the distribution of accuracies over the test set. For the
$N=4$ case, we see that the standard ($m=N$) ansatz performs the best. However,
this is not the case as we go to more qubits. For the 8 and 16 qubit cases,
it is evident that $m<N$ ans\"atze can match the performance of the standard
ansatz. We can also see that the constant choice of $m=4$ can provide a robust
performance with increasing number of qubits (at least up-to $N=16$), matching
our expectations. Training curves of all settings are presented in 
Appendix~\ref{app:classical-training}.

\subsection{Binary classification using a quantum dataset}
\label{sec:ntangled-training}
The binary classification performance of the classical splitting over the
classical datasets provides the first numerical evidence for their advantage
against the standard ans\"atze. It is also important to investigate
if they can be extended to problems where the data consists of quantum states.
Our proof in Section~\ref{sec:bp} assumed the input states to be tensor product
states. Now, we remove this constraint and use a quantum dataset.

For this experiment, we use the NTangled dataset~\cite{schatzki_entangled_2021}.
NTangled dataset provides parameters to produce distributions
of quantum states that are centered around different Concentrable
Entanglement (CE)~\cite{beckey_computable_2021} values. CE is a measure of
entanglement, which is defined as follows,

\begin{equation}
    \mbox{CE}(\ket{\Psi}) = 1 - \frac{1}{2^N} \sum_{\alpha \in Q}
    \mbox{Tr}[\rho_\alpha^2],
\end{equation}

\begin{table*}[!t]
    \centering
    \caption{Binary classification performance of ans\"atze with different
        values of $m$ over different distributions of quantum states from the 
        NTangled dataset~\cite{schatzki_entangled_2021}. Average of 50 runs are presented with errors showing the difference to maximum and minimum observed values. Best average value of each metric for the given task is printed in bold.
    }
    \resizebox{\linewidth}{!}{
    \begin{tabular}{||c|c|c|c|c|c|c|c|c|c||}
        \hline
        \multirow{2}{*}{N} & Task & \multirow{2}{*}{L} & \multirow{2}{*}{m} & \multirow{2}{*}{Train Accuracy ($\%$)} & Avg. epochs to reach & Avg. epochs to reach & \multirow{2}{*}{Test Accuracy ($\%$)} & Avg. epochs to reach & Avg. epochs to reach\\
                           & [CE Values] &                    &                    &  & $90\%$ Train Accuracy & $100\%$ Train Accuracy &  & $90\%$ Test Accuracy & $100\%$ Test Accuracy \\
        \hline
        \hline
        \multirow{3}{*}{4} & \multirow{3}{*}{0.05 vs. 0.35} & \multirow{3}{*}{4} & 1 & $94.6^{+2.5}_{-1.7}$ & $6.7^{+11.3}_{-5.7}$           & N/A &  $94.6^{+3.8}_{-1.8}$ & $6.1^{+11.9}_{-5.1}$        & N/A                     \\
                           &                                &                    & 2 & $\mathbf{100.0}^{+0.0}_{-0.5}$ & $\mathbf{4.9}^{+12.1}_{-3.9}$ & N/A &  $100.0^{+0.0}_{-0.0}$ & $\mathbf{3.9}^{+11.1}_{-2.9}$ & $10.8^{+26.2}_{-9.8}$ \\
                           &                                &                    & 4 & $99.9^{+0.1}_{-1.6}$ & $5.4^{+6.6}_{-4.4}$           & N/A &  $100.0^{+0.0}_{-1.1}$ & $4.1^{+8.9}_{-3.1}$         & N/A                     \\
        \hline
        \multirow{3}{*}{4} & \multirow{3}{*}{0.25 vs. 0.35} & \multirow{3}{*}{4} & 1 & $90.4^{+4.1}_{-3.5}$           & N/A                    & N/A & $86.4^{+6.9}_{-5.9}$           & N/A                    & N/A \\
                           &                                &                    & 2 & $98.2^{+1.5}_{-1.3}$           & $7.7^{+25.3}_{-5.7}$          & N/A & $97.1^{+2.3}_{-1.6}$           & $7.9^{+27.1}_{-6.9}$          & N/A \\
                           &                                &                    & 4 & $\mathbf{100.0}^{+0.0}_{-0.4}$ & $\mathbf{5.1}^{+9.9}_{-4.1}$ & N/A & $\mathbf{100.0}^{+0.0}_{-1.1}$ & $\mathbf{4.5}^{+11.5}_{-3.5}$ & N/A \\
        \hline
        \multirow{4}{*}{8} & \multirow{4}{*}{0.15 vs. 0.45} & \multirow{4}{*}{8} & 1 & $99.9^{+0.1}_{-0.2}$ & $3.3^{+3.7}_{-2.3}$          & N/A                    & $100.0^{+0.0}_{-0.0}$ & $2.4^{+2.6}_{-1.4}$ & $6.1^{+10.9}_{-5.1}$ \\
                           &                                &                    & 2 & $100.0^{+0.0}_{-0.0}$ & $2.5^{+2.5}_{-1.5}$ & $7.1^{+11.9}_{-6.1}$          & $100.0^{+0.0}_{-0.0}$ & $1.5^{+2.5}_{-0.5}$ & $3.2^{+6.8}_{-2.2}$ \\
                           &                                &                    & 4 & $100.0^{+0.0}_{-0.0}$ & $\mathbf{2.4}^{+1.6}_{-1.4}$ & $\mathbf{4.6}^{+4.4}_{-3.6}$ & $100.0^{+0.0}_{-0.0}$ & $\mathbf{1.4}^{+1.6}_{-0.4}$ & $\mathbf{2.9}^{+7.1}_{-1.9}$ \\
                           &                                &                    & 8 & $100.0^{+0.0}_{-0.0}$ & $2.8^{+2.2}_{-0.8}$ & $7.8^{+11.2}_{-5.8}$          & $100.0^{+0.0}_{-0.0}$ & $1.8^{+2.2}_{-0.8}$ & $4.8^{+8.2}_{-3.8}$ \\
        \hline
        \multirow{4}{*}{8} & \multirow{4}{*}{0.40 vs. 0.45} & \multirow{4}{*}{8} & 1 & $99.9^{+0.1}_{-0.4}$ & $3.1^{+2.9}_{-2.1}$          & N/A                    & $99.6^{+0.4}_{-0.7}$ & $2.2^{+2.8}_{-1.2}$ & N/A \\
                           &                                &                    & 2 & $100.0^{+0.0}_{-0.0}$ & $2.8^{+4.2}_{-1.8}$ & $9.2^{+11.8}_{-8.2}$          & $100.0^{+0.0}_{-0.0}$ & $1.9^{+4.1}_{-0.9}$ & $5.2^{+5.8}_{-4.2}$\\
                           &                                &                    & 4 & $100.0^{+0.0}_{-0.0}$ & $\mathbf{2.4}^{+1.6}_{-1.4}$ & $\mathbf{5.3}^{+12.7}_{-4.3}$ & $100.0^{+0.0}_{-0.0}$ & $\mathbf{1.5}^{+1.5}_{-0.5}$ & $\mathbf{3.2}^{+9.8}_{-2.2}$\\
                           &                                &                    & 8 & $100.0^{+0.0}_{-0.0}$ & $2.9^{+3.1}_{-0.9}$ & $8.2^{+9.8}_{-6.2}$          & $100.0^{+0.0}_{-0.0}$ & $1.9^{+3.1}_{-0.9}$ & $5.7^{+5.3}_{-4.7}$ \\
        \hline
    \end{tabular}}
    \label{tab:quantum-trainig}
\end{table*}

\noindent where $Q$ is the power set of the set $\{1,2,...,\mbox{N}\}$,
and $\rho_\alpha$ is the reduced state of subsystems labeled by the elements of $\alpha$ associated to $\ket{\Psi}$. The NTangled dataset provides
three ans\"atze trained for different CE values for N=3, 4 and 8. We choose the
Hardware Efficient Ansatz (Fig.~\ref{fig:all-ansatze}\textcolor{red}{d})
with depth=5, such that the parameters of the first layer of $U3$ gates are sampled
from a unitary distribution $\in [-1,1]$ and the others are provided by the dataset.
Then, we apply the same CS ansatz used in Section~\ref{sec:classical-training} and 
perform binary classification such that the CE values are the labels
of classes. The CE distributions of the produced quantum states are presented in
Appendix~\ref{app:quantum-dataset}.

For the binary classification task, the same training settings are used as in
Section~\ref{sec:classical-training}, except this time models are trained until
50 epochs, as most models were able to reach $100\%$ test accuracy. We report
our results using different pairs of distributions in 
Table~\ref{tab:quantum-trainig}. In the case of $N=4$, we observed that 
classical splitting can perform at similar accuracy, even if the ansatz do not
have any entangling gates ($m=1$). We see that entangling gates are needed for 
better performance if the problem gets harder (e.g.\ 0.25 vs.\ 0.35 case). If we
go to a problem with more qubits, we can safely say that the CS
ansatz can match the performance of the standard ansatz and converge faster. 

\subsection{Practical remarks on classical splitting}
\label{sec:practical}
The efficacy of classical splitting relies on the parts of the circuit before and after the set of gates that undergo classical splitting. This can be seen most clearly if we set $m=1$ and apply classical splitting to the entire circuit after a possible initialization. In this case, we only perform single qubit operations after initialization. Hence, if the initialization produces a tensor product state, then the circuit subject to classical splitting with $m=1$ can no longer generate any entanglement. Similarly, if we initialize with the HEA (Fig. \ref{fig:all-ansatze}\textcolor{red}{d}) and apply classical splitting with $m=1$ to the remaining circuit, then no tensor product state can be found. 

More generally, $m=1$ produces a circuit that cannot change the amount of entanglement. For other choices of $m$, the picture becomes more complicated but, generally, the set of states that can be generated by the quantum circuit before classical splitting will be reduced to a subset based on the characteristics of the remaining initialization. 

A na\"ive implementation of classical splitting therefore requires knowledge of the correct initialization such that the final solution can still be reached with the classically split circuit. In generic applications, this knowledge is likely not available. Hence, an adaptive approach to classical splitting should be considered. 

One adaptive approach would be to increase $m$ to check for improvements. After we observe no further training improvement with $m=1$, we could move to $m=2$. This enlarges the set of states the quantum circuit can reach, and thus may lead to further training improvements, at the cost of possibly stronger BP effects. However, if $m=1$ has already converged fairly well, then the state is already fairly close to the $m=2$ solution and it is unlikely to find a BP. With $m=2$ converged, we can then move to $m=4$ and continue the process by doubling $m$ one step at a time. 

If, for example, we consider the $N=4$ ``0.25 vs. 0.3'' case of Table~\ref{tab:quantum-trainig}, we may start training with $m=1$. This training converges to about $90\%$ accuracy. Increasing $m$ to $m=2$ will lead to further improvements that converge to about $98\%$ accuracy. Finally, we can further improve the $98\%$ to $100\%$ accuracy by going to $m=4$. 

In this way, we utilize the efficiency of classical splitting to obtain an approximate solution which we then refine by trading efficiency for circuit expressivity through increasing $m$. At this point, the efficiency reduction should no longer lead to insurmountable complications as we already are close to the optimal solution for the current $m$ value. 

Another adaptive approach would be to use classical splitting to check and bypass plateaus. For example, if a VQE appears to be converged, it may also just be stuck in a plateau. Applying classical splitting at this point would reduce the effect of the plateau. Thus, if the VQE continues optimizing after classically splitting a seemingly converged circuit, we can conclude that this was in fact a plateau. After a suitable number of updates using the classically split circuit, we can then return to the full circuit in the hopes of having passed the plateau. 

Unfortunately, this approach cannot be used to positively distinguish between true local optima and plateaus since the classical splitting reduces expressivity and thus introduces artificial constraints. Hence, if the set of states expressible by the classically split circuit is orthogonal to the gradient in the cost function landscape, then a plateau will be replaced with a local optimum and, thus, no improvements will be obtained. In this case, we therefore cannot conclude that the VQE has converged simply because classical splitting shows no improvements. However, experimenting with different implementations of classical splitting may result in cases that do not replace the plateau with an artificial local optimum.

\subsection{Extending classical splitting to VQE}
\label{sec:vqe}

Until now, we have investigated using classical splitting for binary classification problems. It
succeeded by showing an overall better training performance in Section~\ref{sec:classical-training}
and a competitive performance and faster convergence in Section~\ref{sec:ntangled-training}. In this section, we consider
simulating the ground state of the transverse-field ising hamiltonian (TFIH) on a 1D chain. The TFIH with periodic boundary conditions can be defined as;

\begin{equation}
    H = -J \sum_{i=1}^{N} Z_i Z_{i+1}  -h \sum_{i=1}^{N} X_i,
    \label{eq:ising-hamiltonian}
\end{equation}

for $N$ lattice sites, where $J$ determines the strength of interactions and $h$
determines the strength of the external field. Simulating the TFIH on a 1D chain requires
connectivity of qubits on the 1D chain. This contradicts with the assumption we made, when
we proved absence of BPs for classically split ans\"atze in Section~\ref{sec:bp}, since the 
TFIH does not fit the definition we had for an 
$m$-local observable in Eq.~\eqref{eq:obs-m-local}. Therefore, we need to 
rely on the numerical experiments to talk about BPs under the new constraints.

The CS ans\"atze can only produce local entangled states, for this reason we
need an extension of the ansatz in Fig.~\ref{fig:all-ansatze}\textcolor{red}{a}. 
We propose to extend the classically split
ansatz by adding standard layers at the end. The reason for adding them at
the end is to keep the base of light cones\footnote{A light cone or a causal cone of an ansatz is an abstract concept that illustrates how information spreads as more gates are applied. The types of gates and their connectivity determines the opening angle of the cone. The evidence from the literature suggests that there is a correspondence between the opening angle of the cone, BPs and quantum circuit complexity~\cite{cerezo_cost_2021, haferkamp_linear_2021}.} produced by the classically split layers
constant. Then, when we add the standard layers, the light cones will grow
at a pace that is determined by the newly-added part\footnote{It also
depends on the choice of $m$, but since we already have a constraint on m
(i.e. $m = \mathcal{O}(\log N)$) the newly-added ansatz will be the dominant
component.}. This way, the overall ansatz can still escape BPs as long as the
newly-added part does not exhibit BPs.

We define the extended classically split (ECS) ansatz with two types of layers. First
$L$ layers consist of classically split $m$ qubit gate blocks. Then, there
are $T$ layers of any no-BP ansatz (see Fig.~\ref{fig:all-ansatze}\textcolor{red}{b}). 
Since the first $L$ layers can only produce $m$-local product states (i.e. $m <
\mathcal{O}(\log N)$), the existence of BPs depends only on the remaining $T$ layers. 
This way we can choose very large $L$, but need to keep $T$ small as standard ans\"atze
reach BPs rather rapidly (e.g. $\mathcal{O}(\log N)$ depth for a ladder connected
ansatz~\cite{cerezo_cost_2021}). We provide numerical evidence for avoiding BPs with the
ECS ansatz in Appendix~\ref{app:vqe-ansatz-bp}.

\begin{figure}
    \centering
    \includegraphics[width=\linewidth]{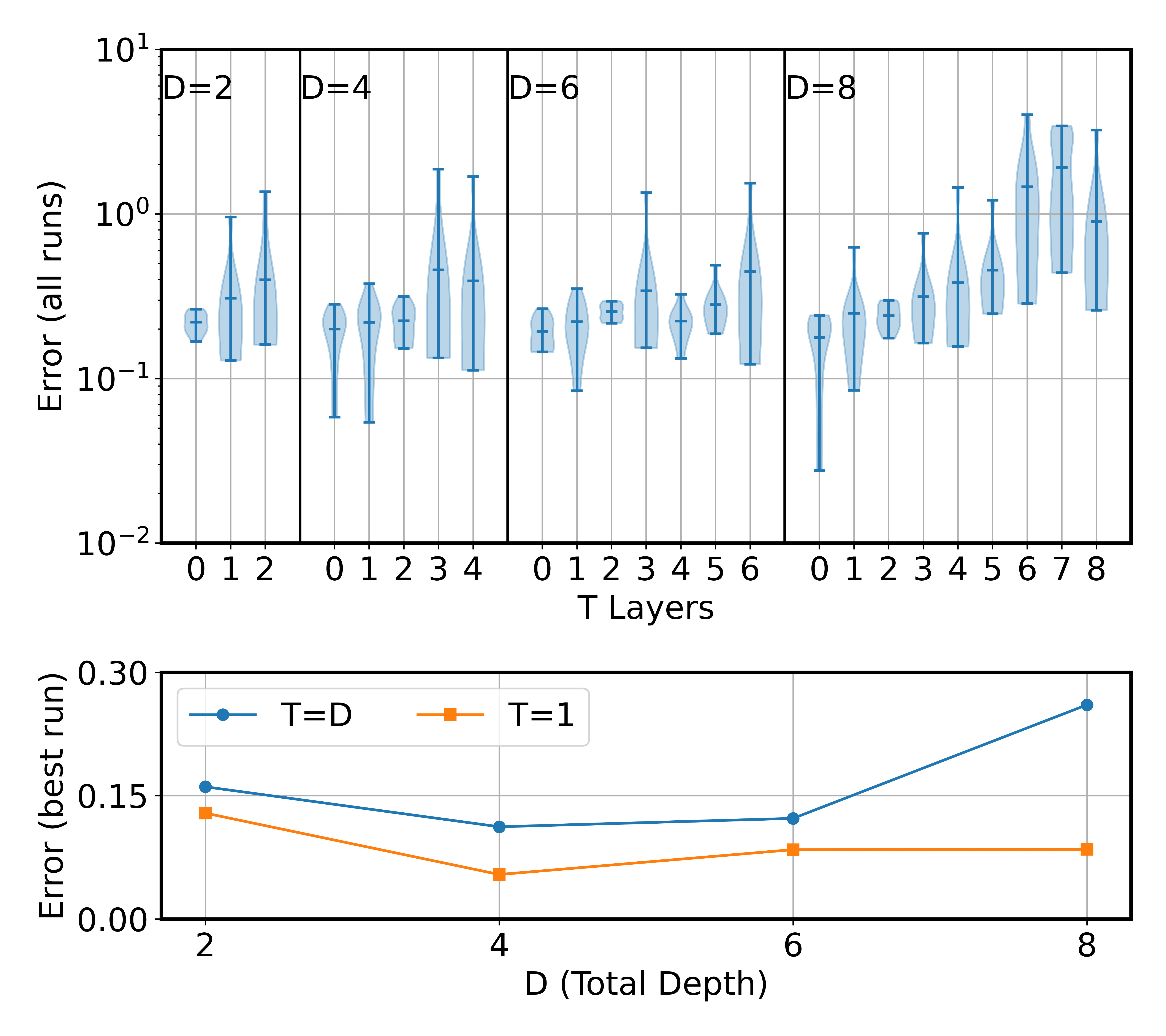}
    \caption{Energy errors of ans\"atze with increasing total depth for $N=12$ TFIH using the
    extended classical splitting (ECS) ansatz with EfficientSU2 sublocks (see
    Fig.~\ref{fig:all-ansatze}\textcolor{red}{b}) and $m=4$. Total depth ($D$) corresponds to
    $L+T$, where $T=0$ is equivalent to the CS ansatz, $T=D$ is equivalent to standard
    EfficientSU2 and other values explore hybrid use cases of the ECS ansatz. Final energy
    measurements of 10 runs are averaged and plotted with their minimum and maximum values
    as the error bars on the upper panel. The lower panel shows the best errors obtained
    for $T=1$ and $T=D$ settings. Energy error is the absolute difference of the energy measurement and the exact ground state energy.}
    \label{fig:vqe}
\end{figure}

For the experiment, we consider the Hamiltonian defined in Eq.~\eqref{eq:ising-hamiltonian}
with $J=1, h=1$. Then, we implement the ECS ansatz with $m=4$ for total depth of 2, 4, 6 and
8. Each side of the ansatz consists of EfficientSU2 layers~\cite{treinish_qiskit_2022} (see
Fig.~\ref{fig:all-ansatze}\textcolor{red}{e}). The first $L$ layers are classically split to
subcircuits of $m$ qubits, while the next $T$ layers do not have any splitting. Total depth ($D$)
corresponds to $L+T$, where $T=0$ is equivalent to the CS ansatz, $T=D$ is equivalent
to the standard EfficientSU2 ansatz and other values explore hybrid use cases of the ECS
ansatz. We report the energy error, which is the absolute difference between the final
energy measurement and the exact ground state energy in Fig.~\ref{fig:vqe}. Results of 10
runs are averaged and plotted with their minimum and maximum values as the error bars.
Experiments are performed under no noise assumption using 10k shots. The SPSA
optimizer~\cite{spsa} is used with 10k iterations. Results with $m=2$ and training curves of
all runs are presented in Appendix~\ref{app:vqe-m2} and \ref{app:vqe-curves}.

The upper panel shows that the mean error increases with increasing total depth in the no
classical splitting setting ($T=D$). This is mainly due to the flattening of the cost
landscape, which makes the optimization process harder. On the other hand, setting $T$ (e.g.
$T=1$) to a low number provides a better error, since it preserves trainability despite the 
increasing total depth. This is a clear indication that the classical splitting allows
deeper ans\"atze. 

The lower panel shows the best error obtained in all the runs for two settings. Here, we
observe that both settings achieve better errors with increasing depth initially. Then, the
no CS setting shows rapidly increasing errors as it looses trainability rather quickly,
compared to the ECS ansatz.

In this experiment, the best error was achieved with the fully classically split ansatz
($T=0$). This is mainly due to the employed EfficientSU2 ansatz not being a very good choice
for this particular problem. This means that by employing other ans\"atze, the observed
behaviour might change, making a larger value of $T$ perform the best. Nevertheless, the
results are still a good indication of how the trainability of the ansatz is affected by
the choice of $L$ and $T$. We plan to draw a more detailed picture of the tradeoff between values of $L$ and $T$ in a future work.

Simulating larger size systems requires a deep ansatz (linear or larger in system size) in
general~\cite{cerezo_variational_2021}. Although a problem-agnostic ansatz can perform well at
small sizes, BPs forbid the scalability. Our results show that the ECS can help circumvent
this issue and allow deeper ans\"atze. Here, we haven't investigated the potential of
classical splitting to obtain the exact ground state energy of the model, but focused on the
trainability aspect. Such a study is left as future work. Our goal here is to  show that
classical splitting can allow one to build wide and deep ans\"atze  without exhibiting
BPs. Typically, faster convergence or a better final energy might be
achieved with a different ansatz or an optimizer, but this is out of scope of this work. 

\section{Discussion}
\label{sec:discussion}
In this work, we showed that the classical splitting of the ans\"atze can be
used to escape BPs both analytically and numerically. Then, we investigated if 
the classical splitting hinders the learning capacity of the ansatz. Our 
experiments showed that this is not the case, and the classically split ansatz
can match the performance at low number of qubits and is potentially superior
at larger number of qubits. 

In general the benefits of classical splitting comes from the reducing the
effective Hilbert Space that the CS ansatz can
explore. Classical splitting only allows the ansatz to produce $m$-qubit tensor 
product states, if the input state is also a tensor product state following 
our assumptions in Section~\ref{sec:bp}. This, as a result, reduces the
expressivity of the ansatz. Nevertheless, this also allows the ansatz to avoid
BPs~\cite{holmes_connecting_2022} by limiting the scaling behavior to the more favorable case of $m$-qubit systems. In the case of the classical splitting, the
exponential increase of the Hilbert Space dimension is prevented and instead a 
polynomial scaling is enforced. For the $m$-local CS ansatz, each local Hilbert Space
have $\dim(H_k)=2^m=N^{\beta\log_2\gamma}$. Although the advantage of using classical
splitting may look trivial, there are many benefits of employing such an ansatz
besides the numerical experiments we  performed in Section~\ref{sec:numerical}.

In our binary classification experiments using a classical dataset, we relied on
single qubit and single rotation gate data encoding. This meant that any
classically split ansatz had  less information in each group. This could in fact
be improved with embedding methods such as data re-uploading, where one can encode
all the data points to each single qubit independently, such that there are alternating
layers of rotation gates that encode the data and parametrized gates that are to be optimized~\cite{perez-salinas_data_2020}. Data re-uploading ans\"atze showed
great classification performance even for low number of qubits. Since the classical
splitting doesn't have a limit on the amount of layers, data re-uploading would
potentially be great way to get a performance increase. 

Classical splitting can provide faster
training when used with gradient based optimizers. In general, the exact gradients of
ans\"atze are computed with the well-known parameter shift
rule~\cite{mitarai_quantum_2018, schuld_evaluating_2019}. However, this requires
2 instances of the same circuit to be executed per parameter. This quickly
results in a bottleneck for the optimization procedure. An ansatz with $L=N$
layers, where each layer has $N$ parameters, requires $\mathcal{O}(N^2)$
circuit executions to compute gradients for a single data sample. On the other
hand, classical splitting provides cost functions that are independent of each
other, as it was shown in Eq.~\eqref{eq:independent-cost}. This allows gradients
to be computed simultaneously across different instances of the classically
split ansatz. As a result, the classically split ansatz optimization requires
$\mathcal{O}(N\log N)$ circuit executions for $m=\mathcal{O}(\log N)$.

The bottleneck in optimization is only one of the challenges of implementing 
scalable VQAs. Another problem that is worth mentioning here is the amount of two-qubit
gates. NISQ hardware provides limited connectivity of qubits. The topology of
the devices plays an essential role in the efficient implementation of quantum
circuits~\cite{weidenfeller_scaling_2022}. Typically, a quantum circuit compilation (or
transpilation) procedure is required to adapt a given circuit to be able to be compatible
with the capabilities of the devices (e.g. converting gates to native gates, applying
SWAP gates to connect qubits which are not physically connected)~\cite{botea_complexity_2018}. 

Classical splitting provides a significant reduction in number of two qubit
gates as it divides a large qubit to many circuits with less qubits. 
To show the scale of the reduction, we can construct a set of hypothetical devices
that has a 2D grid topology (square lattice with no diagonal connections). We start by considering the CS ansatz that consists
the ans\"atze in  Fig.~\ref{fig:all-ansatze}\textcolor{red}{c} and extend it to a
fully entangled architecture. A linear entangled ansatz has $\mathcal{O}(N)$ two
qubit gates, while a fully entangled one has $\mathcal{O}(N^2)$ per layer. Then,
we use Qiskit's transpiler\footnote{Qiskit's transpiler algorithm
is a stochastic algorithm, meaning that it is possible to get better values if
the algorithm is executed many times. Here, we run the algorithm two times and take
the best results using optimization level 3, and sabre-sabre layout and
routing methods. Although, It is possible to obtain better gate counts with more
runs or different transpilation algorithms, the best values obtained wouldn't
change our conclusions.}~\cite{treinish_qiskit_2022} to fit these ans\"atze to
the hypothetical devices and report the two qubit gate counts in Table~\ref{tab:transpiler}.

\begin{table}[!t]
    \centering
    \caption{Two qubit gate counts of different ans\"atze transpiled for
    hypothetical devices that has a 2D grid topology (square lattice with no diagonal connections).} 
    \resizebox{\linewidth}{!}{
    \begin{tabular}{||c|c|c|c|c|c|c|c||}
        \hline
        \multirow{3}{*}{$m$}
        & \multirow{3}{*}{$L$} 
        & \multicolumn{6}{c||}{amount of two qubit gates} \\
        \cline{3-8}
        & 
        & \multicolumn{3}{c|}{linear entanglement} 
        & \multicolumn{3}{c||}{full entanglement} \\
        \cline{3-8}
        & & $N=4$ & $N=16$ & $N=36$ & $N=4$ & $N=16$ & $N=36$ \\
        \hline
        \hline
        \multirow{2}{*}{$N$} & 2 & 6  & 33  & 121  & 24 & 696  & 3601  \\
                           & $N$ & 12 & 240 & 1362 & 46 & 5372 & 65040 \\
        \hline
        \multirow{2}{*}{4} & 2 & 6  & 24  & 54  & 24 & 92  & 250 \\
                           & $N$ & 12 & 192 & 978 & 46 & 964 & 4376 \\
        \hline
        \multirow{2}{*}{2} & 2 & 4 & 16  & 36  & 4 & 16  & 42  \\
                           & $N$ & 8 & 128 & 654 & 8 & 134 & 648 \\
        \hline
    \end{tabular}}
    \label{tab:transpiler}
\end{table}

The amount of gates are not only important to have a better implementation but
also to have a more precise results, since NISQ devices come with noisy
gates. We consider the CX gate errors reported by IBM for their devices, 
which can be taken as $\mathcal{O}(10^{-2})$ on average\footnote{This value is
chosen after a survey of devices listed on
\href{https://quantum-computing.ibm.com}{IBM Quantum Cloud}.}. Then, as a
figure of merit, we can
assume 50\% to be the  limit, in which we can still get meaningful results. 
This would allow us to use 50 CX gates at most. Now, the results from 
Table~\ref{tab:transpiler} implies that it is possible to construct a 36 qubit, 
2 layer ansatz with linear entanglement, if we employ classical splitting. This
would not be possible for the standard case as it comes with more than twice two
qubit gates. The reduction only gets better if we consider a full entanglement
case. Following the same logic, to implement a 36 qubit, 36 layer, fully 
entangled ansatz, a CX gate error of $\mathcal{O}(10^{-6})$ is needed, while
the classically split ansatz only requires a CX gate error of
$\mathcal{O}(10^{-4})$. A similar reduction in noise is also possible for other
types of circuit partitioning methods~\cite{basu_i-qer_2022}.

Classically splitting an ansatz further allows faster implementation on hardware.
A generic ansatz consists of two-qubit gates that follow one and another, matching a 
certain layout. We mentioned some of these as ladder/linear or full. However, this means that
the hardware implementation of such an ansatz requires execution of these gates 
sequentially, taking a significant amount of time. To overcome such obstacles, ans\"atze
such as the HEA (see Fig.~\ref{fig:all-ansatze}\textcolor{red}{d}) are widely used in the
literature~\cite{kandala_hardware-efficient_2017}. Classical splitting an ansatz can reduce the
implementation time significantly since it allows simultaneous two-qubit gates across different
local circuits. This can mean a speed-up of from
$\mathcal{O}(N/ \log N)$ to $\mathcal{O}((N/ \log N)^2)$ depending on the connectivity 
of the original ansatz. 

Finally, the formulation we used in Section~\ref{sec:classical-training} allows the CS ansatz
to be implemented on smaller quantum computers instead of a single large
quantum computer. This means that for similar problems, there are many
implementation options available. These include using one large device, using many small devices (e.g.,
$\mathcal{O}(N/\log N)$ many $\mathcal{O}(\log N)$ qubit devices) and parallelizing the task or using one
small device and performing all computation sequentially. All of these features makes the classical
splitting an ideal approach for Quantum Machine Learning (QML) applications using NISQ devices.

\section{Conclusion}
\label{sec:conclusion}

In this work, we presented some foundational ideas of applying classical splitting
to generic ans\"atze. Our results indicate many benefits of using
classical splitting, such as better trainability, faster hardware implementation,
faster convergence, robustness against noise and parallelization under certain conditions.
These suggest that classical splitting or variations of this idea might play an essential 
role in how we are designing ans\"atze for QML problems. We also presented an extension
to the initial classical splitting idea so that these types of ans\"atze can be used in VQE.
The initial results that we presented in this work suggest that classical splitting
can help improve the trainability and reach better error values.
However, it is still an open question to what extent VQE can benefit from classical
splitting. Our results encourages employing approaches that are based upon classically
splitting or partitioning parametrized quantum circuits~\cite{bravyi_trading_2016,
peng_simulating_2020, tang_cutqc_2021, perlin_quantum_2021, eddins_doubling_2021,
saleem_quantum_2021, fujii_deep_2022, marshall_high_2022}, as they are in general
more robust against hardware noise. We consider in-depth analysis and applications
with VQE and QAOA as future directions for this work.

\begin{acknowledgments}
C.T. and A.C.  are supported in part by the Helmholtz
Association - “Innopool Project Variational Quantum Computer Simulations
(VQCS)”. S.K. acknowledges financial support from the
Cyprus Research and Innovation Foundation under project ``Future-proofing
Scientific Applications for the Supercomputers of Tomorrow (FAST)'', contract
no.\ COMPLEMENTARY/0916/0048, and ``Quantum Computing for Lattice Gauge Theories (QC4LGT)'', contract no.\ EXCELLENCE/0421/0019. We thank Lena Funcke for valuable discussions. 
\end{acknowledgments}
\bibliography{main}

\begin{thebibliography}{51}%
\makeatletter
\providecommand \@ifxundefined [1]{%
 \@ifx{#1\undefined}
}%
\providecommand \@ifnum [1]{%
 \ifnum #1\expandafter \@firstoftwo
 \else \expandafter \@secondoftwo
 \fi
}%
\providecommand \@ifx [1]{%
 \ifx #1\expandafter \@firstoftwo
 \else \expandafter \@secondoftwo
 \fi
}%
\providecommand \natexlab [1]{#1}%
\providecommand \enquote  [1]{``#1''}%
\providecommand \bibnamefont  [1]{#1}%
\providecommand \bibfnamefont [1]{#1}%
\providecommand \citenamefont [1]{#1}%
\providecommand \href@noop [0]{\@secondoftwo}%
\providecommand \href [0]{\begingroup \@sanitize@url \@href}%
\providecommand \@href[1]{\@@startlink{#1}\@@href}%
\providecommand \@@href[1]{\endgroup#1\@@endlink}%
\providecommand \@sanitize@url [0]{\catcode `\\12\catcode `\$12\catcode
  `\&12\catcode `\#12\catcode `\^12\catcode `\_12\catcode `\%12\relax}%
\providecommand \@@startlink[1]{}%
\providecommand \@@endlink[0]{}%
\providecommand \url  [0]{\begingroup\@sanitize@url \@url }%
\providecommand \@url [1]{\endgroup\@href {#1}{\urlprefix }}%
\providecommand \urlprefix  [0]{URL }%
\providecommand \Eprint [0]{\href }%
\providecommand \doibase [0]{https://doi.org/}%
\providecommand \selectlanguage [0]{\@gobble}%
\providecommand \bibinfo  [0]{\@secondoftwo}%
\providecommand \bibfield  [0]{\@secondoftwo}%
\providecommand \translation [1]{[#1]}%
\providecommand \BibitemOpen [0]{}%
\providecommand \bibitemStop [0]{}%
\providecommand \bibitemNoStop [0]{.\EOS\space}%
\providecommand \EOS [0]{\spacefactor3000\relax}%
\providecommand \BibitemShut  [1]{\csname bibitem#1\endcsname}%
\let\auto@bib@innerbib\@empty
\bibitem [{\citenamefont {Cerezo}\ \emph
  {et~al.}(2021{\natexlab{a}})\citenamefont {Cerezo}, \citenamefont
  {Arrasmith}, \citenamefont {Babbush}, \citenamefont {Benjamin}, \citenamefont
  {Endo}, \citenamefont {Fujii}, \citenamefont {McClean}, \citenamefont
  {Mitarai}, \citenamefont {Yuan}, \citenamefont {Cincio},\ and\ \citenamefont
  {Coles}}]{cerezo_variational_2021}%
  \BibitemOpen
  \bibfield  {author} {\bibinfo {author} {\bibfnamefont {M.}~\bibnamefont
  {Cerezo}}, \bibinfo {author} {\bibfnamefont {A.}~\bibnamefont {Arrasmith}},
  \bibinfo {author} {\bibfnamefont {R.}~\bibnamefont {Babbush}}, \bibinfo
  {author} {\bibfnamefont {S.~C.}\ \bibnamefont {Benjamin}}, \bibinfo {author}
  {\bibfnamefont {S.}~\bibnamefont {Endo}}, \bibinfo {author} {\bibfnamefont
  {K.}~\bibnamefont {Fujii}}, \bibinfo {author} {\bibfnamefont {J.~R.}\
  \bibnamefont {McClean}}, \bibinfo {author} {\bibfnamefont {K.}~\bibnamefont
  {Mitarai}}, \bibinfo {author} {\bibfnamefont {X.}~\bibnamefont {Yuan}},
  \bibinfo {author} {\bibfnamefont {L.}~\bibnamefont {Cincio}},\ and\ \bibinfo
  {author} {\bibfnamefont {P.~J.}\ \bibnamefont {Coles}},\ }\bibfield  {title}
  {\bibinfo {title} {Variational quantum algorithms},\ }\href
  {https://doi.org/10.1038/s42254-021-00348-9} {\bibfield  {journal} {\bibinfo
  {journal} {Nature Reviews Physics}\ ,\ \bibinfo {pages} {625}} (\bibinfo
  {year} {2021}{\natexlab{a}})}\BibitemShut {NoStop}%
\bibitem [{\citenamefont {Peruzzo}\ \emph {et~al.}(2014)\citenamefont
  {Peruzzo}, \citenamefont {McClean}, \citenamefont {Shadbolt}, \citenamefont
  {Yung}, \citenamefont {Zhou}, \citenamefont {Love}, \citenamefont
  {Aspuru-Guzik},\ and\ \citenamefont {O’Brien}}]{Peruzzo2014}%
  \BibitemOpen
  \bibfield  {author} {\bibinfo {author} {\bibfnamefont {A.}~\bibnamefont
  {Peruzzo}}, \bibinfo {author} {\bibfnamefont {J.}~\bibnamefont {McClean}},
  \bibinfo {author} {\bibfnamefont {P.}~\bibnamefont {Shadbolt}}, \bibinfo
  {author} {\bibfnamefont {M.-H.}\ \bibnamefont {Yung}}, \bibinfo {author}
  {\bibfnamefont {X.-Q.}\ \bibnamefont {Zhou}}, \bibinfo {author}
  {\bibfnamefont {P.~J.}\ \bibnamefont {Love}}, \bibinfo {author}
  {\bibfnamefont {A.}~\bibnamefont {Aspuru-Guzik}},\ and\ \bibinfo {author}
  {\bibfnamefont {J.~L.}\ \bibnamefont {O’Brien}},\ }\bibfield  {title}
  {\bibinfo {title} {A variational eigenvalue solver on a photonic quantum
  processor},\ }\href {https://doi.org/10.1038/ncomms5213} {\bibfield
  {journal} {\bibinfo  {journal} {Nature Communications}\ }\textbf {\bibinfo
  {volume} {5}},\ \bibinfo {pages} {4213} (\bibinfo {year} {2014})}\BibitemShut
  {NoStop}%
\bibitem [{\citenamefont {Farhi}\ \emph {et~al.}(2014)\citenamefont {Farhi},
  \citenamefont {Goldstone},\ and\ \citenamefont
  {Gutmann}}]{farhi_quantum_2014}%
  \BibitemOpen
  \bibfield  {author} {\bibinfo {author} {\bibfnamefont {E.}~\bibnamefont
  {Farhi}}, \bibinfo {author} {\bibfnamefont {J.}~\bibnamefont {Goldstone}},\
  and\ \bibinfo {author} {\bibfnamefont {S.}~\bibnamefont {Gutmann}},\
  }\bibfield  {title} {\bibinfo {title} {A {Quantum} {Approximate}
  {Optimization} {Algorithm}},\ }\href {https://arxiv.org/abs/1411.4028}
  {\bibfield  {journal} {\bibinfo  {journal} {arXiv:1411.4028}\ } (\bibinfo
  {year} {2014})}\BibitemShut {NoStop}%
\bibitem [{\citenamefont {Farhi}\ and\ \citenamefont
  {Neven}(2018)}]{farhi_classification_2018}%
  \BibitemOpen
  \bibfield  {author} {\bibinfo {author} {\bibfnamefont {E.}~\bibnamefont
  {Farhi}}\ and\ \bibinfo {author} {\bibfnamefont {H.}~\bibnamefont {Neven}},\
  }\bibfield  {title} {\bibinfo {title} {Classification with {Quantum} {Neural}
  {Networks} on {Near} {Term} {Processors}},\ }\href
  {https://arxiv.org/abs/1802.06002} {\bibfield  {journal} {\bibinfo  {journal}
  {arXiv:1802.06002}\ } (\bibinfo {year} {2018})}\BibitemShut {NoStop}%
\bibitem [{\citenamefont {Preskill}(2018)}]{Preskill2018}%
  \BibitemOpen
  \bibfield  {author} {\bibinfo {author} {\bibfnamefont {J.}~\bibnamefont
  {Preskill}},\ }\bibfield  {title} {\bibinfo {title} {Quantum computing in the
  {NISQ} era and beyond},\ }\href {https://doi.org/10.22331/q-2018-08-06-79}
  {\bibfield  {journal} {\bibinfo  {journal} {Quantum}\ }\textbf {\bibinfo
  {volume} {2}},\ \bibinfo {pages} {1} (\bibinfo {year} {2018})}\BibitemShut
  {NoStop}%
\bibitem [{\citenamefont {McClean}\ \emph {et~al.}(2018)\citenamefont
  {McClean}, \citenamefont {Boixo}, \citenamefont {Smelyanskiy}, \citenamefont
  {Babbush},\ and\ \citenamefont {Neven}}]{mcclean_barren_2018}%
  \BibitemOpen
  \bibfield  {author} {\bibinfo {author} {\bibfnamefont {J.~R.}\ \bibnamefont
  {McClean}}, \bibinfo {author} {\bibfnamefont {S.}~\bibnamefont {Boixo}},
  \bibinfo {author} {\bibfnamefont {V.~N.}\ \bibnamefont {Smelyanskiy}},
  \bibinfo {author} {\bibfnamefont {R.}~\bibnamefont {Babbush}},\ and\ \bibinfo
  {author} {\bibfnamefont {H.}~\bibnamefont {Neven}},\ }\bibfield  {title}
  {\bibinfo {title} {Barren plateaus in quantum neural network training
  landscapes},\ }\href {https://doi.org/10.1038/s41467-018-07090-4} {\bibfield
  {journal} {\bibinfo  {journal} {Nature Communications}\ }\textbf {\bibinfo
  {volume} {9}},\ \bibinfo {pages} {4812} (\bibinfo {year} {2018})}\BibitemShut
  {NoStop}%
\bibitem [{\citenamefont {Wang}\ \emph {et~al.}(2021)\citenamefont {Wang},
  \citenamefont {Fontana}, \citenamefont {Cerezo}, \citenamefont {Sharma},
  \citenamefont {Sone}, \citenamefont {Cincio},\ and\ \citenamefont
  {Coles}}]{wang_noise-induced_2021}%
  \BibitemOpen
  \bibfield  {author} {\bibinfo {author} {\bibfnamefont {S.}~\bibnamefont
  {Wang}}, \bibinfo {author} {\bibfnamefont {E.}~\bibnamefont {Fontana}},
  \bibinfo {author} {\bibfnamefont {M.}~\bibnamefont {Cerezo}}, \bibinfo
  {author} {\bibfnamefont {K.}~\bibnamefont {Sharma}}, \bibinfo {author}
  {\bibfnamefont {A.}~\bibnamefont {Sone}}, \bibinfo {author} {\bibfnamefont
  {L.}~\bibnamefont {Cincio}},\ and\ \bibinfo {author} {\bibfnamefont {P.~J.}\
  \bibnamefont {Coles}},\ }\bibfield  {title} {\bibinfo {title} {Noise-induced
  barren plateaus in variational quantum algorithms},\ }\href
  {https://doi.org/10.1038/s41467-021-27045-6} {\bibfield  {journal} {\bibinfo
  {journal} {Nature Communications}\ }\textbf {\bibinfo {volume} {12}},\
  \bibinfo {pages} {6961} (\bibinfo {year} {2021})}\BibitemShut {NoStop}%
\bibitem [{\citenamefont {Ortiz~Marrero}\ \emph {et~al.}(2021)\citenamefont
  {Ortiz~Marrero}, \citenamefont {Kieferová},\ and\ \citenamefont
  {Wiebe}}]{ortiz_marrero_entanglement-induced_2021}%
  \BibitemOpen
  \bibfield  {author} {\bibinfo {author} {\bibfnamefont {C.}~\bibnamefont
  {Ortiz~Marrero}}, \bibinfo {author} {\bibfnamefont {M.}~\bibnamefont
  {Kieferová}},\ and\ \bibinfo {author} {\bibfnamefont {N.}~\bibnamefont
  {Wiebe}},\ }\bibfield  {title} {\bibinfo {title} {Entanglement-{Induced}
  {Barren} {Plateaus}},\ }\href {https://doi.org/10.1103/PRXQuantum.2.040316}
  {\bibfield  {journal} {\bibinfo  {journal} {PRX Quantum}\ }\textbf {\bibinfo
  {volume} {2}},\ \bibinfo {pages} {040316} (\bibinfo {year}
  {2021})}\BibitemShut {NoStop}%
\bibitem [{\citenamefont {Cerezo}\ \emph
  {et~al.}(2021{\natexlab{b}})\citenamefont {Cerezo}, \citenamefont {Sone},
  \citenamefont {Volkoff}, \citenamefont {Cincio},\ and\ \citenamefont
  {Coles}}]{cerezo_cost_2021}%
  \BibitemOpen
  \bibfield  {author} {\bibinfo {author} {\bibfnamefont {M.}~\bibnamefont
  {Cerezo}}, \bibinfo {author} {\bibfnamefont {A.}~\bibnamefont {Sone}},
  \bibinfo {author} {\bibfnamefont {T.}~\bibnamefont {Volkoff}}, \bibinfo
  {author} {\bibfnamefont {L.}~\bibnamefont {Cincio}},\ and\ \bibinfo {author}
  {\bibfnamefont {P.~J.}\ \bibnamefont {Coles}},\ }\bibfield  {title} {\bibinfo
  {title} {Cost function dependent barren plateaus in shallow parametrized
  quantum circuits},\ }\href {https://doi.org/10.1038/s41467-021-21728-w}
  {\bibfield  {journal} {\bibinfo  {journal} {Nature Communications}\ }\textbf
  {\bibinfo {volume} {12}},\ \bibinfo {pages} {1791} (\bibinfo {year}
  {2021}{\natexlab{b}})}\BibitemShut {NoStop}%
\bibitem [{\citenamefont {Holmes}\ \emph {et~al.}(2022)\citenamefont {Holmes},
  \citenamefont {Sharma}, \citenamefont {Cerezo},\ and\ \citenamefont
  {Coles}}]{holmes_connecting_2022}%
  \BibitemOpen
  \bibfield  {author} {\bibinfo {author} {\bibfnamefont {Z.}~\bibnamefont
  {Holmes}}, \bibinfo {author} {\bibfnamefont {K.}~\bibnamefont {Sharma}},
  \bibinfo {author} {\bibfnamefont {M.}~\bibnamefont {Cerezo}},\ and\ \bibinfo
  {author} {\bibfnamefont {P.~J.}\ \bibnamefont {Coles}},\ }\bibfield  {title}
  {\bibinfo {title} {Connecting {Ansatz} {Expressibility} to {Gradient}
  {Magnitudes} and {Barren} {Plateaus}},\ }\href
  {https://doi.org/10.1103/PRXQuantum.3.010313} {\bibfield  {journal} {\bibinfo
   {journal} {PRX Quantum}\ }\textbf {\bibinfo {volume} {3}},\ \bibinfo {pages}
  {010313} (\bibinfo {year} {2022})}\BibitemShut {NoStop}%
\bibitem [{\citenamefont {Cong}\ \emph {et~al.}(2019)\citenamefont {Cong},
  \citenamefont {Choi},\ and\ \citenamefont {Lukin}}]{cong_quantum_2019}%
  \BibitemOpen
  \bibfield  {author} {\bibinfo {author} {\bibfnamefont {I.}~\bibnamefont
  {Cong}}, \bibinfo {author} {\bibfnamefont {S.}~\bibnamefont {Choi}},\ and\
  \bibinfo {author} {\bibfnamefont {M.~D.}\ \bibnamefont {Lukin}},\ }\bibfield
  {title} {\bibinfo {title} {Quantum convolutional neural networks},\ }\href
  {https://doi.org/10.1038/s41567-019-0648-8} {\bibfield  {journal} {\bibinfo
  {journal} {Nature Physics}\ }\textbf {\bibinfo {volume} {15}},\ \bibinfo
  {pages} {1273–1278} (\bibinfo {year} {2019})}\BibitemShut {NoStop}%
\bibitem [{\citenamefont {Pesah}\ \emph {et~al.}(2021)\citenamefont {Pesah},
  \citenamefont {Cerezo}, \citenamefont {Wang}, \citenamefont {Volkoff},
  \citenamefont {Sornborger},\ and\ \citenamefont {Coles}}]{Pesah2020}%
  \BibitemOpen
  \bibfield  {author} {\bibinfo {author} {\bibfnamefont {A.}~\bibnamefont
  {Pesah}}, \bibinfo {author} {\bibfnamefont {M.}~\bibnamefont {Cerezo}},
  \bibinfo {author} {\bibfnamefont {S.}~\bibnamefont {Wang}}, \bibinfo {author}
  {\bibfnamefont {T.}~\bibnamefont {Volkoff}}, \bibinfo {author} {\bibfnamefont
  {A.~T.}\ \bibnamefont {Sornborger}},\ and\ \bibinfo {author} {\bibfnamefont
  {P.~J.}\ \bibnamefont {Coles}},\ }\bibfield  {title} {\bibinfo {title}
  {Absence of {Barren} {Plateaus} in {Quantum} {Convolutional} {Neural}
  {Networks}},\ }\href {https://doi.org/10.1103/PhysRevX.11.041011} {\bibfield
  {journal} {\bibinfo  {journal} {Physical Review X}\ }\textbf {\bibinfo
  {volume} {11}},\ \bibinfo {pages} {041011} (\bibinfo {year}
  {2021})}\BibitemShut {NoStop}%
\bibitem [{\citenamefont {Grant}\ \emph {et~al.}(2018)\citenamefont {Grant},
  \citenamefont {Benedetti}, \citenamefont {Cao}, \citenamefont {Hallam},
  \citenamefont {Lockhart}, \citenamefont {Stojevic}, \citenamefont {Green},\
  and\ \citenamefont {Severini}}]{Grant2018}%
  \BibitemOpen
  \bibfield  {author} {\bibinfo {author} {\bibfnamefont {E.}~\bibnamefont
  {Grant}}, \bibinfo {author} {\bibfnamefont {M.}~\bibnamefont {Benedetti}},
  \bibinfo {author} {\bibfnamefont {S.}~\bibnamefont {Cao}}, \bibinfo {author}
  {\bibfnamefont {A.}~\bibnamefont {Hallam}}, \bibinfo {author} {\bibfnamefont
  {J.}~\bibnamefont {Lockhart}}, \bibinfo {author} {\bibfnamefont
  {V.}~\bibnamefont {Stojevic}}, \bibinfo {author} {\bibfnamefont {A.~G.}\
  \bibnamefont {Green}},\ and\ \bibinfo {author} {\bibfnamefont
  {S.}~\bibnamefont {Severini}},\ }\bibfield  {title} {\bibinfo {title}
  {Hierarchical quantum classifiers},\ }\href
  {https://doi.org/10.1038/s41534-018-0116-9} {\bibfield  {journal} {\bibinfo
  {journal} {npj Quantum Information}\ }\textbf {\bibinfo {volume} {4}},\
  \bibinfo {pages} {17} (\bibinfo {year} {2018})}\BibitemShut {NoStop}%
\bibitem [{\citenamefont {Zhao}\ and\ \citenamefont
  {Gao}(2021)}]{zhao_analyzing_2021}%
  \BibitemOpen
  \bibfield  {author} {\bibinfo {author} {\bibfnamefont {C.}~\bibnamefont
  {Zhao}}\ and\ \bibinfo {author} {\bibfnamefont {X.-S.}\ \bibnamefont {Gao}},\
  }\bibfield  {title} {\bibinfo {title} {Analyzing the barren plateau
  phenomenon in training quantum neural networks with the {ZX}-calculus},\
  }\href {https://doi.org/10.22331/q-2021-06-04-466} {\bibfield  {journal}
  {\bibinfo  {journal} {Quantum}\ }\textbf {\bibinfo {volume} {5}},\ \bibinfo
  {pages} {466} (\bibinfo {year} {2021})}\BibitemShut {NoStop}%
\bibitem [{\citenamefont {Kandala}\ \emph {et~al.}(2017)\citenamefont
  {Kandala}, \citenamefont {Mezzacapo}, \citenamefont {Temme}, \citenamefont
  {Takita}, \citenamefont {Brink}, \citenamefont {Chow},\ and\ \citenamefont
  {Gambetta}}]{kandala_hardware-efficient_2017}%
  \BibitemOpen
  \bibfield  {author} {\bibinfo {author} {\bibfnamefont {A.}~\bibnamefont
  {Kandala}}, \bibinfo {author} {\bibfnamefont {A.}~\bibnamefont {Mezzacapo}},
  \bibinfo {author} {\bibfnamefont {K.}~\bibnamefont {Temme}}, \bibinfo
  {author} {\bibfnamefont {M.}~\bibnamefont {Takita}}, \bibinfo {author}
  {\bibfnamefont {M.}~\bibnamefont {Brink}}, \bibinfo {author} {\bibfnamefont
  {J.~M.}\ \bibnamefont {Chow}},\ and\ \bibinfo {author} {\bibfnamefont
  {J.~M.}\ \bibnamefont {Gambetta}},\ }\bibfield  {title} {\bibinfo {title}
  {Hardware-efficient variational quantum eigensolver for small molecules and
  quantum magnets},\ }\href {https://doi.org/10.1038/nature23879} {\bibfield
  {journal} {\bibinfo  {journal} {Nature}\ }\textbf {\bibinfo {volume} {549}},\
  \bibinfo {pages} {242} (\bibinfo {year} {2017})}\BibitemShut {NoStop}%
\bibitem [{\citenamefont {Arrasmith}\ \emph
  {et~al.}(2021{\natexlab{a}})\citenamefont {Arrasmith}, \citenamefont
  {Holmes}, \citenamefont {Cerezo},\ and\ \citenamefont
  {Coles}}]{arrasmith_equivalence_2021}%
  \BibitemOpen
  \bibfield  {author} {\bibinfo {author} {\bibfnamefont {A.}~\bibnamefont
  {Arrasmith}}, \bibinfo {author} {\bibfnamefont {Z.}~\bibnamefont {Holmes}},
  \bibinfo {author} {\bibfnamefont {M.}~\bibnamefont {Cerezo}},\ and\ \bibinfo
  {author} {\bibfnamefont {P.~J.}\ \bibnamefont {Coles}},\ }\bibfield  {title}
  {\bibinfo {title} {Equivalence of quantum barren plateaus to cost
  concentration and narrow gorges},\ }\href {http://arxiv.org/abs/2104.05868}
  {\bibfield  {journal} {\bibinfo  {journal} {arXiv:2104.05868}\ } (\bibinfo
  {year} {2021}{\natexlab{a}})}\BibitemShut {NoStop}%
\bibitem [{\citenamefont {Arrasmith}\ \emph
  {et~al.}(2021{\natexlab{b}})\citenamefont {Arrasmith}, \citenamefont
  {Cerezo}, \citenamefont {Czarnik}, \citenamefont {Cincio},\ and\
  \citenamefont {Coles}}]{arrasmith_effect_2021}%
  \BibitemOpen
  \bibfield  {author} {\bibinfo {author} {\bibfnamefont {A.}~\bibnamefont
  {Arrasmith}}, \bibinfo {author} {\bibfnamefont {M.}~\bibnamefont {Cerezo}},
  \bibinfo {author} {\bibfnamefont {P.}~\bibnamefont {Czarnik}}, \bibinfo
  {author} {\bibfnamefont {L.}~\bibnamefont {Cincio}},\ and\ \bibinfo {author}
  {\bibfnamefont {P.~J.}\ \bibnamefont {Coles}},\ }\bibfield  {title} {\bibinfo
  {title} {Effect of barren plateaus on gradient-free optimization},\ }\href
  {https://doi.org/10.22331/q-2021-10-05-558} {\bibfield  {journal} {\bibinfo
  {journal} {Quantum}\ }\textbf {\bibinfo {volume} {5}},\ \bibinfo {pages}
  {558} (\bibinfo {year} {2021}{\natexlab{b}})}\BibitemShut {NoStop}%
\bibitem [{\citenamefont {Wu}\ \emph {et~al.}(2021)\citenamefont {Wu},
  \citenamefont {Li}, \citenamefont {Ding},\ and\ \citenamefont
  {Xie}}]{wu_mitigating_2021}%
  \BibitemOpen
  \bibfield  {author} {\bibinfo {author} {\bibfnamefont {A.}~\bibnamefont
  {Wu}}, \bibinfo {author} {\bibfnamefont {G.}~\bibnamefont {Li}}, \bibinfo
  {author} {\bibfnamefont {Y.}~\bibnamefont {Ding}},\ and\ \bibinfo {author}
  {\bibfnamefont {Y.}~\bibnamefont {Xie}},\ }\bibfield  {title} {\bibinfo
  {title} {Mitigating {Noise}-{Induced} {Gradient} {Vanishing} in {Variational}
  {Quantum} {Algorithm} {Training}},\ }\href {https://arxiv.org/abs/2111.13209}
  {\bibfield  {journal} {\bibinfo  {journal} {arXiv:2111.13209}\ } (\bibinfo
  {year} {2021})}\BibitemShut {NoStop}%
\bibitem [{\citenamefont {Zhang}\ \emph {et~al.}(2021)\citenamefont {Zhang},
  \citenamefont {Hsieh}, \citenamefont {Liu},\ and\ \citenamefont
  {Tao}}]{zhang_toward_2021}%
  \BibitemOpen
  \bibfield  {author} {\bibinfo {author} {\bibfnamefont {K.}~\bibnamefont
  {Zhang}}, \bibinfo {author} {\bibfnamefont {M.-H.}\ \bibnamefont {Hsieh}},
  \bibinfo {author} {\bibfnamefont {L.}~\bibnamefont {Liu}},\ and\ \bibinfo
  {author} {\bibfnamefont {D.}~\bibnamefont {Tao}},\ }\bibfield  {title}
  {\bibinfo {title} {Toward {Trainability} of {Deep} {Quantum} {Neural}
  {Networks}},\ }\href {http://arxiv.org/abs/2112.15002} {\bibfield  {journal}
  {\bibinfo  {journal} {arXiv:2112.15002}\ } (\bibinfo {year}
  {2021})}\BibitemShut {NoStop}%
\bibitem [{\citenamefont {Grant}\ \emph {et~al.}(2019)\citenamefont {Grant},
  \citenamefont {Ostaszewski}, \citenamefont {Wossnig},\ and\ \citenamefont
  {Benedetti}}]{Grant2019}%
  \BibitemOpen
  \bibfield  {author} {\bibinfo {author} {\bibfnamefont {E.}~\bibnamefont
  {Grant}}, \bibinfo {author} {\bibfnamefont {M.}~\bibnamefont {Ostaszewski}},
  \bibinfo {author} {\bibfnamefont {L.}~\bibnamefont {Wossnig}},\ and\ \bibinfo
  {author} {\bibfnamefont {M.}~\bibnamefont {Benedetti}},\ }\bibfield  {title}
  {\bibinfo {title} {An initialization strategy for addressing barren plateaus
  in parametrized quantum circuits},\ }\href
  {https://doi.org/10.22331/q-2019-12-09-214} {\bibfield  {journal} {\bibinfo
  {journal} {Quantum}\ }\textbf {\bibinfo {volume} {3}},\ \bibinfo {pages}
  {214} (\bibinfo {year} {2019})}\BibitemShut {NoStop}%
\bibitem [{\citenamefont {Liu}\ \emph {et~al.}(2021)\citenamefont {Liu},
  \citenamefont {Sun}, \citenamefont {Wu}, \citenamefont {Han},\ and\
  \citenamefont {Guo}}]{liu_parameter_2021}%
  \BibitemOpen
  \bibfield  {author} {\bibinfo {author} {\bibfnamefont {H.-Y.}\ \bibnamefont
  {Liu}}, \bibinfo {author} {\bibfnamefont {T.-P.}\ \bibnamefont {Sun}},
  \bibinfo {author} {\bibfnamefont {Y.-C.}\ \bibnamefont {Wu}}, \bibinfo
  {author} {\bibfnamefont {Y.-J.}\ \bibnamefont {Han}},\ and\ \bibinfo {author}
  {\bibfnamefont {G.-P.}\ \bibnamefont {Guo}},\ }\bibfield  {title} {\bibinfo
  {title} {A {Parameter} {Initialization} {Method} for {Variational} {Quantum}
  {Algorithms} to {Mitigate} {Barren} {Plateaus} {Based} on {Transfer}
  {Learning}},\ }\href {http://arxiv.org/abs/2112.10952} {\bibfield  {journal}
  {\bibinfo  {journal} {arXiv:2112.10952}\ } (\bibinfo {year}
  {2021})}\BibitemShut {NoStop}%
\bibitem [{\citenamefont {Rad}\ \emph {et~al.}(2022)\citenamefont {Rad},
  \citenamefont {Seif},\ and\ \citenamefont {Linke}}]{rad_surviving_2022}%
  \BibitemOpen
  \bibfield  {author} {\bibinfo {author} {\bibfnamefont {A.}~\bibnamefont
  {Rad}}, \bibinfo {author} {\bibfnamefont {A.}~\bibnamefont {Seif}},\ and\
  \bibinfo {author} {\bibfnamefont {N.~M.}\ \bibnamefont {Linke}},\ }\bibfield
  {title} {\bibinfo {title} {Surviving {The} {Barren} {Plateau} in
  {Variational} {Quantum} {Circuits} with {Bayesian} {Learning}
  {Initialization}},\ }\href {http://arxiv.org/abs/2203.02464} {\bibfield
  {journal} {\bibinfo  {journal} {arXiv:2203.02464}\ } (\bibinfo {year}
  {2022})}\BibitemShut {NoStop}%
\bibitem [{\citenamefont {Zhang}\ \emph {et~al.}(2022)\citenamefont {Zhang},
  \citenamefont {Hsieh}, \citenamefont {Liu},\ and\ \citenamefont
  {Tao}}]{zhang_gaussian_2022}%
  \BibitemOpen
  \bibfield  {author} {\bibinfo {author} {\bibfnamefont {K.}~\bibnamefont
  {Zhang}}, \bibinfo {author} {\bibfnamefont {M.-H.}\ \bibnamefont {Hsieh}},
  \bibinfo {author} {\bibfnamefont {L.}~\bibnamefont {Liu}},\ and\ \bibinfo
  {author} {\bibfnamefont {D.}~\bibnamefont {Tao}},\ }\bibfield  {title}
  {\bibinfo {title} {Gaussian initializations help deep variational quantum
  circuits escape from the barren plateau},\ }\href
  {http://arxiv.org/abs/2203.09376} {\bibfield  {journal} {\bibinfo  {journal}
  {arXiv:2203.09376}\ } (\bibinfo {year} {2022})}\BibitemShut {NoStop}%
\bibitem [{\citenamefont {Sack}\ \emph {et~al.}(2022)\citenamefont {Sack},
  \citenamefont {Medina}, \citenamefont {Michailidis}, \citenamefont {Kueng},\
  and\ \citenamefont {Serbyn}}]{sack_avoiding_2022}%
  \BibitemOpen
  \bibfield  {author} {\bibinfo {author} {\bibfnamefont {S.~H.}\ \bibnamefont
  {Sack}}, \bibinfo {author} {\bibfnamefont {R.~A.}\ \bibnamefont {Medina}},
  \bibinfo {author} {\bibfnamefont {A.~A.}\ \bibnamefont {Michailidis}},
  \bibinfo {author} {\bibfnamefont {R.}~\bibnamefont {Kueng}},\ and\ \bibinfo
  {author} {\bibfnamefont {M.}~\bibnamefont {Serbyn}},\ }\bibfield  {title}
  {\bibinfo {title} {Avoiding barren plateaus using classical shadows},\ }\href
  {http://arxiv.org/abs/2201.08194} {\bibfield  {journal} {\bibinfo  {journal}
  {arXiv:2201.08194}\ } (\bibinfo {year} {2022})}\BibitemShut {NoStop}%
\bibitem [{\citenamefont {Volkoff}\ and\ \citenamefont
  {Coles}(2021)}]{volkoff_large_2021}%
  \BibitemOpen
  \bibfield  {author} {\bibinfo {author} {\bibfnamefont {T.}~\bibnamefont
  {Volkoff}}\ and\ \bibinfo {author} {\bibfnamefont {P.~J.}\ \bibnamefont
  {Coles}},\ }\bibfield  {title} {\bibinfo {title} {Large gradients via
  correlation in random parameterized quantum circuits},\ }\href
  {https://doi.org/10.1088/2058-9565/abd891} {\bibfield  {journal} {\bibinfo
  {journal} {Quantum Science and Technology}\ }\textbf {\bibinfo {volume}
  {6}},\ \bibinfo {pages} {025008} (\bibinfo {year} {2021})},\ \bibinfo {note}
  {arXiv: 2005.12200}\BibitemShut {NoStop}%
\bibitem [{\citenamefont {Patti}\ \emph {et~al.}(2021)\citenamefont {Patti},
  \citenamefont {Najafi}, \citenamefont {Gao},\ and\ \citenamefont
  {Yelin}}]{patti_entanglement_2021}%
  \BibitemOpen
  \bibfield  {author} {\bibinfo {author} {\bibfnamefont {T.~L.}\ \bibnamefont
  {Patti}}, \bibinfo {author} {\bibfnamefont {K.}~\bibnamefont {Najafi}},
  \bibinfo {author} {\bibfnamefont {X.}~\bibnamefont {Gao}},\ and\ \bibinfo
  {author} {\bibfnamefont {S.~F.}\ \bibnamefont {Yelin}},\ }\bibfield  {title}
  {\bibinfo {title} {Entanglement devised barren plateau mitigation},\ }\href
  {https://doi.org/10.1103/PhysRevResearch.3.033090} {\bibfield  {journal}
  {\bibinfo  {journal} {Physical Review Research}\ }\textbf {\bibinfo {volume}
  {3}},\ \bibinfo {pages} {033090} (\bibinfo {year} {2021})}\BibitemShut
  {NoStop}%
\bibitem [{\citenamefont {Broers}\ and\ \citenamefont
  {Mathey}(2021)}]{broers_optimization_2021}%
  \BibitemOpen
  \bibfield  {author} {\bibinfo {author} {\bibfnamefont {L.}~\bibnamefont
  {Broers}}\ and\ \bibinfo {author} {\bibfnamefont {L.}~\bibnamefont
  {Mathey}},\ }\bibfield  {title} {\bibinfo {title} {Optimization of {Quantum}
  {Algorithm} {Protocols} without {Barren} {Plateaus}},\ }\href
  {https://arxiv.org/abs/2111.08085} {\bibfield  {journal} {\bibinfo  {journal}
  {arXiv:2111.08085}\ } (\bibinfo {year} {2021})}\BibitemShut {NoStop}%
\bibitem [{\citenamefont {Bravyi}\ \emph {et~al.}(2016)\citenamefont {Bravyi},
  \citenamefont {Smith},\ and\ \citenamefont {Smolin}}]{bravyi_trading_2016}%
  \BibitemOpen
  \bibfield  {author} {\bibinfo {author} {\bibfnamefont {S.}~\bibnamefont
  {Bravyi}}, \bibinfo {author} {\bibfnamefont {G.}~\bibnamefont {Smith}},\ and\
  \bibinfo {author} {\bibfnamefont {J.~A.}\ \bibnamefont {Smolin}},\ }\bibfield
   {title} {\bibinfo {title} {Trading {Classical} and {Quantum} {Computational}
  {Resources}},\ }\href {https://doi.org/10.1103/PhysRevX.6.021043} {\bibfield
  {journal} {\bibinfo  {journal} {Physical Review X}\ }\textbf {\bibinfo
  {volume} {6}},\ \bibinfo {pages} {021043} (\bibinfo {year}
  {2016})}\BibitemShut {NoStop}%
\bibitem [{\citenamefont {Peng}\ \emph {et~al.}(2020)\citenamefont {Peng},
  \citenamefont {Harrow}, \citenamefont {Ozols},\ and\ \citenamefont
  {Wu}}]{peng_simulating_2020}%
  \BibitemOpen
  \bibfield  {author} {\bibinfo {author} {\bibfnamefont {T.}~\bibnamefont
  {Peng}}, \bibinfo {author} {\bibfnamefont {A.~W.}\ \bibnamefont {Harrow}},
  \bibinfo {author} {\bibfnamefont {M.}~\bibnamefont {Ozols}},\ and\ \bibinfo
  {author} {\bibfnamefont {X.}~\bibnamefont {Wu}},\ }\bibfield  {title}
  {\bibinfo {title} {Simulating {Large} {Quantum} {Circuits} on a {Small}
  {Quantum} {Computer}},\ }\href
  {https://doi.org/10.1103/PhysRevLett.125.150504} {\bibfield  {journal}
  {\bibinfo  {journal} {Physical Review Letters}\ }\textbf {\bibinfo {volume}
  {125}},\ \bibinfo {pages} {150504} (\bibinfo {year} {2020})}\BibitemShut
  {NoStop}%
\bibitem [{\citenamefont {Tang}\ \emph {et~al.}(2021)\citenamefont {Tang},
  \citenamefont {Tomesh}, \citenamefont {Suchara}, \citenamefont {Larson},\
  and\ \citenamefont {Martonosi}}]{tang_cutqc_2021}%
  \BibitemOpen
  \bibfield  {author} {\bibinfo {author} {\bibfnamefont {W.}~\bibnamefont
  {Tang}}, \bibinfo {author} {\bibfnamefont {T.}~\bibnamefont {Tomesh}},
  \bibinfo {author} {\bibfnamefont {M.}~\bibnamefont {Suchara}}, \bibinfo
  {author} {\bibfnamefont {J.}~\bibnamefont {Larson}},\ and\ \bibinfo {author}
  {\bibfnamefont {M.}~\bibnamefont {Martonosi}},\ }\bibfield  {title} {\bibinfo
  {title} {{CutQC}: {Using} {Small} {Quantum} {Computers} for {Large} {Quantum}
  {Circuit} {Evaluations}},\ }\href {https://doi.org/10.1145/3445814.3446758}
  {\bibfield  {journal} {\bibinfo  {journal} {Proceedings of the 26th ACM
  International Conference on Architectural Support for Programming Languages
  and Operating Systems}\ ,\ \bibinfo {pages} {473}} (\bibinfo {year}
  {2021})},\ \bibinfo {note} {arXiv: 2012.02333}\BibitemShut {NoStop}%
\bibitem [{\citenamefont {Perlin}\ \emph {et~al.}(2021)\citenamefont {Perlin},
  \citenamefont {Saleem}, \citenamefont {Suchara},\ and\ \citenamefont
  {Osborn}}]{perlin_quantum_2021}%
  \BibitemOpen
  \bibfield  {author} {\bibinfo {author} {\bibfnamefont {M.~A.}\ \bibnamefont
  {Perlin}}, \bibinfo {author} {\bibfnamefont {Z.~H.}\ \bibnamefont {Saleem}},
  \bibinfo {author} {\bibfnamefont {M.}~\bibnamefont {Suchara}},\ and\ \bibinfo
  {author} {\bibfnamefont {J.~C.}\ \bibnamefont {Osborn}},\ }\bibfield  {title}
  {\bibinfo {title} {Quantum circuit cutting with maximum-likelihood
  tomography},\ }\href {https://doi.org/10.1038/s41534-021-00390-6} {\bibfield
  {journal} {\bibinfo  {journal} {npj Quantum Information}\ }\textbf {\bibinfo
  {volume} {7}},\ \bibinfo {pages} {1} (\bibinfo {year} {2021})}\BibitemShut
  {NoStop}%
\bibitem [{\citenamefont {Eddins}\ \emph {et~al.}(2021)\citenamefont {Eddins},
  \citenamefont {Motta}, \citenamefont {Gujarati}, \citenamefont {Bravyi},
  \citenamefont {Mezzacapo}, \citenamefont {Hadfield},\ and\ \citenamefont
  {Sheldon}}]{eddins_doubling_2021}%
  \BibitemOpen
  \bibfield  {author} {\bibinfo {author} {\bibfnamefont {A.}~\bibnamefont
  {Eddins}}, \bibinfo {author} {\bibfnamefont {M.}~\bibnamefont {Motta}},
  \bibinfo {author} {\bibfnamefont {T.~P.}\ \bibnamefont {Gujarati}}, \bibinfo
  {author} {\bibfnamefont {S.}~\bibnamefont {Bravyi}}, \bibinfo {author}
  {\bibfnamefont {A.}~\bibnamefont {Mezzacapo}}, \bibinfo {author}
  {\bibfnamefont {C.}~\bibnamefont {Hadfield}},\ and\ \bibinfo {author}
  {\bibfnamefont {S.}~\bibnamefont {Sheldon}},\ }\bibfield  {title} {\bibinfo
  {title} {Doubling the size of quantum simulators by entanglement forging},\
  }\href {https://arxiv.org/abs/2104.10220} {\bibfield  {journal} {\bibinfo
  {journal} {arXiv:2104.10220}\ } (\bibinfo {year} {2021})}\BibitemShut
  {NoStop}%
\bibitem [{\citenamefont {Saleem}\ \emph {et~al.}(2021)\citenamefont {Saleem},
  \citenamefont {Tomesh}, \citenamefont {Perlin}, \citenamefont {Gokhale},\
  and\ \citenamefont {Suchara}}]{saleem_quantum_2021}%
  \BibitemOpen
  \bibfield  {author} {\bibinfo {author} {\bibfnamefont {Z.~H.}\ \bibnamefont
  {Saleem}}, \bibinfo {author} {\bibfnamefont {T.}~\bibnamefont {Tomesh}},
  \bibinfo {author} {\bibfnamefont {M.~A.}\ \bibnamefont {Perlin}}, \bibinfo
  {author} {\bibfnamefont {P.}~\bibnamefont {Gokhale}},\ and\ \bibinfo {author}
  {\bibfnamefont {M.}~\bibnamefont {Suchara}},\ }\bibfield  {title} {\bibinfo
  {title} {Quantum {Divide} and {Conquer} for {Combinatorial} {Optimization}
  and {Distributed} {Computing}},\ }\href {http://arxiv.org/abs/2107.07532}
  {\bibfield  {journal} {\bibinfo  {journal} {arXiv:2107.07532}\ } (\bibinfo
  {year} {2021})}\BibitemShut {NoStop}%
\bibitem [{\citenamefont {Fujii}\ \emph {et~al.}(2022)\citenamefont {Fujii},
  \citenamefont {Mizuta}, \citenamefont {Ueda}, \citenamefont {Mitarai},
  \citenamefont {Mizukami},\ and\ \citenamefont {Nakagawa}}]{fujii_deep_2022}%
  \BibitemOpen
  \bibfield  {author} {\bibinfo {author} {\bibfnamefont {K.}~\bibnamefont
  {Fujii}}, \bibinfo {author} {\bibfnamefont {K.}~\bibnamefont {Mizuta}},
  \bibinfo {author} {\bibfnamefont {H.}~\bibnamefont {Ueda}}, \bibinfo {author}
  {\bibfnamefont {K.}~\bibnamefont {Mitarai}}, \bibinfo {author} {\bibfnamefont
  {W.}~\bibnamefont {Mizukami}},\ and\ \bibinfo {author} {\bibfnamefont
  {Y.~O.}\ \bibnamefont {Nakagawa}},\ }\bibfield  {title} {\bibinfo {title}
  {Deep {Variational} {Quantum} {Eigensolver}: {A} {Divide}-{And}-{Conquer}
  {Method} for {Solving} a {Larger} {Problem} with {Smaller} {Size} {Quantum}
  {Computers}},\ }\href {https://doi.org/10.1103/PRXQuantum.3.010346}
  {\bibfield  {journal} {\bibinfo  {journal} {PRX Quantum}\ }\textbf {\bibinfo
  {volume} {3}},\ \bibinfo {pages} {010346} (\bibinfo {year}
  {2022})}\BibitemShut {NoStop}%
\bibitem [{\citenamefont {Marshall}\ \emph {et~al.}(2022)\citenamefont
  {Marshall}, \citenamefont {Gyurik},\ and\ \citenamefont
  {Dunjko}}]{marshall_high_2022}%
  \BibitemOpen
  \bibfield  {author} {\bibinfo {author} {\bibfnamefont {S.~C.}\ \bibnamefont
  {Marshall}}, \bibinfo {author} {\bibfnamefont {C.}~\bibnamefont {Gyurik}},\
  and\ \bibinfo {author} {\bibfnamefont {V.}~\bibnamefont {Dunjko}},\
  }\bibfield  {title} {\bibinfo {title} {High {Dimensional} {Quantum}
  {Learning} {With} {Small} {Quantum} {Computers}},\ }\href
  {http://arxiv.org/abs/2203.13739} {\bibfield  {journal} {\bibinfo  {journal}
  {arXiv:2203.13739}\ } (\bibinfo {year} {2022})}\BibitemShut {NoStop}%
\bibitem [{\citenamefont {Schatzki}\ \emph {et~al.}(2021)\citenamefont
  {Schatzki}, \citenamefont {Arrasmith}, \citenamefont {Coles},\ and\
  \citenamefont {Cerezo}}]{schatzki_entangled_2021}%
  \BibitemOpen
  \bibfield  {author} {\bibinfo {author} {\bibfnamefont {L.}~\bibnamefont
  {Schatzki}}, \bibinfo {author} {\bibfnamefont {A.}~\bibnamefont {Arrasmith}},
  \bibinfo {author} {\bibfnamefont {P.~J.}\ \bibnamefont {Coles}},\ and\
  \bibinfo {author} {\bibfnamefont {M.}~\bibnamefont {Cerezo}},\ }\bibfield
  {title} {\bibinfo {title} {Entangled {Datasets} for {Quantum} {Machine}
  {Learning}},\ }\href {http://arxiv.org/abs/2109.03400} {\bibfield  {journal}
  {\bibinfo  {journal} {arXiv:2109.03400}\ } (\bibinfo {year}
  {2021})}\BibitemShut {NoStop}%
\bibitem [{\citenamefont {Treinish}\ \emph {et~al.}(2022)\citenamefont
  {Treinish}, \citenamefont {Gambetta}, \citenamefont {Nation}, \citenamefont
  {Kassebaum}, \citenamefont {qiskit bot}, \citenamefont {Rodríguez},
  \citenamefont {González}, \citenamefont {Hu}, \citenamefont {Krsulich},
  \citenamefont {Zdanski}, \citenamefont {Garrison}, \citenamefont {Yu},
  \citenamefont {Gacon}, \citenamefont {McKay}, \citenamefont {Gomez},
  \citenamefont {Capelluto}, \citenamefont {Travis-S-IBM}, \citenamefont
  {Marques}, \citenamefont {Panigrahi}, \citenamefont {Lishman}, \citenamefont
  {lerongil}, \citenamefont {Rahman}, \citenamefont {Wood}, \citenamefont
  {Bello}, \citenamefont {Itoko}, \citenamefont {Singh}, \citenamefont {Drew},
  \citenamefont {Arbel}, \citenamefont {Schwarm},\ and\ \citenamefont
  {Daniel}}]{treinish_qiskit_2022}%
  \BibitemOpen
  \bibfield  {author} {\bibinfo {author} {\bibfnamefont {M.}~\bibnamefont
  {Treinish}}, \bibinfo {author} {\bibfnamefont {J.}~\bibnamefont {Gambetta}},
  \bibinfo {author} {\bibfnamefont {P.}~\bibnamefont {Nation}}, \bibinfo
  {author} {\bibfnamefont {P.}~\bibnamefont {Kassebaum}}, \bibinfo {author}
  {\bibnamefont {qiskit bot}}, \bibinfo {author} {\bibfnamefont {D.~M.}\
  \bibnamefont {Rodríguez}}, \bibinfo {author} {\bibfnamefont {S.~d. l.~P.}\
  \bibnamefont {González}}, \bibinfo {author} {\bibfnamefont {S.}~\bibnamefont
  {Hu}}, \bibinfo {author} {\bibfnamefont {K.}~\bibnamefont {Krsulich}},
  \bibinfo {author} {\bibfnamefont {L.}~\bibnamefont {Zdanski}}, \bibinfo
  {author} {\bibfnamefont {J.}~\bibnamefont {Garrison}}, \bibinfo {author}
  {\bibfnamefont {J.}~\bibnamefont {Yu}}, \bibinfo {author} {\bibfnamefont
  {J.}~\bibnamefont {Gacon}}, \bibinfo {author} {\bibfnamefont
  {D.}~\bibnamefont {McKay}}, \bibinfo {author} {\bibfnamefont
  {J.}~\bibnamefont {Gomez}}, \bibinfo {author} {\bibfnamefont
  {L.}~\bibnamefont {Capelluto}}, \bibinfo {author} {\bibnamefont
  {Travis-S-IBM}}, \bibinfo {author} {\bibfnamefont {M.}~\bibnamefont
  {Marques}}, \bibinfo {author} {\bibfnamefont {A.}~\bibnamefont {Panigrahi}},
  \bibinfo {author} {\bibfnamefont {J.}~\bibnamefont {Lishman}}, \bibinfo
  {author} {\bibnamefont {lerongil}}, \bibinfo {author} {\bibfnamefont {R.~I.}\
  \bibnamefont {Rahman}}, \bibinfo {author} {\bibfnamefont {S.}~\bibnamefont
  {Wood}}, \bibinfo {author} {\bibfnamefont {L.}~\bibnamefont {Bello}},
  \bibinfo {author} {\bibfnamefont {T.}~\bibnamefont {Itoko}}, \bibinfo
  {author} {\bibfnamefont {D.}~\bibnamefont {Singh}}, \bibinfo {author}
  {\bibnamefont {Drew}}, \bibinfo {author} {\bibfnamefont {E.}~\bibnamefont
  {Arbel}}, \bibinfo {author} {\bibfnamefont {J.}~\bibnamefont {Schwarm}},\
  and\ \bibinfo {author} {\bibfnamefont {J.}~\bibnamefont {Daniel}},\ }\href
  {https://doi.org/10.5281/zenodo.6403335} {\bibinfo {title} {Qiskit: {An}
  {Open}-source {Framework} for {Quantum} {Computing}}} (\bibinfo {year}
  {2022})\BibitemShut {NoStop}%
\bibitem [{\citenamefont {Bergholm}\ \emph {et~al.}(2020)\citenamefont
  {Bergholm}, \citenamefont {Izaac}, \citenamefont {Schuld}, \citenamefont
  {Gogolin}, \citenamefont {Alam}, \citenamefont {Ahmed}, \citenamefont
  {Arrazola}, \citenamefont {Blank}, \citenamefont {Delgado}, \citenamefont
  {Jahangiri}, \citenamefont {McKiernan}, \citenamefont {Meyer}, \citenamefont
  {Niu}, \citenamefont {Száva},\ and\ \citenamefont
  {Killoran}}]{bergholm_pennylane_2020}%
  \BibitemOpen
  \bibfield  {author} {\bibinfo {author} {\bibfnamefont {V.}~\bibnamefont
  {Bergholm}}, \bibinfo {author} {\bibfnamefont {J.}~\bibnamefont {Izaac}},
  \bibinfo {author} {\bibfnamefont {M.}~\bibnamefont {Schuld}}, \bibinfo
  {author} {\bibfnamefont {C.}~\bibnamefont {Gogolin}}, \bibinfo {author}
  {\bibfnamefont {M.~S.}\ \bibnamefont {Alam}}, \bibinfo {author}
  {\bibfnamefont {S.}~\bibnamefont {Ahmed}}, \bibinfo {author} {\bibfnamefont
  {J.~M.}\ \bibnamefont {Arrazola}}, \bibinfo {author} {\bibfnamefont
  {C.}~\bibnamefont {Blank}}, \bibinfo {author} {\bibfnamefont
  {A.}~\bibnamefont {Delgado}}, \bibinfo {author} {\bibfnamefont
  {S.}~\bibnamefont {Jahangiri}}, \bibinfo {author} {\bibfnamefont
  {K.}~\bibnamefont {McKiernan}}, \bibinfo {author} {\bibfnamefont {J.~J.}\
  \bibnamefont {Meyer}}, \bibinfo {author} {\bibfnamefont {Z.}~\bibnamefont
  {Niu}}, \bibinfo {author} {\bibfnamefont {A.}~\bibnamefont {Száva}},\ and\
  \bibinfo {author} {\bibfnamefont {N.}~\bibnamefont {Killoran}},\ }\bibfield
  {title} {\bibinfo {title} {{PennyLane}: {Automatic} differentiation of hybrid
  quantum-classical computations},\ }\href {http://arxiv.org/abs/1811.04968}
  {\bibfield  {journal} {\bibinfo  {journal} {arXiv:1811.04968}\ } (\bibinfo
  {year} {2020})}\BibitemShut {NoStop}%
\bibitem [{\citenamefont {Paszke}\ \emph {et~al.}(2019)\citenamefont {Paszke},
  \citenamefont {Gross}, \citenamefont {Massa}, \citenamefont {Lerer},
  \citenamefont {Bradbury}, \citenamefont {Chanan}, \citenamefont {Killeen},
  \citenamefont {Lin}, \citenamefont {Gimelshein}, \citenamefont {Antiga},
  \citenamefont {Desmaison}, \citenamefont {Kopf}, \citenamefont {Yang},
  \citenamefont {DeVito}, \citenamefont {Raison}, \citenamefont {Tejani},
  \citenamefont {Chilamkurthy}, \citenamefont {Steiner}, \citenamefont {Fang},
  \citenamefont {Bai},\ and\ \citenamefont {Chintala}}]{paszke_pytorch_2019}%
  \BibitemOpen
  \bibfield  {author} {\bibinfo {author} {\bibfnamefont {A.}~\bibnamefont
  {Paszke}}, \bibinfo {author} {\bibfnamefont {S.}~\bibnamefont {Gross}},
  \bibinfo {author} {\bibfnamefont {F.}~\bibnamefont {Massa}}, \bibinfo
  {author} {\bibfnamefont {A.}~\bibnamefont {Lerer}}, \bibinfo {author}
  {\bibfnamefont {J.}~\bibnamefont {Bradbury}}, \bibinfo {author}
  {\bibfnamefont {G.}~\bibnamefont {Chanan}}, \bibinfo {author} {\bibfnamefont
  {T.}~\bibnamefont {Killeen}}, \bibinfo {author} {\bibfnamefont
  {Z.}~\bibnamefont {Lin}}, \bibinfo {author} {\bibfnamefont {N.}~\bibnamefont
  {Gimelshein}}, \bibinfo {author} {\bibfnamefont {L.}~\bibnamefont {Antiga}},
  \bibinfo {author} {\bibfnamefont {A.}~\bibnamefont {Desmaison}}, \bibinfo
  {author} {\bibfnamefont {A.}~\bibnamefont {Kopf}}, \bibinfo {author}
  {\bibfnamefont {E.}~\bibnamefont {Yang}}, \bibinfo {author} {\bibfnamefont
  {Z.}~\bibnamefont {DeVito}}, \bibinfo {author} {\bibfnamefont
  {M.}~\bibnamefont {Raison}}, \bibinfo {author} {\bibfnamefont
  {A.}~\bibnamefont {Tejani}}, \bibinfo {author} {\bibfnamefont
  {S.}~\bibnamefont {Chilamkurthy}}, \bibinfo {author} {\bibfnamefont
  {B.}~\bibnamefont {Steiner}}, \bibinfo {author} {\bibfnamefont
  {L.}~\bibnamefont {Fang}}, \bibinfo {author} {\bibfnamefont {J.}~\bibnamefont
  {Bai}},\ and\ \bibinfo {author} {\bibfnamefont {S.}~\bibnamefont
  {Chintala}},\ }\bibfield  {title} {\bibinfo {title} {{PyTorch}: {An}
  {Imperative} {Style}, {High}-{Performance} {Deep} {Learning} {Library}},\
  }in\ \href
  {https://proceedings.neurips.cc/paper/2019/file/bdbca288fee7f92f2bfa9f7012727740-Paper.pdf}
  {\emph {\bibinfo {booktitle} {Advances in {Neural} {Information} {Processing}
  {Systems}}}},\ Vol.~\bibinfo {volume} {32},\ \bibinfo {editor} {edited by\
  \bibinfo {editor} {\bibfnamefont {H.}~\bibnamefont {Wallach}}, \bibinfo
  {editor} {\bibfnamefont {H.}~\bibnamefont {Larochelle}}, \bibinfo {editor}
  {\bibfnamefont {A.}~\bibnamefont {Beygelzimer}}, \bibinfo {editor}
  {\bibfnamefont {F.~d.}\ \bibnamefont {Alché-Buc}}, \bibinfo {editor}
  {\bibfnamefont {E.}~\bibnamefont {Fox}},\ and\ \bibinfo {editor}
  {\bibfnamefont {R.}~\bibnamefont {Garnett}}}\ (\bibinfo  {publisher} {Curran
  Associates, Inc.},\ \bibinfo {year} {2019})\BibitemShut {NoStop}%
\bibitem [{\citenamefont {Pedregosa}\ \emph {et~al.}(2011)\citenamefont
  {Pedregosa}, \citenamefont {Varoquaux}, \citenamefont {Gramfort},
  \citenamefont {Michel}, \citenamefont {Thirion}, \citenamefont {Grisel},
  \citenamefont {Blondel}, \citenamefont {Prettenhofer}, \citenamefont {Weiss},
  \citenamefont {Dubourg}, \citenamefont {Vanderplas}, \citenamefont {Passos},
  \citenamefont {Cournapeau}, \citenamefont {Brucher}, \citenamefont {Perrot},\
  and\ \citenamefont {Duchesnay}}]{pedregosa_scikit-learn_2011}%
  \BibitemOpen
  \bibfield  {author} {\bibinfo {author} {\bibfnamefont {F.}~\bibnamefont
  {Pedregosa}}, \bibinfo {author} {\bibfnamefont {G.}~\bibnamefont
  {Varoquaux}}, \bibinfo {author} {\bibfnamefont {A.}~\bibnamefont {Gramfort}},
  \bibinfo {author} {\bibfnamefont {V.}~\bibnamefont {Michel}}, \bibinfo
  {author} {\bibfnamefont {B.}~\bibnamefont {Thirion}}, \bibinfo {author}
  {\bibfnamefont {O.}~\bibnamefont {Grisel}}, \bibinfo {author} {\bibfnamefont
  {M.}~\bibnamefont {Blondel}}, \bibinfo {author} {\bibfnamefont
  {P.}~\bibnamefont {Prettenhofer}}, \bibinfo {author} {\bibfnamefont
  {R.}~\bibnamefont {Weiss}}, \bibinfo {author} {\bibfnamefont
  {V.}~\bibnamefont {Dubourg}}, \bibinfo {author} {\bibfnamefont
  {J.}~\bibnamefont {Vanderplas}}, \bibinfo {author} {\bibfnamefont
  {A.}~\bibnamefont {Passos}}, \bibinfo {author} {\bibfnamefont
  {D.}~\bibnamefont {Cournapeau}}, \bibinfo {author} {\bibfnamefont
  {M.}~\bibnamefont {Brucher}}, \bibinfo {author} {\bibfnamefont
  {M.}~\bibnamefont {Perrot}},\ and\ \bibinfo {author} {\bibfnamefont
  {E.}~\bibnamefont {Duchesnay}},\ }\bibfield  {title} {\bibinfo {title}
  {Scikit-learn: {Machine} {Learning} in {Python}},\ }\href
  {http://jmlr.org/papers/v12/pedregosa11a.html} {\bibfield  {journal}
  {\bibinfo  {journal} {Journal of Machine Learning Research}\ }\textbf
  {\bibinfo {volume} {12}},\ \bibinfo {pages} {2825} (\bibinfo {year}
  {2011})}\BibitemShut {NoStop}%
\bibitem [{\citenamefont {Kingma}\ and\ \citenamefont
  {Ba}(2017)}]{kingma_adam_2017}%
  \BibitemOpen
  \bibfield  {author} {\bibinfo {author} {\bibfnamefont {D.~P.}\ \bibnamefont
  {Kingma}}\ and\ \bibinfo {author} {\bibfnamefont {J.}~\bibnamefont {Ba}},\
  }\bibfield  {title} {\bibinfo {title} {Adam: {A} {Method} for {Stochastic}
  {Optimization}},\ }\href {http://arxiv.org/abs/1412.6980} {\bibfield
  {journal} {\bibinfo  {journal} {arXiv:1412.6980}\ } (\bibinfo {year}
  {2017})}\BibitemShut {NoStop}%
\bibitem [{\citenamefont {Beckey}\ \emph {et~al.}(2021)\citenamefont {Beckey},
  \citenamefont {Gigena}, \citenamefont {Coles},\ and\ \citenamefont
  {Cerezo}}]{beckey_computable_2021}%
  \BibitemOpen
  \bibfield  {author} {\bibinfo {author} {\bibfnamefont {J.~L.}\ \bibnamefont
  {Beckey}}, \bibinfo {author} {\bibfnamefont {N.}~\bibnamefont {Gigena}},
  \bibinfo {author} {\bibfnamefont {P.~J.}\ \bibnamefont {Coles}},\ and\
  \bibinfo {author} {\bibfnamefont {M.}~\bibnamefont {Cerezo}},\ }\bibfield
  {title} {\bibinfo {title} {Computable and {Operationally} {Meaningful}
  {Multipartite} {Entanglement} {Measures}},\ }\href
  {https://doi.org/10.1103/PhysRevLett.127.140501} {\bibfield  {journal}
  {\bibinfo  {journal} {Physical Review Letters}\ }\textbf {\bibinfo {volume}
  {127}},\ \bibinfo {pages} {140501} (\bibinfo {year} {2021})}\BibitemShut
  {NoStop}%
\bibitem [{\citenamefont {Haferkamp}\ \emph {et~al.}(2021)\citenamefont
  {Haferkamp}, \citenamefont {Faist}, \citenamefont {Kothakonda}, \citenamefont
  {Eisert},\ and\ \citenamefont {Halpern}}]{haferkamp_linear_2021}%
  \BibitemOpen
  \bibfield  {author} {\bibinfo {author} {\bibfnamefont {J.}~\bibnamefont
  {Haferkamp}}, \bibinfo {author} {\bibfnamefont {P.}~\bibnamefont {Faist}},
  \bibinfo {author} {\bibfnamefont {N.~B.~T.}\ \bibnamefont {Kothakonda}},
  \bibinfo {author} {\bibfnamefont {J.}~\bibnamefont {Eisert}},\ and\ \bibinfo
  {author} {\bibfnamefont {N.~Y.}\ \bibnamefont {Halpern}},\ }\bibfield
  {title} {\bibinfo {title} {Linear growth of quantum circuit complexity},\
  }\href {http://arxiv.org/abs/2106.05305} {\bibfield  {journal} {\bibinfo
  {journal} {arXiv:2106.05305}\ } (\bibinfo {year} {2021})}\BibitemShut
  {NoStop}%
\bibitem [{\citenamefont {Spall}(1998)}]{spsa}%
  \BibitemOpen
  \bibfield  {author} {\bibinfo {author} {\bibfnamefont {J.~C.}\ \bibnamefont
  {Spall}},\ }\bibfield  {title} {\bibinfo {title} {Overview of the
  simultaneous perturbation method for efficient optimization},\ }\href
  {https://www.jhuapl.edu/SPSA/PDF-SPSA/Spall_An_Overview.PDF} {\bibfield
  {journal} {\bibinfo  {journal} {Johns Hopkins {APL} Technical Digest}\
  }\textbf {\bibinfo {volume} {19}},\ \bibinfo {pages} {482} (\bibinfo {year}
  {1998})}\BibitemShut {NoStop}%
\bibitem [{\citenamefont {Pérez-Salinas}\ \emph {et~al.}(2020)\citenamefont
  {Pérez-Salinas}, \citenamefont {Cervera-Lierta}, \citenamefont
  {Gil-Fuster},\ and\ \citenamefont {Latorre}}]{perez-salinas_data_2020}%
  \BibitemOpen
  \bibfield  {author} {\bibinfo {author} {\bibfnamefont {A.}~\bibnamefont
  {Pérez-Salinas}}, \bibinfo {author} {\bibfnamefont {A.}~\bibnamefont
  {Cervera-Lierta}}, \bibinfo {author} {\bibfnamefont {E.}~\bibnamefont
  {Gil-Fuster}},\ and\ \bibinfo {author} {\bibfnamefont {J.~I.}\ \bibnamefont
  {Latorre}},\ }\bibfield  {title} {\bibinfo {title} {Data re-uploading for a
  universal quantum classifier},\ }\href
  {https://doi.org/10.22331/q-2020-02-06-226} {\bibfield  {journal} {\bibinfo
  {journal} {Quantum}\ }\textbf {\bibinfo {volume} {4}},\ \bibinfo {pages}
  {226} (\bibinfo {year} {2020})}\BibitemShut {NoStop}%
\bibitem [{\citenamefont {Mitarai}\ \emph {et~al.}(2018)\citenamefont
  {Mitarai}, \citenamefont {Negoro}, \citenamefont {Kitagawa},\ and\
  \citenamefont {Fujii}}]{mitarai_quantum_2018}%
  \BibitemOpen
  \bibfield  {author} {\bibinfo {author} {\bibfnamefont {K.}~\bibnamefont
  {Mitarai}}, \bibinfo {author} {\bibfnamefont {M.}~\bibnamefont {Negoro}},
  \bibinfo {author} {\bibfnamefont {M.}~\bibnamefont {Kitagawa}},\ and\
  \bibinfo {author} {\bibfnamefont {K.}~\bibnamefont {Fujii}},\ }\bibfield
  {title} {\bibinfo {title} {Quantum circuit learning},\ }\href
  {https://doi.org/10.1103/PhysRevA.98.032309} {\bibfield  {journal} {\bibinfo
  {journal} {Physical Review A}\ }\textbf {\bibinfo {volume} {98}},\ \bibinfo
  {pages} {032309} (\bibinfo {year} {2018})}\BibitemShut {NoStop}%
\bibitem [{\citenamefont {Schuld}\ \emph {et~al.}(2019)\citenamefont {Schuld},
  \citenamefont {Bergholm}, \citenamefont {Gogolin}, \citenamefont {Izaac},\
  and\ \citenamefont {Killoran}}]{schuld_evaluating_2019}%
  \BibitemOpen
  \bibfield  {author} {\bibinfo {author} {\bibfnamefont {M.}~\bibnamefont
  {Schuld}}, \bibinfo {author} {\bibfnamefont {V.}~\bibnamefont {Bergholm}},
  \bibinfo {author} {\bibfnamefont {C.}~\bibnamefont {Gogolin}}, \bibinfo
  {author} {\bibfnamefont {J.}~\bibnamefont {Izaac}},\ and\ \bibinfo {author}
  {\bibfnamefont {N.}~\bibnamefont {Killoran}},\ }\bibfield  {title} {\bibinfo
  {title} {Evaluating analytic gradients on quantum hardware},\ }\href
  {https://doi.org/10.1103/PhysRevA.99.032331} {\bibfield  {journal} {\bibinfo
  {journal} {Physical Review A}\ }\textbf {\bibinfo {volume} {99}},\ \bibinfo
  {pages} {1} (\bibinfo {year} {2019})}\BibitemShut {NoStop}%
\bibitem [{\citenamefont {Weidenfeller}\ \emph {et~al.}(2022)\citenamefont
  {Weidenfeller}, \citenamefont {Valor}, \citenamefont {Gacon}, \citenamefont
  {Tornow}, \citenamefont {Bello}, \citenamefont {Woerner},\ and\ \citenamefont
  {Egger}}]{weidenfeller_scaling_2022}%
  \BibitemOpen
  \bibfield  {author} {\bibinfo {author} {\bibfnamefont {J.}~\bibnamefont
  {Weidenfeller}}, \bibinfo {author} {\bibfnamefont {L.~C.}\ \bibnamefont
  {Valor}}, \bibinfo {author} {\bibfnamefont {J.}~\bibnamefont {Gacon}},
  \bibinfo {author} {\bibfnamefont {C.}~\bibnamefont {Tornow}}, \bibinfo
  {author} {\bibfnamefont {L.}~\bibnamefont {Bello}}, \bibinfo {author}
  {\bibfnamefont {S.}~\bibnamefont {Woerner}},\ and\ \bibinfo {author}
  {\bibfnamefont {D.~J.}\ \bibnamefont {Egger}},\ }\bibfield  {title} {\bibinfo
  {title} {Scaling of the quantum approximate optimization algorithm on
  superconducting qubit based hardware},\ }\bibfield  {journal} {\bibinfo
  {journal} {arXiv:2202.03459}\ }\href
  {https://doi.org/10.48550/arXiv.2202.03459} {10.48550/arXiv.2202.03459}
  (\bibinfo {year} {2022})\BibitemShut {NoStop}%
\bibitem [{\citenamefont {Botea}\ \emph {et~al.}(2018)\citenamefont {Botea},
  \citenamefont {Kishimoto},\ and\ \citenamefont
  {Marinescu}}]{botea_complexity_2018}%
  \BibitemOpen
  \bibfield  {author} {\bibinfo {author} {\bibfnamefont {A.}~\bibnamefont
  {Botea}}, \bibinfo {author} {\bibfnamefont {A.}~\bibnamefont {Kishimoto}},\
  and\ \bibinfo {author} {\bibfnamefont {R.}~\bibnamefont {Marinescu}},\
  }\bibfield  {title} {\bibinfo {title} {On the {Complexity} of {Quantum}
  {Circuit} {Compilation}},\ }\href
  {https://ojs.aaai.org/index.php/SOCS/article/view/18463} {\bibfield
  {journal} {\bibinfo  {journal} {Proceedings of the International Symposium on
  Combinatorial Search}\ }\textbf {\bibinfo {volume} {9}},\ \bibinfo {pages}
  {138} (\bibinfo {year} {2018})}\BibitemShut {NoStop}%
\bibitem [{\citenamefont {Basu}\ \emph {et~al.}(2022)\citenamefont {Basu},
  \citenamefont {Saha}, \citenamefont {Chakrabarti},\ and\ \citenamefont
  {Sur-Kolay}}]{basu_i-qer_2022}%
  \BibitemOpen
  \bibfield  {author} {\bibinfo {author} {\bibfnamefont {S.}~\bibnamefont
  {Basu}}, \bibinfo {author} {\bibfnamefont {A.}~\bibnamefont {Saha}}, \bibinfo
  {author} {\bibfnamefont {A.}~\bibnamefont {Chakrabarti}},\ and\ \bibinfo
  {author} {\bibfnamefont {S.}~\bibnamefont {Sur-Kolay}},\ }\bibfield  {title}
  {\bibinfo {title} {$i$-{QER}: {An} {Intelligent} {Approach} towards {Quantum}
  {Error} {Reduction}},\ }\href {https://arxiv.org/abs/2110.06347} {\bibfield
  {journal} {\bibinfo  {journal} {arXiv:2110.06347}\ } (\bibinfo {year}
  {2022})}\BibitemShut {NoStop}%
\bibitem [{\citenamefont {Miszczak}\ and\ \citenamefont
  {Puchała}(2017)}]{miszczak_symbolic_2017}%
  \BibitemOpen
  \bibfield  {author} {\bibinfo {author} {\bibfnamefont {J.~A.}\ \bibnamefont
  {Miszczak}}\ and\ \bibinfo {author} {\bibfnamefont {Z.}~\bibnamefont
  {Puchała}},\ }\bibfield  {title} {\bibinfo {title} {Symbolic integration
  with respect to the {Haar} measure on the unitary groups},\ }\href
  {https://journals.pan.pl/dlibra/publication/121307/edition/105697} {\bibfield
   {journal} {\bibinfo  {journal} {Bulletin of the Polish Academy of Sciences:
  Technical Sciences; 2017; 65; No 1; 21-27}\ } (\bibinfo {year}
  {2017})}\BibitemShut {NoStop}%
\end{thebibliography}%


%

 
\onecolumngrid
\newpage
\appendix

\section{\label{app:bp}}
When analyzing the size of the gradients of an ansatz we need tools that allows
integration over all states allowed by the ansatz over the d-dimensional
Hilbert Space. This can be achieved by using the Haar measure. Haar measure is
an invariant measure over the SU(d) group. An ensemble of unitary
operators $U$ is called as a unitary t-design if they are equal to the 
Haar measure $\mu(U)$ up-to polynomial order t. Then, the expectation of
ensemble $U$, where unitary $V_i$ can be sampled with probability $p_i$ is
given as,

\begin{equation}
    \mathbb{E}_{H}^t(\rho) = \int U^{\otimes t} \rho (U^{\otimes t})^{\dagger}
    \,dU = \sum_i p_iV_i^{\otimes t}\rho(V_i^{\otimes t})^\dagger.
    \label{app-eq:haar}
\end{equation}

Then, to perform symbolic integration over the Haar measure we will need to use
some properties of the measure~\cite{miszczak_symbolic_2017}. For the first moment we have,

\begin{equation}
    \int d\mu(U)U_{ij}U_{km}^* = \frac{\delta_{ik}\delta_{jm}}{d},
    \label{eq:unitary-1}
\end{equation}

where $d$ is the dimension of the Unitary, such that $d=2^N$ and N is number of qubits. 
Then, for the second moment we have,

\begin{multline}
    \int d\mu(U)U_{i_1j_1}U_{i_2j_2}U_{k_1m_1}^*U_{k_2m_2}^* = \\
    = \frac{
    \delta_{i_1k_1}\delta_{j_1m_1}\delta_{i_2k_2}\delta_{j_2m_2} +
    \delta_{i_1k_2}\delta_{i_2k_1}\delta_{j_1m_2}\delta_{j_2m_1}
    }{d^2+1}
    - 
    \frac{
    \delta_{i_1k_1}\delta_{j_2m_2}\delta_{j_1m_2}\delta_{j_2m_1} +
    \delta_{i_1k_2}\delta_{i_2k_1}\delta_{j_1m_1}\delta_{j_2m_2}
    }{d(d^2+1)}
\end{multline}

Then one can derive the following identities for integrals over the Haar measure~\cite{mcclean_barren_2018, cerezo_cost_2021, holmes_connecting_2022},

\begin{equation}
    \int d\mu(U)\mbox{Tr}[UAU^\dagger B] = \frac{\mbox{Tr}[A]\mbox{Tr}[B]}{d}.
    \label{eq:one-trace}
\end{equation}

We can extend this to the second moment to obtain the following identity,

\begin{multline}
    \int d\mu(U)\mbox{Tr}[UAU^\dagger BUCU^\dagger D] = \\
    = \frac{\mbox{Tr}[A]\mbox{Tr}[C]\mbox{Tr}[BD]+\mbox{Tr}[AC]\mbox{Tr}[B]\mbox{Tr}[D]}{d^2-1}
    -
    \frac{\mbox{Tr}[AC]\mbox{Tr}[BD]+\mbox{Tr}[A]\mbox{Tr}[B]\mbox{Tr}[C]\mbox{Tr}[D]}{d(d^2-1)}.
    \label{eq:long-trace}
\end{multline}

We also have,
\begin{multline}
    \int d\mu(U)\mbox{Tr}[UAU^\dagger B]\mbox{Tr}[UCU^\dagger D] = \\
    = \frac{\mbox{Tr}[AC]\mbox{Tr}[B]\mbox{Tr}[D]+\mbox{Tr}[AC]\mbox{Tr}[BD]}{d^2-1}
    -
    \frac{\mbox{Tr}[AC]\mbox{Tr}[B]\mbox{Tr}[D]+\mbox{Tr}[A]\mbox{Tr}[C]\mbox{Tr}[BD]}{d(d^2-1)}.
    \label{eq:two-trace}
\end{multline}

Now, we can use these identities to compute the average value of the gradients. 
Let's start by reminding ourselves the definitions we used before. The ansatz is 
composed of consecutive parametrized ($V$) and non-parametrized entangling ($W$)
layers. We define  $U_l(\theta_l) = \mbox{exp}(-i\theta_l V_l)$, where $V_l$ is a
Hermitian operator and $W_l$ is a generic unitary operator. Then, the curcuit ansatz can
be expressed with a multiplication of layers,

\begin{equation}
    U(\boldsymbol{\theta}) = \prod_{l=1}^{L} U_l(\theta_l)W_l
\end{equation}

For an observable $O$ and an input state $\rho$, the cost function is given as

\begin{equation}
    C(\boldsymbol{\theta}) = \mbox{Tr}[OU(\boldsymbol{\theta})\rho 
    U^\dagger(\boldsymbol{\theta})]
\end{equation}

The ansatz can be separated into two parts to investigate a certain layer, such
that $U_- \equiv \prod_{l=1}^{j-1} U_l(\theta_l)W_l$ and
$U_+ \equiv \prod_{l=j}^{L} U_l(\theta_l)W_l$. Then, the gradient of the
$j^{th}$ parameter can be expressed as~\cite{mcclean_barren_2018}

\begin{align}
    \partial_{j} C(\boldsymbol{\theta}) = 
    \frac{\partial C(\boldsymbol{\theta})}{\partial{\theta_j}}
    &= i\,\mbox{Tr}[[V_j,U_+^\dagger OU_+]U_- \rho U_-^\dagger]
\end{align}

Then the expected value of the gradient can be computed by using the Haar integral such that, 

\begin{align}
    \langle \partial_{j}C(\boldsymbol{\theta}) \rangle &=
    i \int d\mu(U_-)d\mu(U_+) \mbox{Tr}[[V_j,U_+^\dagger OU_+]U_- \rho U_-^\dagger] \label{eq:avg-grad-1}\\
    &= \frac{i\mbox{Tr}[\rho]}{d} \int d\mu(U_+) \mbox{Tr}[[V_j,U_+^\dagger OU_+] = 0, \label{eq:avg-grad-2}
\end{align}

where we use Eq.~\eqref{eq:one-trace} to obtain \eqref{eq:avg-grad-1} and use the fact that trace of the commutator is zero in \eqref{eq:avg-grad-2}. This proves that the gradients are centered around zero. Then, the variance of the gradient can inform us
about the size of the gradients. The variance is defined as,

\begin{equation}
    \mbox{Var}[\partial_{j}C(\boldsymbol{\theta})] 
    = \langle (\partial_{j}C(\boldsymbol{\theta}))^2 \rangle
    - \langle \partial_{j}C(\boldsymbol{\theta}) \rangle^2
    = \langle (\partial_{j}C(\boldsymbol{\theta}))^2 \rangle
\end{equation}

We can compute the expected value of the variance using the same logic. Then we have,

\begin{align}
    \mbox{Var}[\partial_{j}C(\boldsymbol{\theta})]  &=
    - \int d\mu(U_-)d\mu(U_+) \mbox{Tr}[[V_j,U_+^\dagger OU_+]U_- \rho U_-^\dagger ]^2 \\
    &= -\frac{1}{d^2-1} \left(\int d\mu(U_+) \mbox{Tr}[\rho]^2\mbox{Tr}[[V_j,U_+^\dagger OU_+]]^2+
    \mbox{Tr}[\rho^2]\mbox{Tr}[[V_j,U_+^\dagger OU_+]^2]\right) \label{eq:line-2} \\
    &+ \frac{1}{d(d^2-1)} \left(\int d\mu(U_+)
    \mbox{Tr}[\rho^2]\mbox{Tr}[[V_j,U_+^\dagger OU_+]^2]+\mbox{Tr}[\rho]^2\mbox{Tr}[[V_j,U_+^\dagger OU_+]]^2\right) \\
    &= - \left(\mbox{Tr}[\rho^2] - \frac{1}{d} \right) \frac{1}{d^2-1} \int d\mu(U_+) \mbox{Tr}[[V_j,U_+^\dagger OU_+]^2]
    \label{eq:var-expected}
\end{align}

We use Eq.~\eqref{eq:two-trace} to obtain Eq.~\eqref{eq:line-2}.
Then, use the fact that commutator being traceless to obtain
Eq.~\eqref{eq:var-expected}. To compute the integral of
Eq.~\eqref{eq:var-expected} we need another identity such
that~\cite{holmes_connecting_2022},

\begin{equation}
    \mbox{Tr}[[V_j,U_+^\dagger OU_+]^2] 
    = 2\mbox{Tr}[U_+ V_j U_+^\dagger O U_+ V_j U_+^\dagger O] 
    - 2\mbox{Tr}[U_+ V_j^2 U_+^\dagger O^2].
\end{equation}

Then, the variance becomes,

\begin{align}
    \mbox{Var}[\partial_{j}C(\boldsymbol{\theta})] =
    - \left(\mbox{Tr}[\rho^2] - \frac{1}{d} \right)\frac{2}{d^2-1} \left( 
    \int d\mu(U_+)\mbox{Tr}[U_+ V_j U_+^\dagger O U_+ V_j U_+^\dagger O] 
    +\int d\mu(U_+)\mbox{Tr}[U_+ V_j^2 U_+^\dagger O^2] 
    \right).
\end{align}

The first integral can be computed using Eq.~\eqref{eq:long-trace}
and the second can be computed using Eq.~\eqref{eq:one-trace}. Then
we obtain,

\begin{multline}
    \mbox{Var}[\partial_{j}C(\boldsymbol{\theta})] =
    - \left(\mbox{Tr}[\rho^2] - \frac{1}{d} \right)\frac{2}{d^2-1} \left( \right.
    \frac{1}{d^2-1}(\mbox{Tr}[V]^2 \mbox{Tr}[O]^2 + \mbox{Tr}[V^2] \mbox{Tr}[O^2]) \\
    - \frac{1}{d(d^2-1)}(\mbox{Tr}[V]^2 \mbox{Tr}[O]^2 + \mbox{Tr}[V^2] \mbox{Tr}[O^2])
    - \frac{1}{d}\mbox{Tr}[V^2] \mbox{Tr}[O^2] \left. \right) \\
    = - \left(\mbox{Tr}[\rho^2] - \frac{1}{d} \right)\frac{2\mbox{Tr}[V^2] \mbox{Tr}[O^2]}{d^2-1}
    \left( \frac{d-1}{d(d^2-1)}(1+\mbox{Tr}[V]^2 \mbox{Tr}[O]^2) - \frac{1}{d} \right).
\end{multline}

Finally, the asymptotic behaviour of the variance can be expressed
as

\begin{equation}
    \mbox{Var}[\partial_{j}C(\boldsymbol{\theta})] \approx \mathcal{O} \left( \frac{1}{d^6} \right) \approx \mathcal{O} \left( \frac{1}{2^{6N}} \right),
\end{equation}

where $d=2^N$. Thus, the variance vanishes exponentially with
respect to N.

\newpage
\section{}
\label{app:bp-numerics}
\begin{figure}[!h]
    \centering
    \includegraphics[width=.7\linewidth]{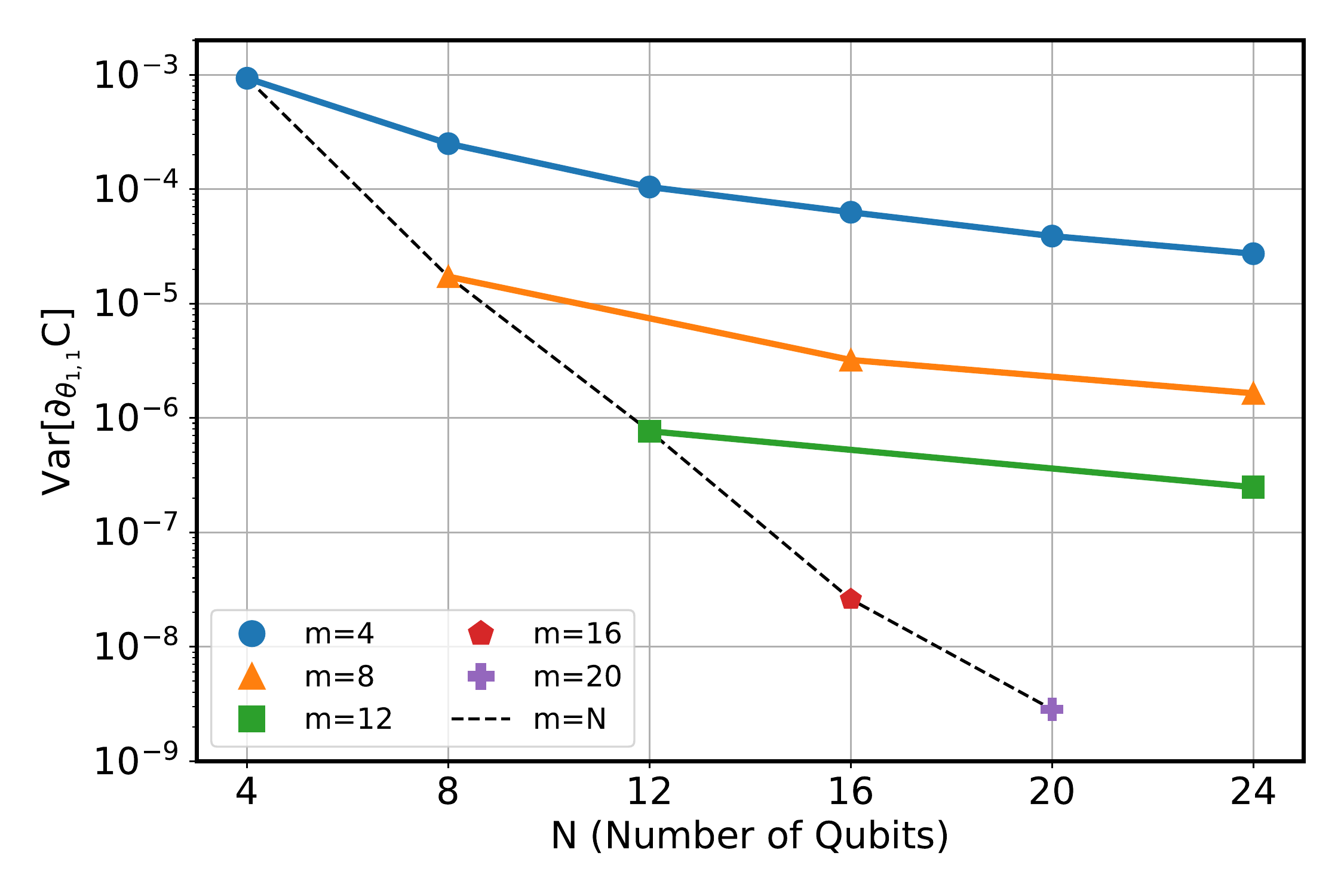}
    \caption{The variance of the gradients of the
    first parameter of the ansatz as a function of the number of qubits for varying
    values of $m$. Each color/marker represents a certain value of $m$
    and data points of the standard ansatz ($m=N$) is plotted with 
    a dashed black line.
    }
    \label{fig:app-b1}
\end{figure}

\begin{figure}[!h]
    \centering
    \includegraphics[width=.7\linewidth]{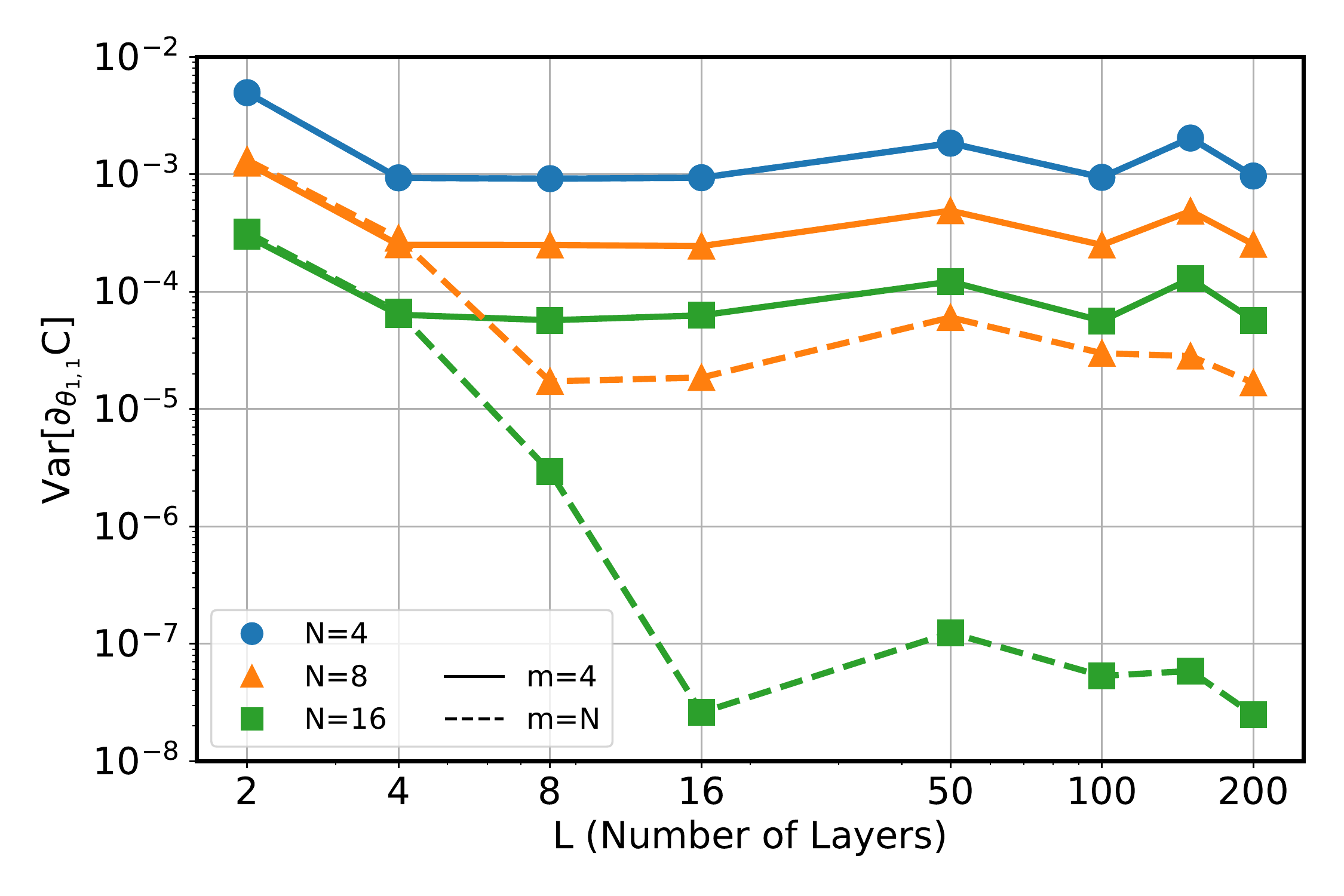}
    \caption{The log plot of variance of the gradients of the
    first parameter of the ansatz vs. number of layers for $m=4$
    (solid lines) and $m=N$ (dashed lines) with varying number 
    of qubits.}
    \label{fig:app-b2}
\end{figure}

\begin{figure*}[!h]
    \centering
    \includegraphics[width=.75\linewidth]{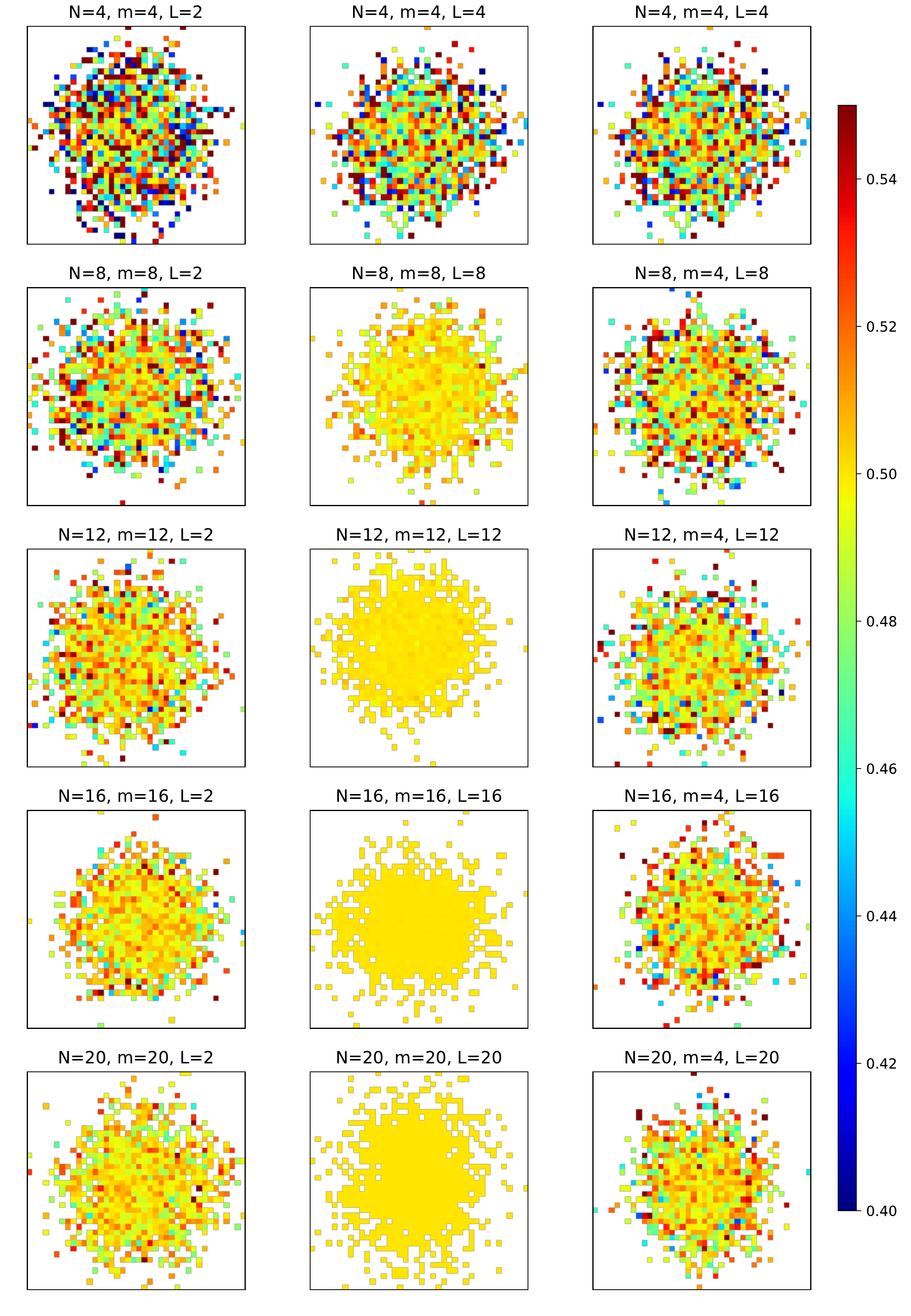}
    \caption{Cost landscapes of ans\"atze with different settings.
    Parameters of the ansatz are reduced down to two using PCA and
    the $x$ and $y$ axis of the plots represents the PCA variables
    in same scale but with arbitrary units. The
    cost values (shown with the colormap) are obtained using the
    definitions in
    Section~\ref{sec:classical-training}. First column shows cost
    values of an $L=2$ standard ansatz for increasing number of
    qubits. Second column shows the results for the same ansatz but
    with $L=N$ layers. As, expected the landscape flattens with more
    qubits and we see a single color for $N>12$. Third column shows
    results for splitting (for $m=4$) of the ansatz in the case of
    $L=N$ layers. We see that the landscape does not become flatter
    with more qubits.}
    \label{fig:cost_landscape}
\end{figure*}

\newpage
\section{}
\label{app:classical-dataset}

\begin{figure*}[!h]
    \centering
    \includegraphics[width=.9\linewidth]{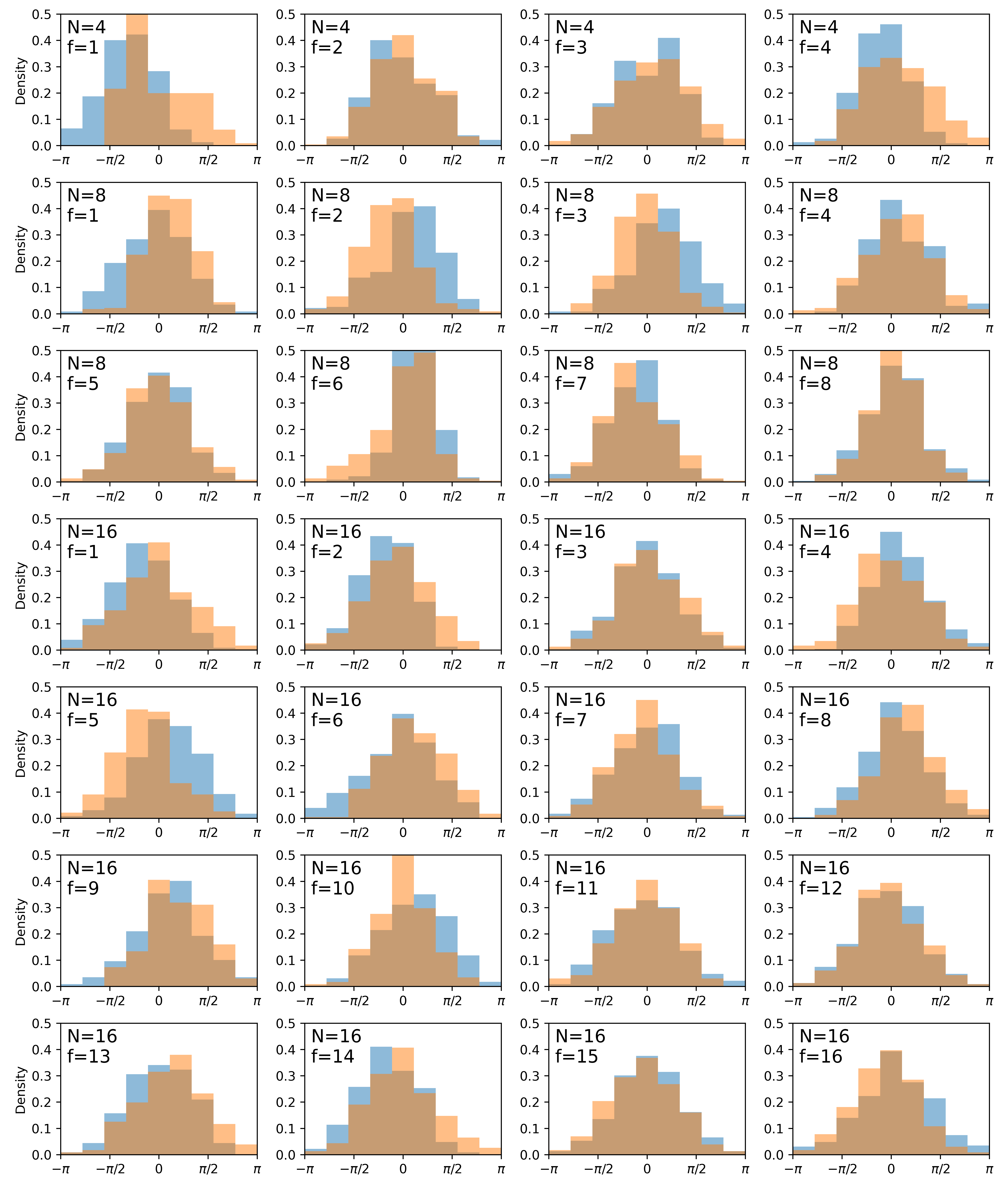}
    \caption{Distributions of the ad-hoc dataset used in Section \ref{sec:classical-training}. Each panel shows distribution of a single feature from one of three datasets.
    $N$ denotes the size of the dataset (number of features), while $f$ denotes the feature number. There exists 600 samples of $N$ features for a size $N$ dataset. Colors represent two classes. During training, data samples are divided with a 420/180 train/test ratio.  The dataset is produced using make\_classification function of scikit-learn \cite{pedregosa_scikit-learn_2011} with a class separation value of 1.0, 2\% class assignment error and no redundant or repeated features.}
    \label{fig:classical_dataset}
\end{figure*}

\newpage
\section{}
\label{app:classical-training}

\begin{figure*}[!h]
    \centering
    \includegraphics[width=\linewidth]{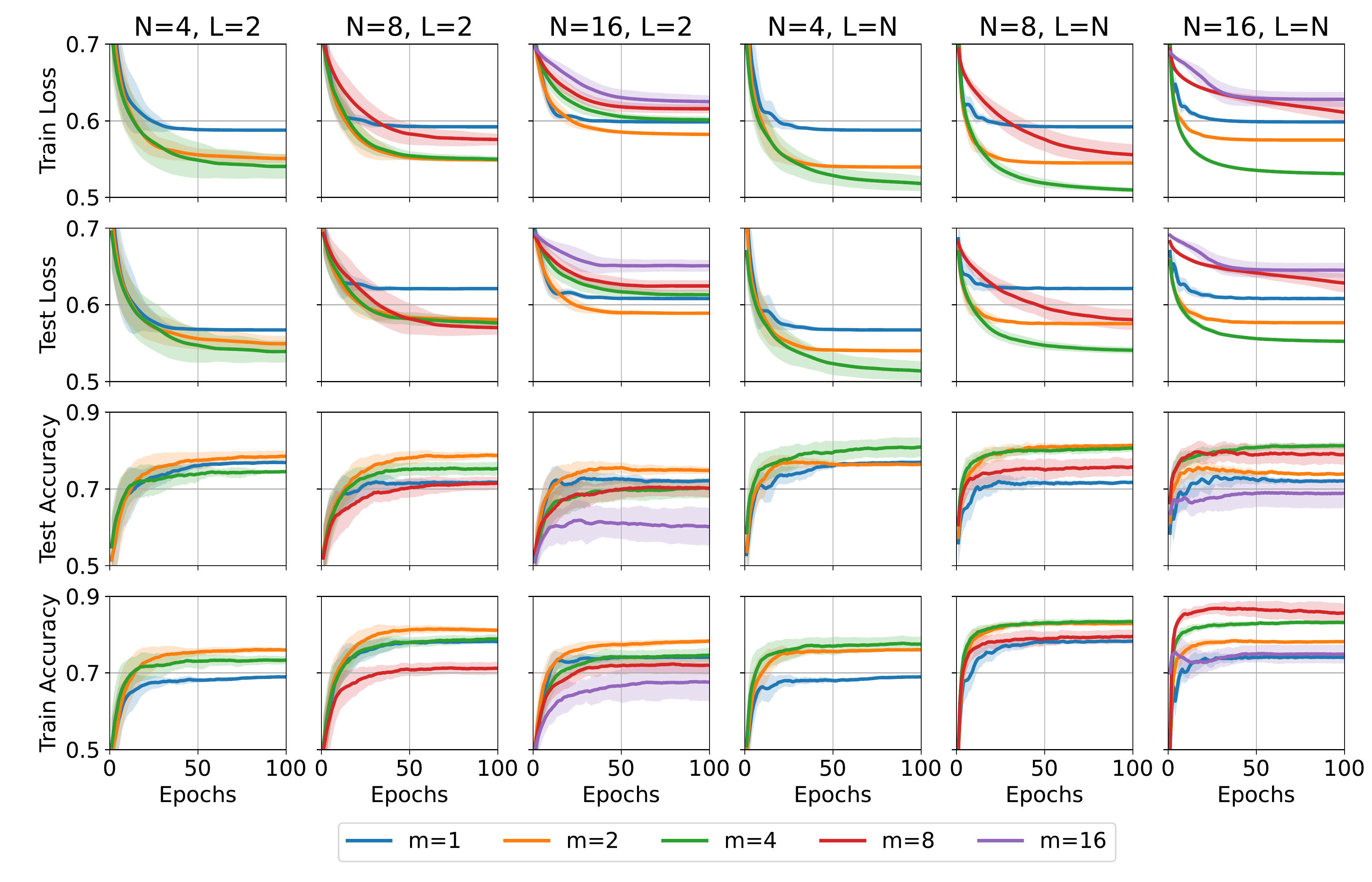}
    \caption{Training curves showing four different metrics for the
    problem described in Section~\ref{sec:classical-training}.
    Panels of each row show a different metric. First three columns
    show training results from $L=2$ ans\"atze, the last three
    columns show training results from $L=N$ ans\"atze for
    $N\in$\{4,8,16\}. Each value of $m$ is plotted with a different
    color. Lines are obtained by averaging 50 runs and their
    standard deviation is shown with shades.}
    \label{fig:training_curves_classical}
\end{figure*}

\begin{figure*}[!h]
    \centering
    \includegraphics[width=.95\linewidth]{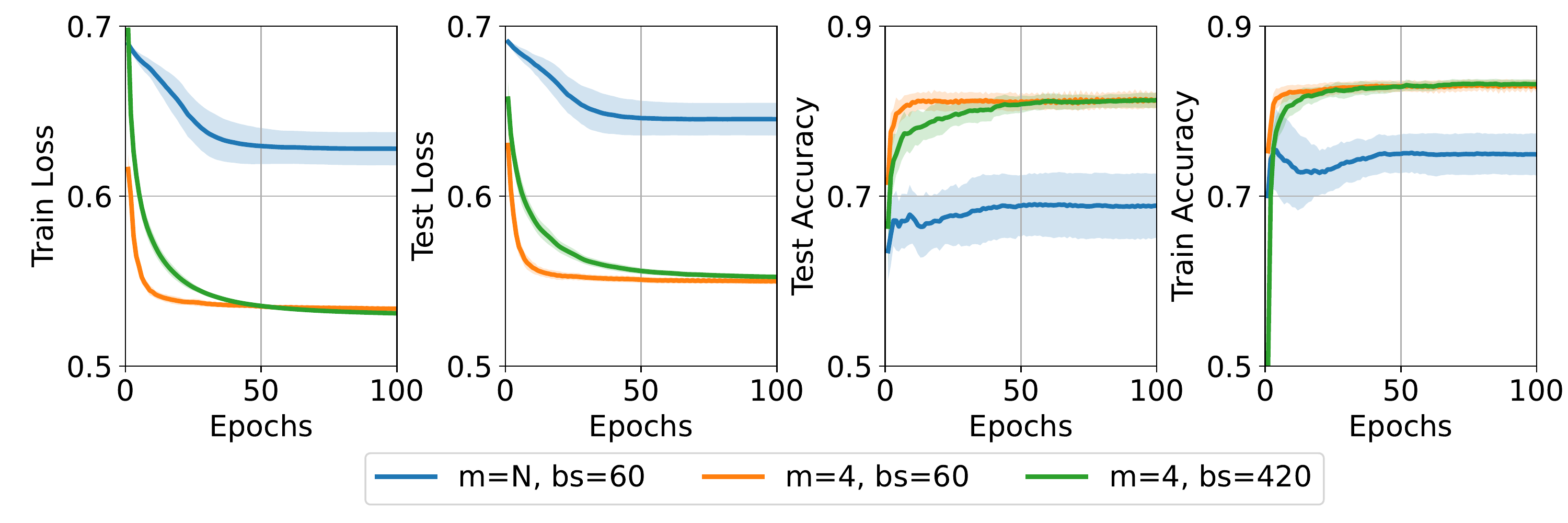}
    \caption{Batch size comparison for the training of $N=16$, $m=16$
    and $m=4$. Training the $N=L=16$ model requires vast computational
    resources, especially memory. This restricted us from using a
    full batch size during the training of $N=m=L=16$ setting.
    Therefore, we presented results from a training that used a
    batch size of 60 instead of 420 (full). Here, we show training
    curves for $m=4$ on addition to $m=16$ for two different batch size
    (bs). Behaviour of the curves show that the gain in performance
    has nothing to do with the batch size difference.}
    \label{fig:batch_size_comparison}
\end{figure*}

\newpage
\section{}
\label{app:quantum-dataset}
\begin{figure*}[!h]
    \centering
    \includegraphics[width=\linewidth]{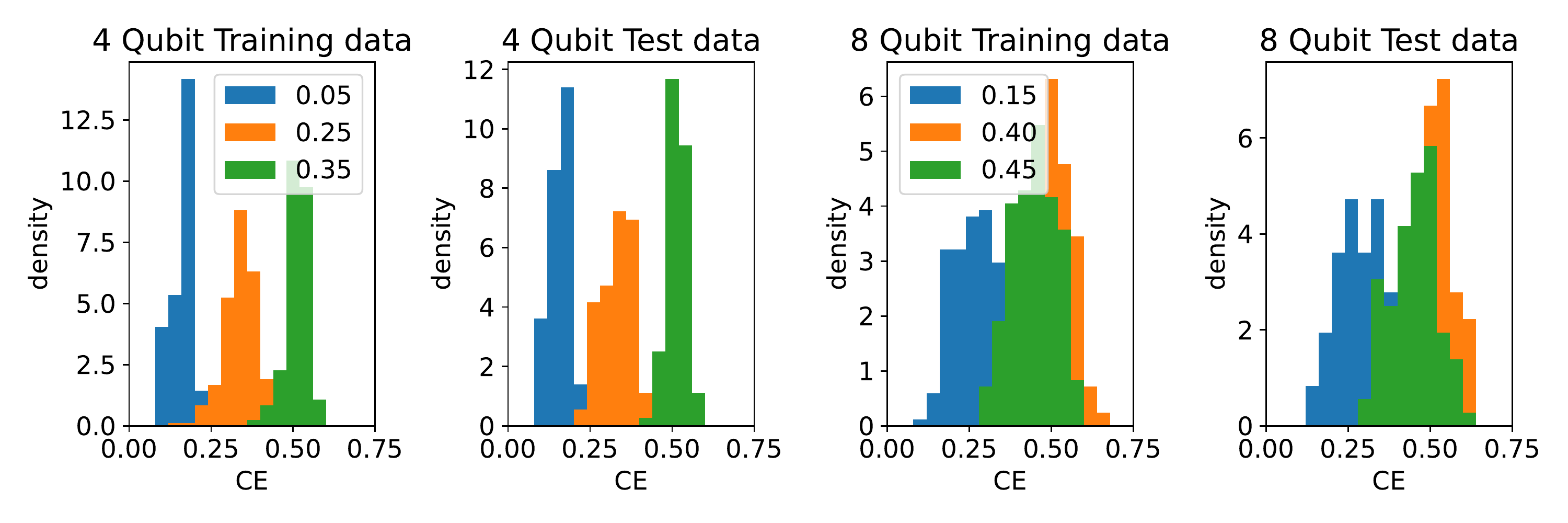}
    \caption{Distributions of the
    NTangled~\cite{schatzki_entangled_2021} dataset with respect to
    the $CE$ values described in Section \ref{sec:ntangled-training}.
    The HEA ansatz (Fig.~\ref{fig:all-ansatze}\textcolor{red}{d}) is used to produce the
    distributions. Each training set has 420 and each test set has
    180 data samples. We see a mismatch for $CE\in$\{0.40,0.45\} in
    the 8 qubit case. We are not sure what causes this, but it is
    not an issue for our problem as we are not interested in the
    $CE$ values themselves but the quantum states as a whole. So,
    they are valid quantum state distributions as long as they can
    be separated with a given metric for our problem. Our results
    show that this is in fact true.}
    \label{fig:quantum_dataset}
\end{figure*}

\begin{figure*}[!h]
    \centering
    \includegraphics[width=.85\linewidth]{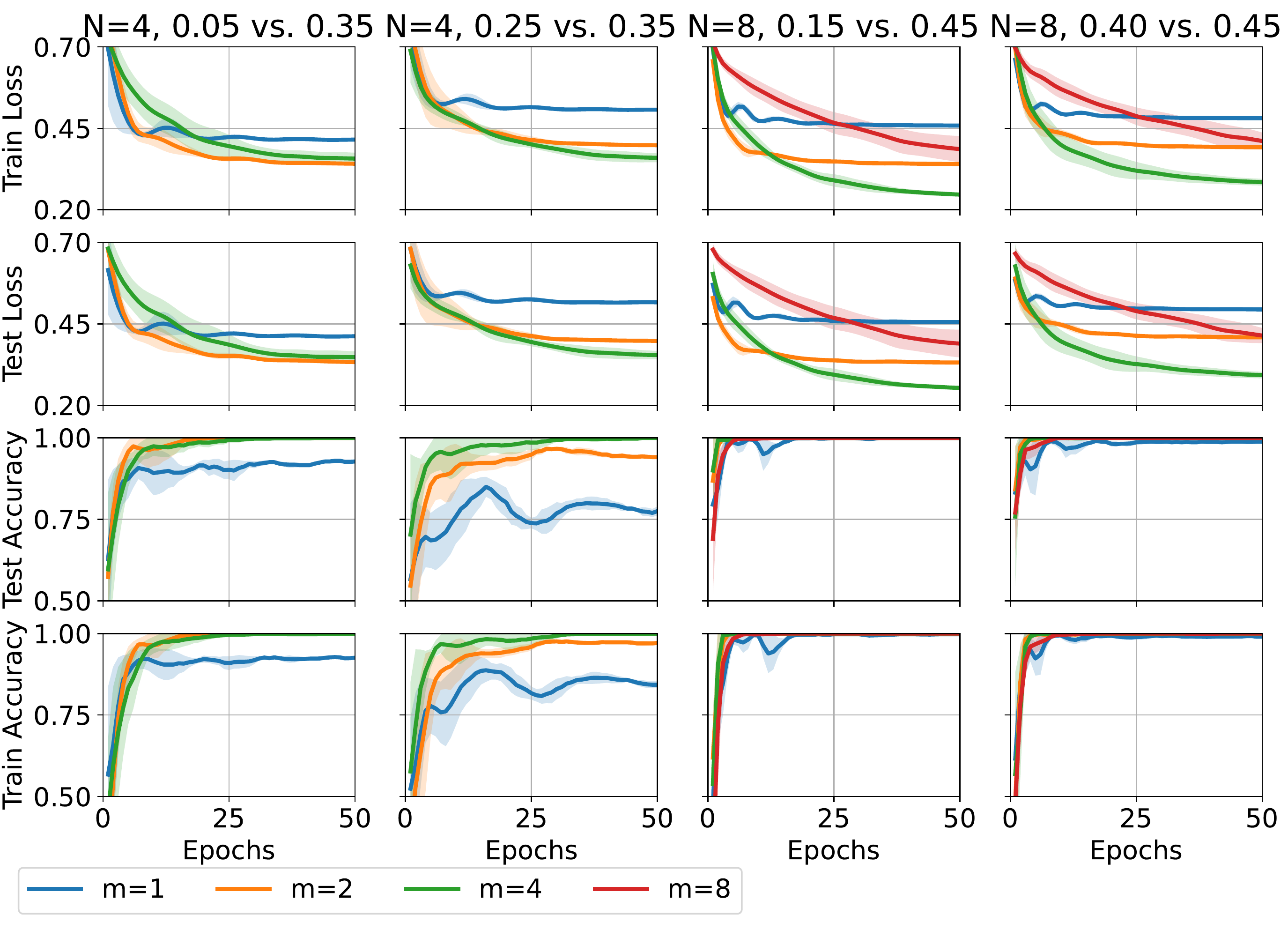}
    \caption{Training curves showing four different metrics for the
    problem described in Section~\ref{sec:ntangled-training}.
    Panels of each row show a different metric. Each column
    presents a different task, where $N$ determines the problem
    size and the $CE$ values are the labels of the classes. Each value of m is plotted with a different
    color. Lines are obtained by averaging 50 runs and their
    standard deviation is shown with shades.}
    \label{fig:training_curves_ntangled}
\end{figure*}

\newpage
\section{}
\label{app:vqe-ansatz-bp}

\begin{figure*}[!h]
    \centering
    \includegraphics[width=.45\linewidth]{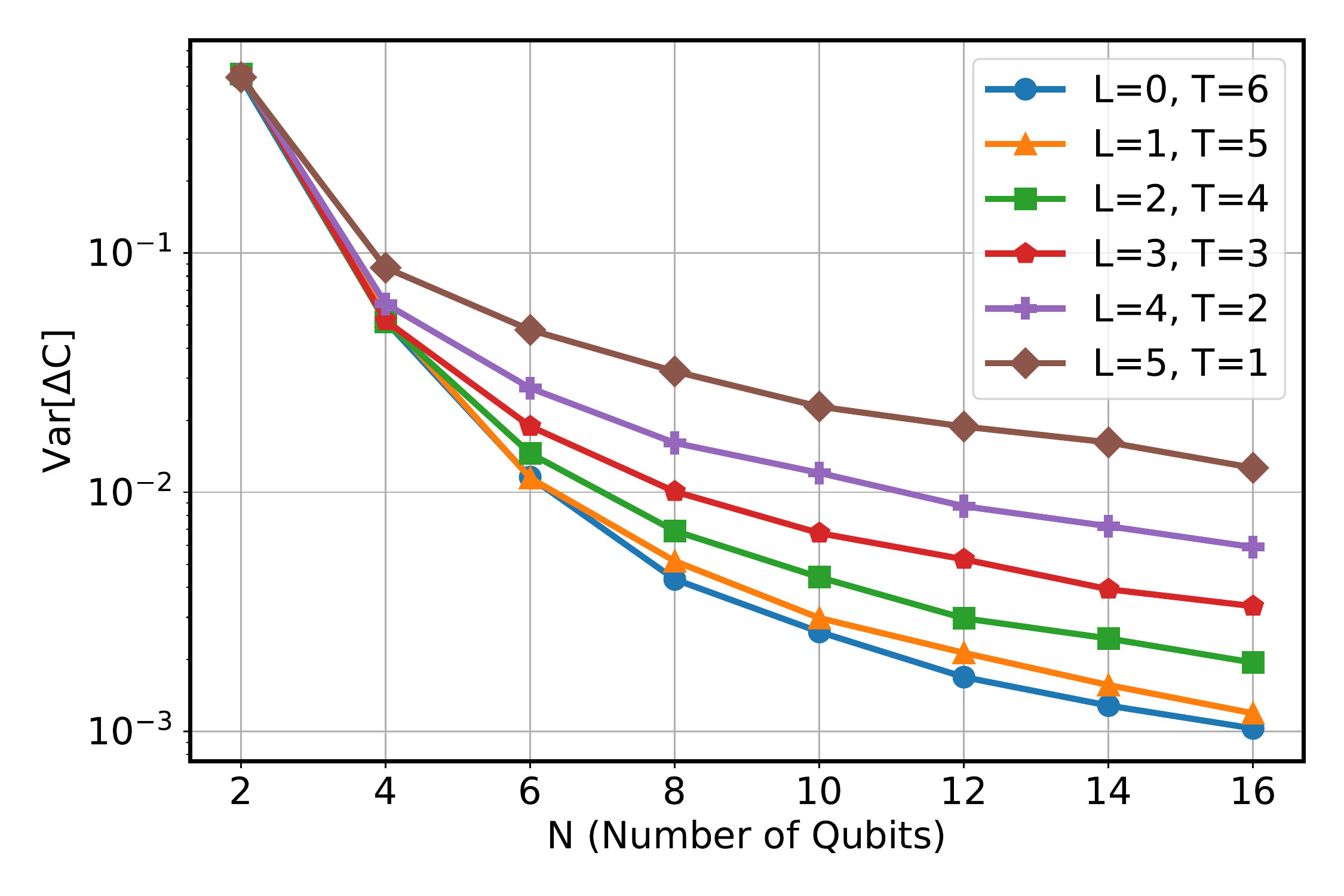}%
    \includegraphics[width=.45\linewidth]{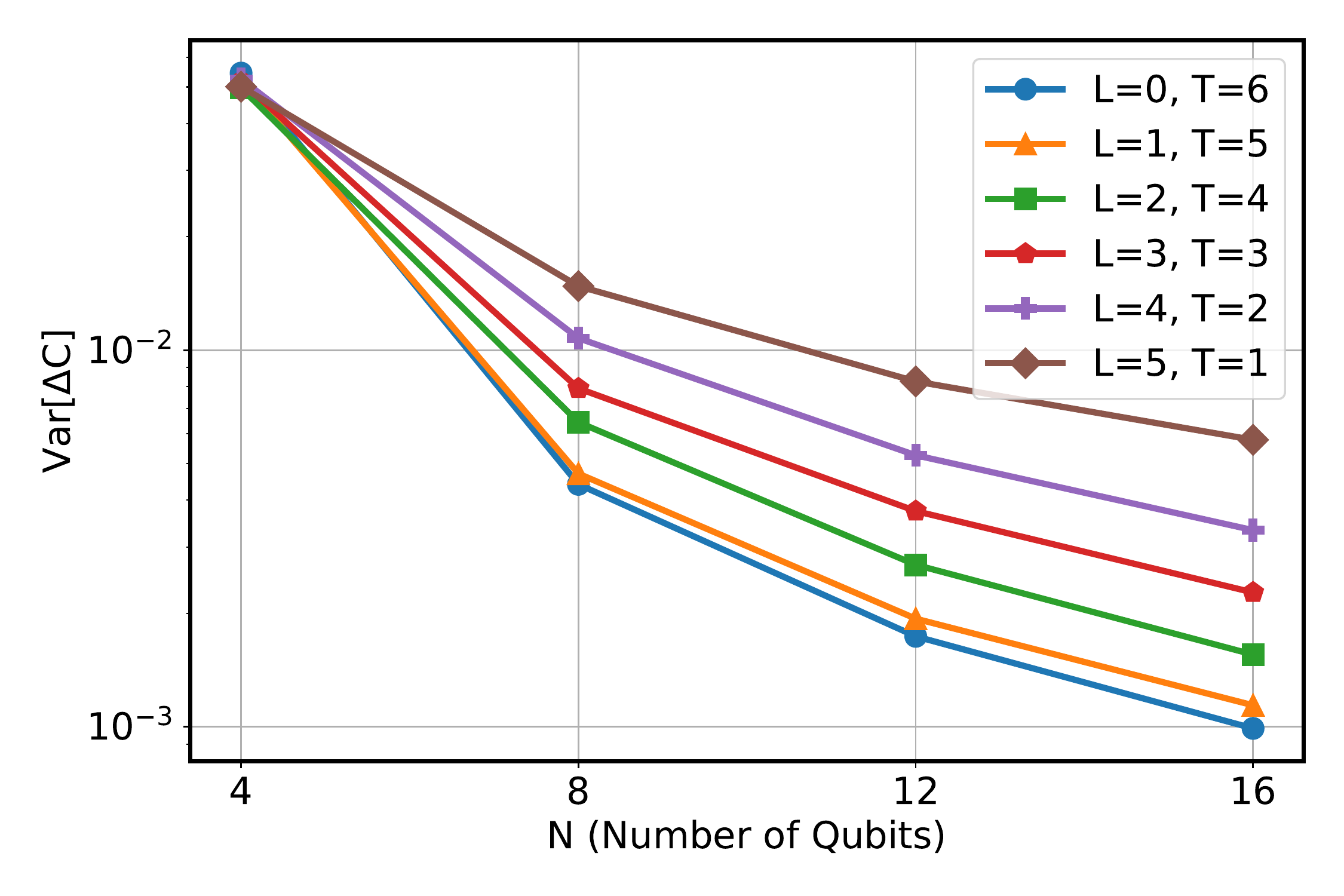}
    \caption{The variance of the change in cost as a function of the number of qubits for varying values of $L$ and $T$ for $L+T=D=6$. The cost function is the TFIH hamiltonian defined in Eq.~\ref{eq:ising-hamiltonian} and the ansatz is the ECS ansatz with EfficientSU2 sublocks (see Fig.~\ref{fig:all-ansatze}). Each line depicts a different value of $T$. The left panels show results for $m=2$ and the right panel shows results for $m=4$. Variances are obtained over 2000 cost samples, where $\Delta C$ is the difference of any arbitrary two cost samples.}
    \label{fig:vqe-ansatz-bp}
\end{figure*}

\section{}
\label{app:vqe-m2}
\begin{figure*}[!h]
    \centering
    \includegraphics[width=.6\linewidth]{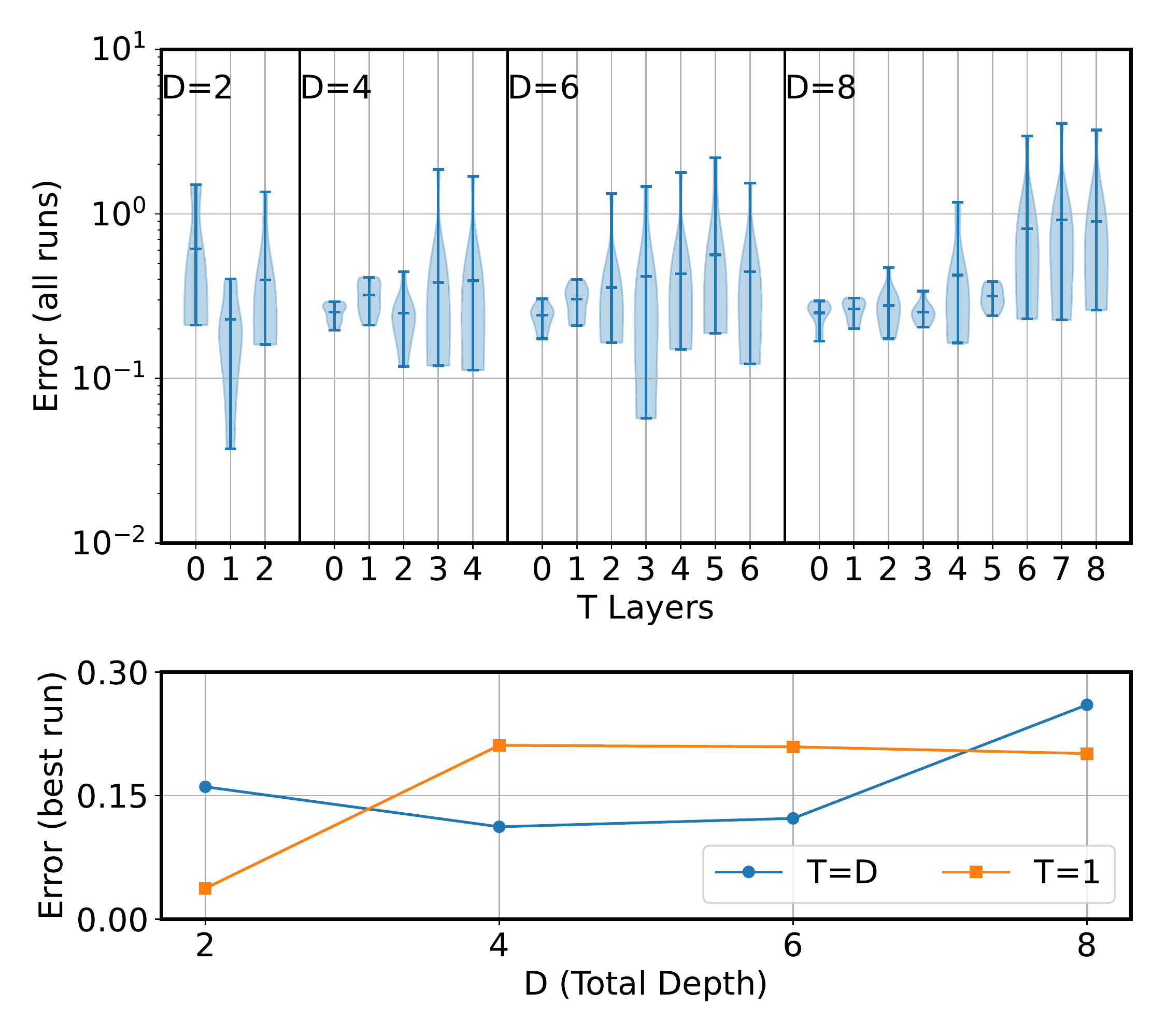}
    \caption{Energy errors of ans\"atze with increasing total depth for $N=12$ TFIH using the
    extended classical splitting (ECS) ansatz with EfficientSU2 sublocks (see
    Fig.~\ref{fig:all-ansatze}\textcolor{red}{b}) and $m=2$. Total depth (D) corresponds to
    L+T, where $T=0$ is equivalent to the CS ansatz, $T=D$ is equivalent to standard
    EfficientSU2 and other values explore hybrid use cases of the ECS ansatz. Final energy
    measurements of 10 runs are averaged and plotted with their minimum and maximum values
    as the error bars on the upper panel. The lower panel shows the best errors obtained
    for $T=1$ and $T=D$ settings. Energy error is the absolute difference of the energy measurement and the exact ground state energy.}
    \label{fig:vqe-m2}
\end{figure*}

\newpage 
\section{}
\label{app:vqe-curves}

\begin{figure*}[!b]
    \centering
    \includegraphics[width=.40\linewidth]{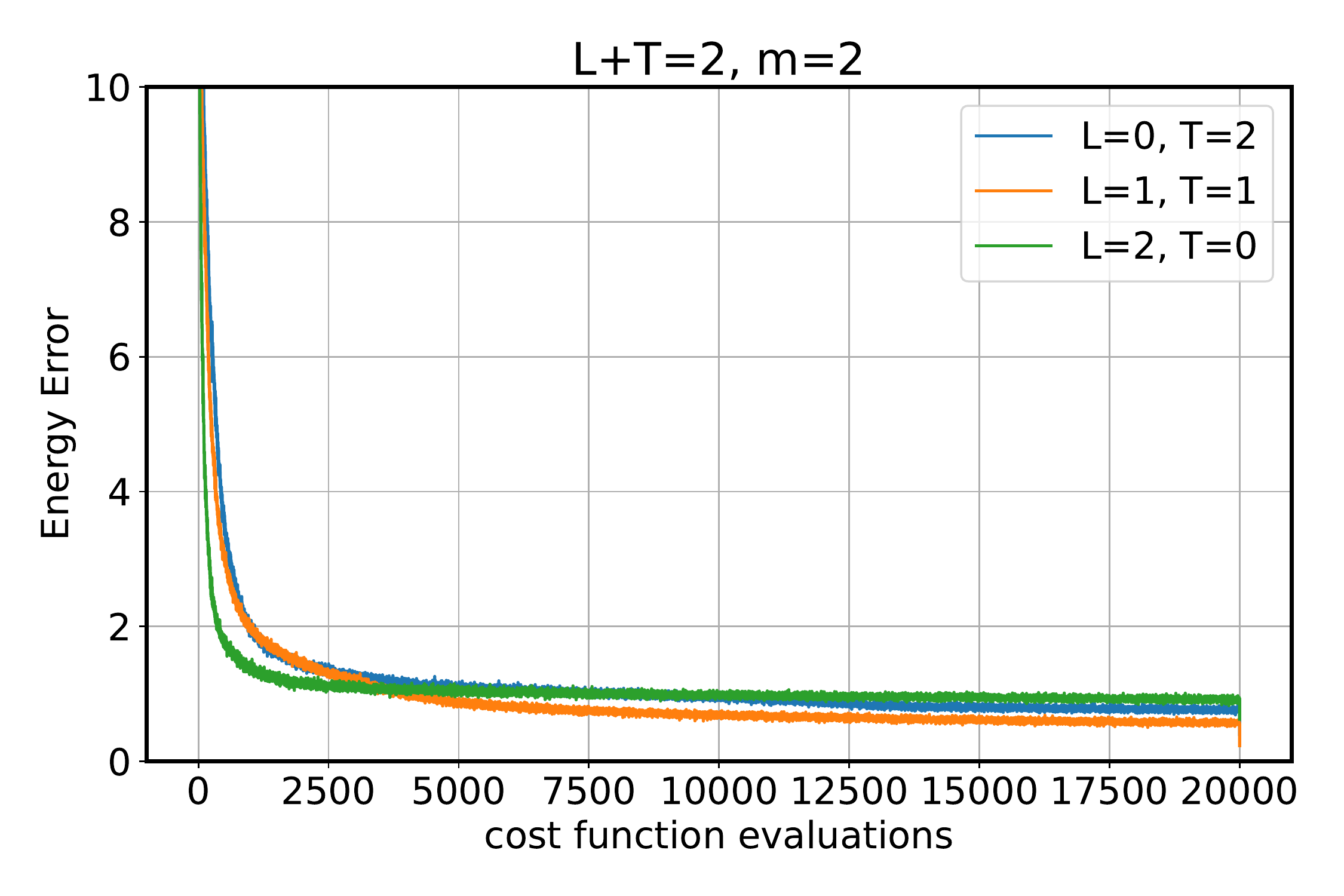}%
    \includegraphics[width=.40\linewidth]{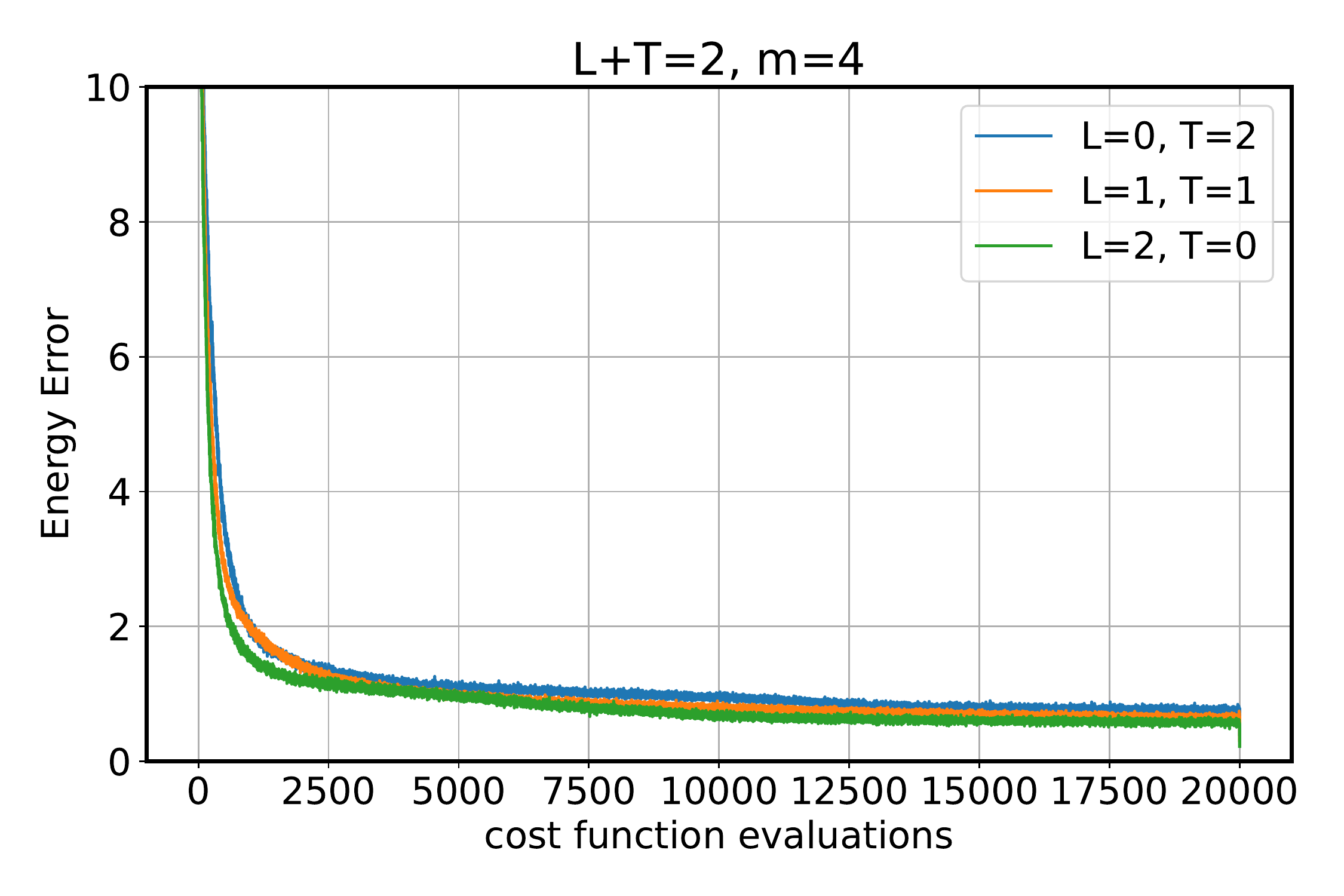}
    \includegraphics[width=.40\linewidth]{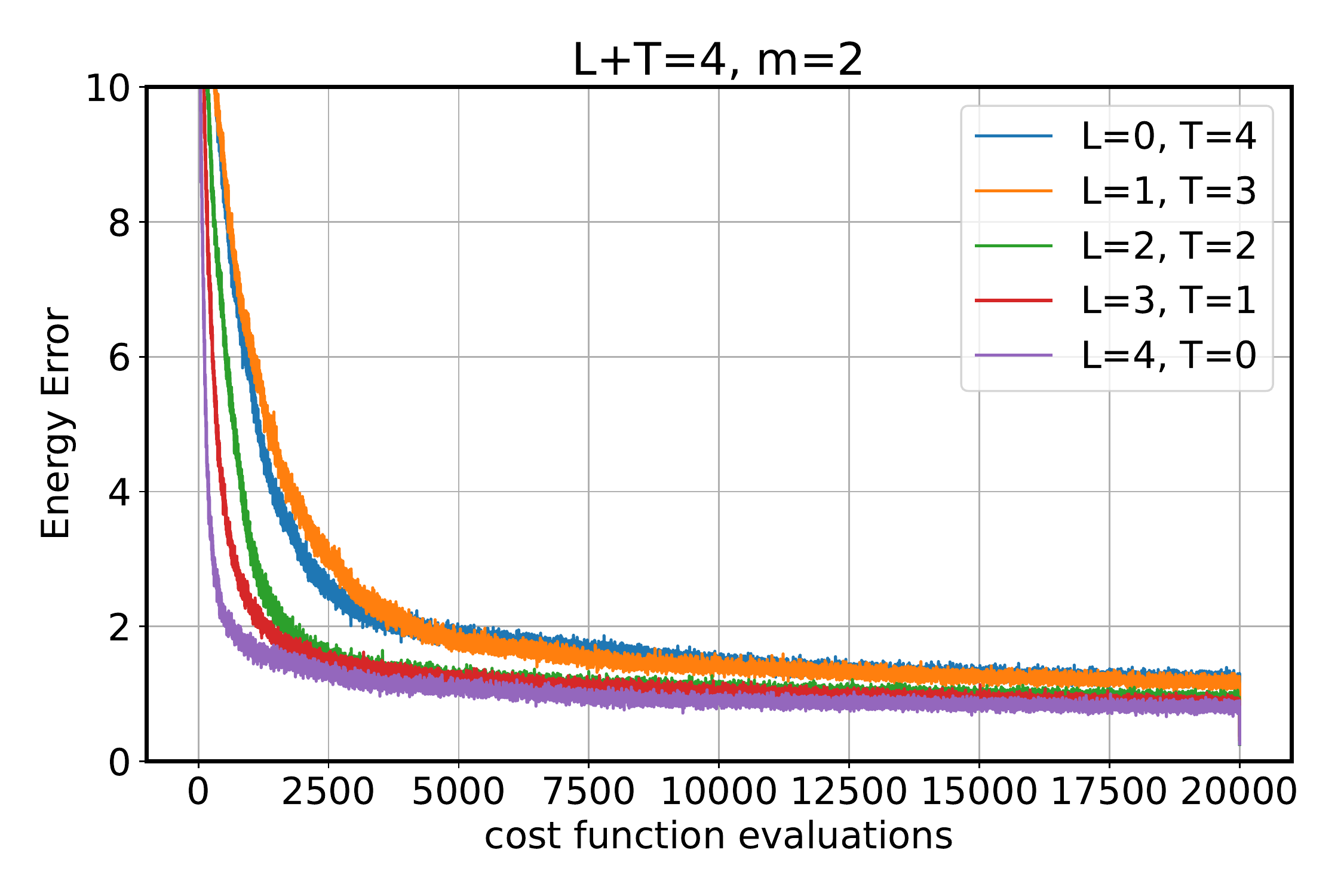}%
    \includegraphics[width=.40\linewidth]{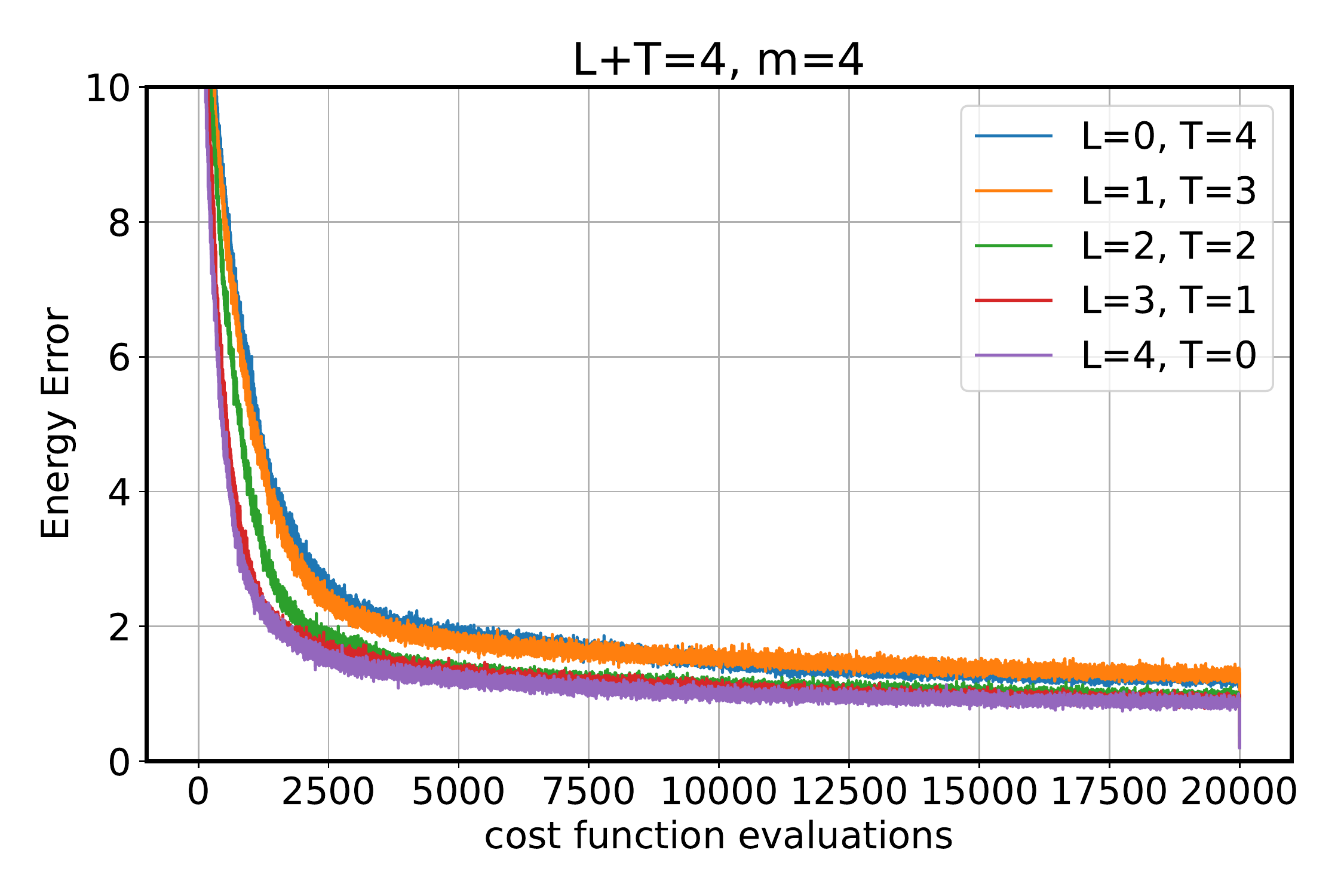}
    \includegraphics[width=.40\linewidth]{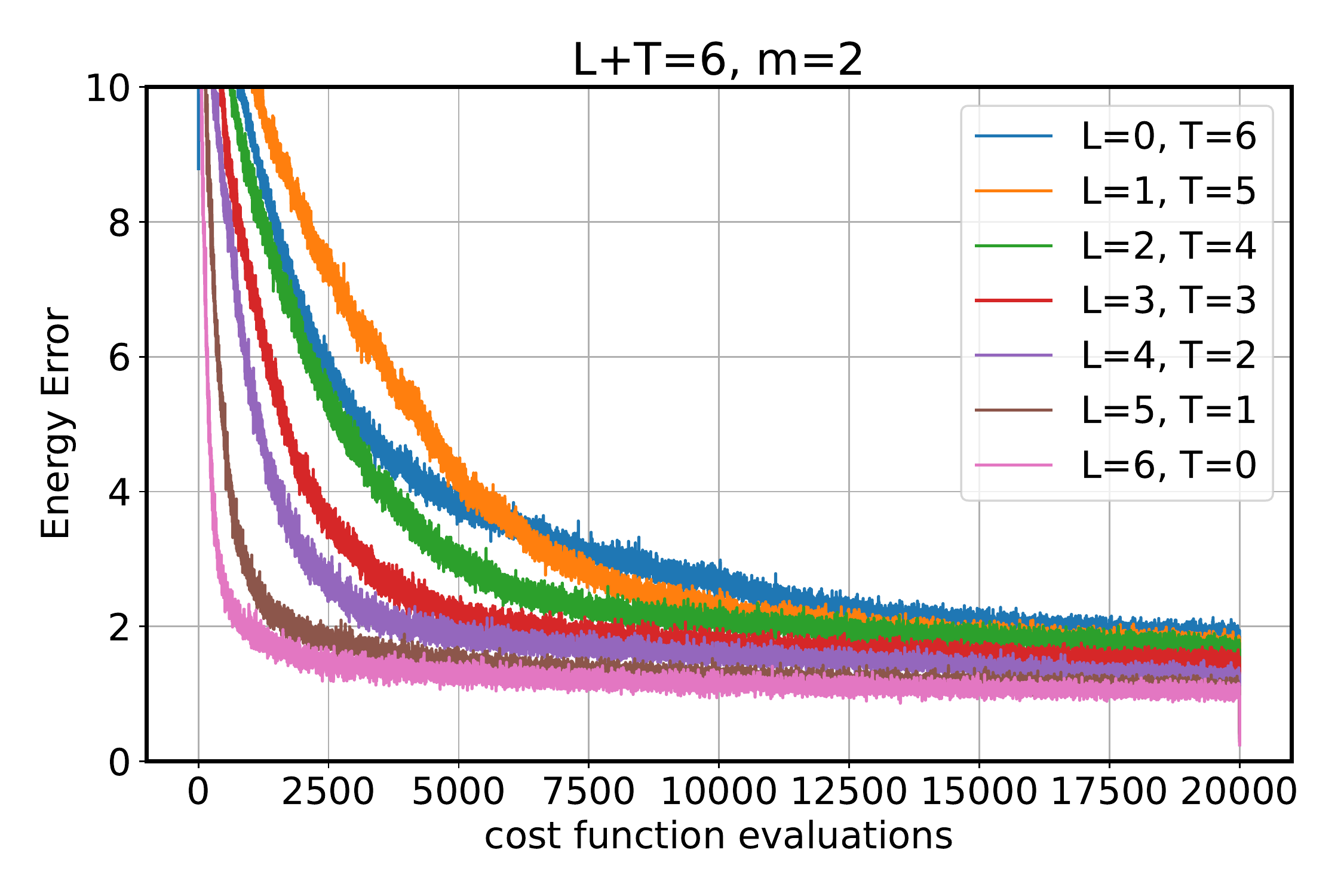}%
    \includegraphics[width=.40\linewidth]{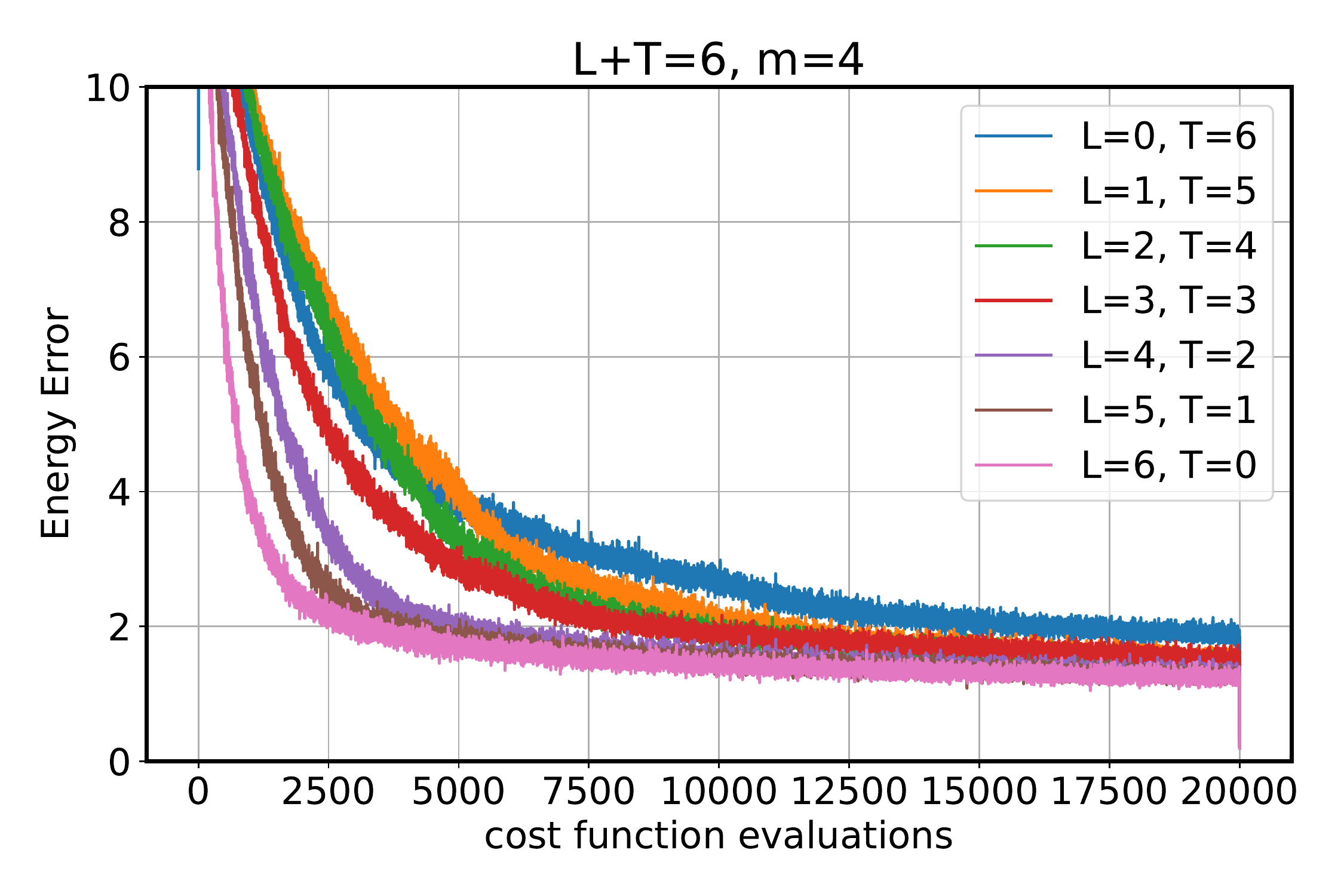}
    \includegraphics[width=.40\linewidth]{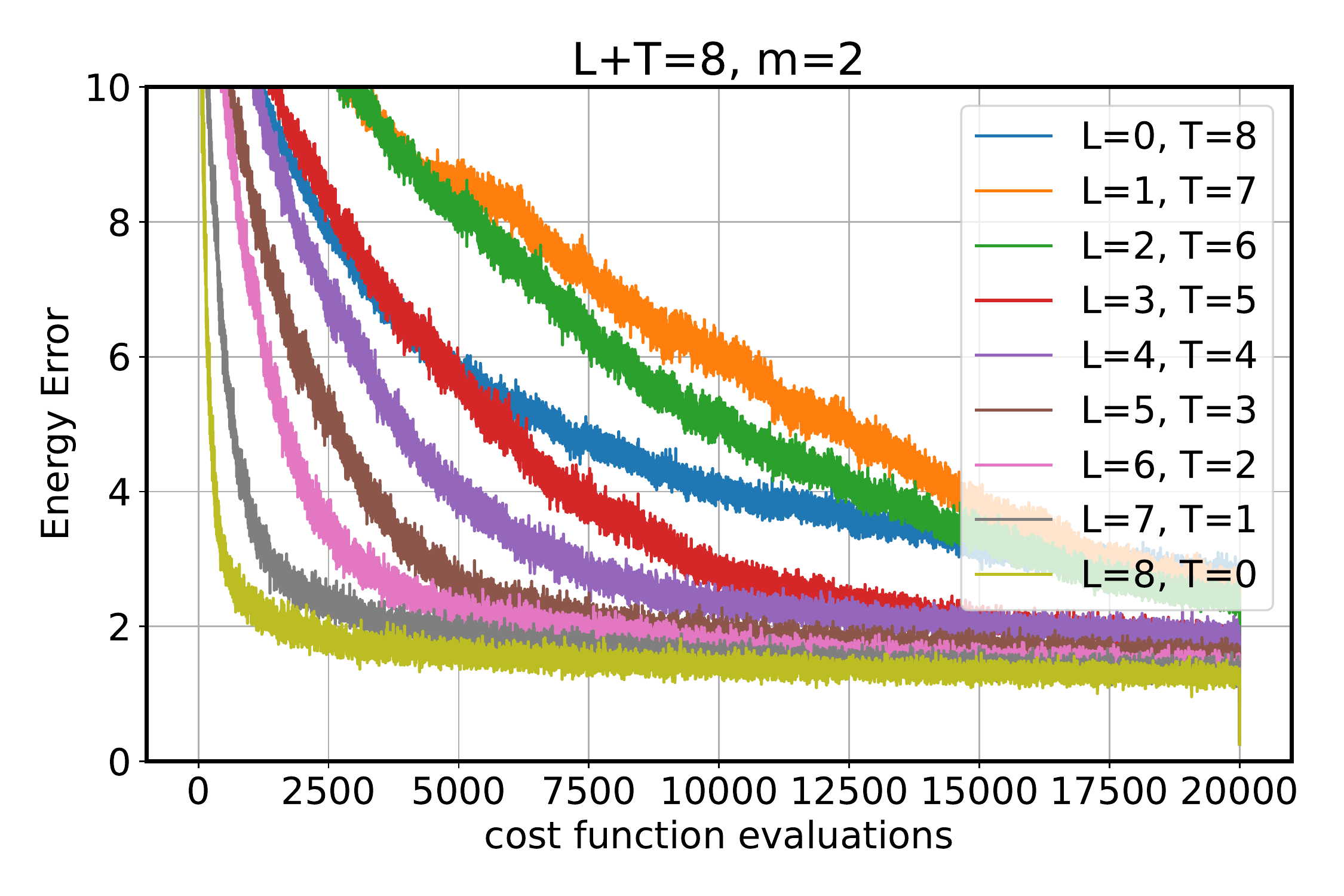}%
    \includegraphics[width=.40\linewidth]{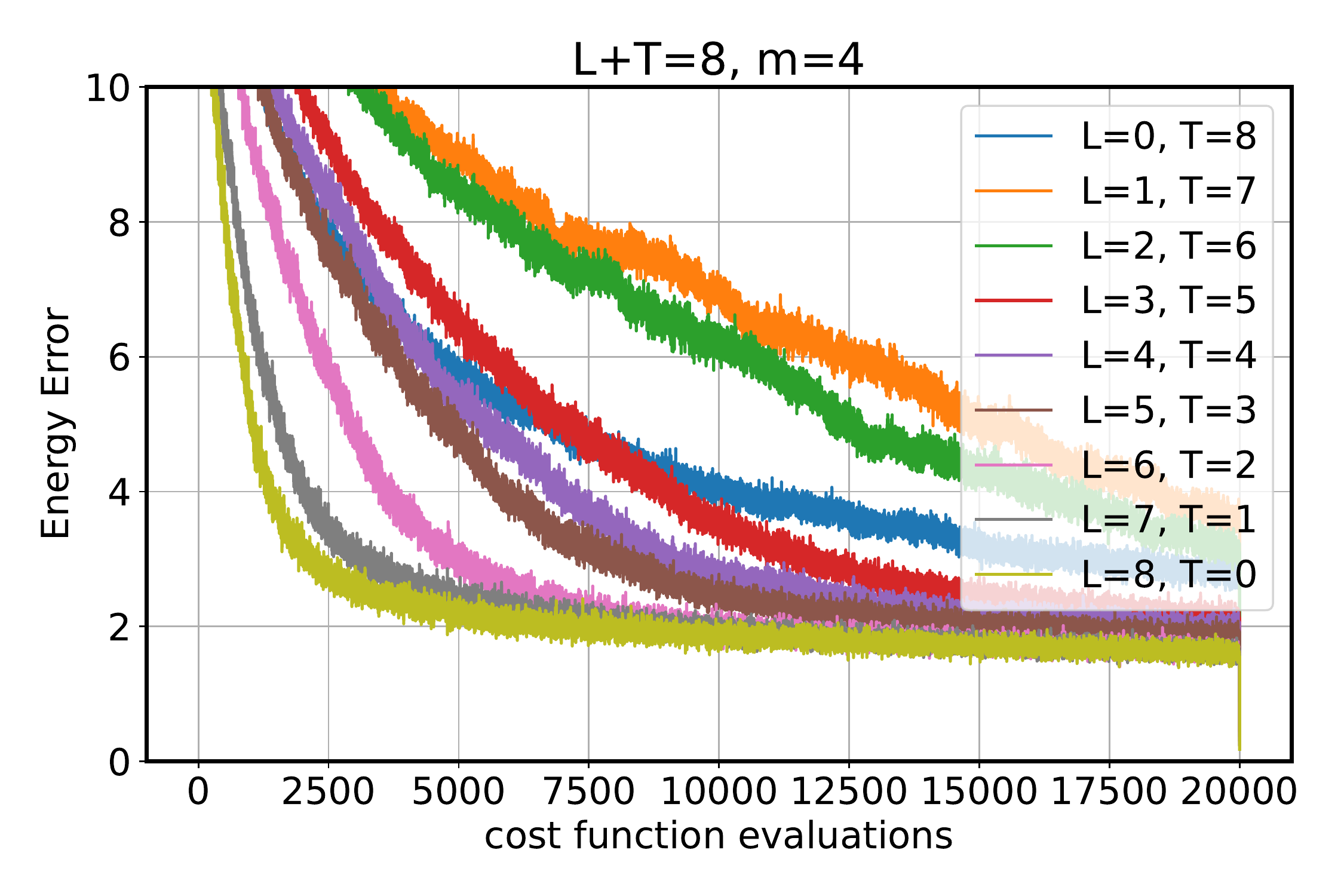}
    \caption{Energy error curves of ans\"atze TFIH for $N=12$ using the
    extended classical splitting (ECS) ansatz with EfficientSU2 sublocks (see
    Fig.~\ref{fig:all-ansatze}\textcolor{red}{b}). Columns corresponds to ans\"atze with $m=2$ and $m=4$ respectively. Each row shows results
    with increasing total depth ($D$), such that $L+T=D$. Energy errors of 10 runs are averaged and their mean is presented. Energy error is the absolute difference of the energy measurement and the exact ground state energy. It becomes harder to optimize an ansatz with no classical splitting ($L=0$) as depth increases. However, we see that the optimization does not get as hard if we set $T$ to a small value, e.g. to 1, and employ classical splitting. We observe similar conclusions with $m=2$ and $m=4$.}
    \label{fig:vqe-curves}
\end{figure*}

\end{document}